\documentclass[aps,pre,superscriptaddress,reprint,longbibliography,floatfix]{revtex4-2}
\usepackage{graphicx}
\usepackage{amsmath}
\usepackage{amssymb}
\usepackage{bm}
\usepackage{color}
\usepackage{float}
\usepackage[utf8]{inputenc}
\usepackage[T1]{fontenc}
\usepackage{newunicodechar}
\usepackage{hyperref}
\usepackage{footmisc}
\DeclareMathOperator\Arg{Arg}
\DeclareMathOperator{\diag}{diag}

\newcommand{\Wint}{\Omega_{\rm int.}}

\begin{document}

\title{Collective behavior in the nonreciprocal multi-species Vicsek model}

\author{Chul-Ung Woo}
\affiliation{Department of Physics, University of Seoul, Seoul 02504, Korea}
\affiliation{Department of Theoretical Physics and Center for Biophysics, Saarland University, Saarbr\"ucken, Germany}
\author{Heiko Rieger}
\affiliation{Department of Theoretical Physics and Center for Biophysics, Saarland University, Saarbr\"ucken, Germany}
\author{Jae Dong Noh}
\affiliation{Department of Physics, University of Seoul, Seoul 02504, Korea}

\date{\today}

\begin{abstract}
We investigate collective behavior in a $Q$-species Vicsek model with a
nonreciprocal velocity alignment interaction. This system is characterized
by a constant phase shift $\alpha$ in the inter-species velocity alignment
rule. While the phase shift renders the interaction nonreciprocal, the
system is globally invariant under any permutations of particle species,
possessing Potts symmetry. The combination of Potts symmetry and
nonreciprocity gives rise to a rich phase diagram. The nonreciprocal phase
shift generates either counter-clockwise or clockwise chirality. Potts
symmetry can be broken spontaneously. Consequently, the system exhibits four
distinct phases:  A species-mixed chiral phase where particles perform
counter-clockwise chiral motion with quasi-long-range order, a species
separation phase where Potts symmetry is broken and species-separated
particles form vortex cells with clockwise chirality, a coexistence phase,
and a disordered phase{, for $0<\alpha<\pi$}. 
We derive a Boltzmann equation and a hydrodynamic equation describing the system in the continuum limit, and present analytic arguments for the emergence of chirality and species separation. 
\end{abstract}

\maketitle

\section{introduction}

Advances in active matter physics reveal that collective behavior of many-body systems is characterized not only by the nature of interactions among constituent particles but also by their self-propulsion~\cite{Marchetti.2013, Cates.2014, Bechinger.2016, Bowick.2022, Shaebani.2020, Chate.2020}. A comparison between the equilibrium $XY$ model in two dimensions~(2D)~\cite{Kosterlitz.1973} and the Vicsek model~\cite{Vicsek.1995} exemplifies the relevance of active motility. Both models possess continuous rotational symmetry. However, the $XY$ model cannot exhibit long-range order due to the Mermin-Wagner theorem~\cite{Mermin.1966}, while the Vicsek model can. 

Recently, active systems with nonreciprocal interactions have attracted
growing interest. A nonreciprocal interaction, violating the action-reaction
principle, is ubiquitous in active matter systems. A bird in a flock can
follow a leader but not vice versa~\cite{Nagy.2010, Barberis.2016,
Chen.2017lo, Bastien.2020, Negi.2024}. In an ecological system, a predator tends to pursue a prey, but a prey tends to evade a predator~\cite{Tsyganov.2003}. In these examples, asymmetry in the leader-follower or prey-predator relationship is the origin of nonreciprocity. 

Fruchart {\it et al.} formulated a nonreciprocal active system model consisting of two species, A and B, of self-propelled particles with asymmetric roles~\cite{fruchart2021non}: A species particles~(pursuers) tend to align their velocity with B species particles while B species particles~(evaders) tend to anti-align with A species particles. It turned out that the system can resolve the dynamic frustration by forming a chiral state: both species particles perform chiral motion, counter-clockwise or clockwise, with a relative phase difference. 
{However, recently it has been shown that this non-reciprocal Vicsek
    model has a finite wave-length instability that drives it for large
system size into extensive spatio-temporal chaos~\cite{Woo.2026}.}
Nonreciprocal active systems have been investigated further in the context
of self-propelled particles with pursuer-evader type asymmetric
interactions\cite{fruchart2021non, Knezevic.2022, Maity.2023,
Kreienkamp.2024, Kreienkamp.2025, Duan.2025, Mangeat.2025} and with random
interactions~\cite{Lardet.2024, Choi.2025}, and in the context of continuous
field theories~\cite{Saha.2020, Dinelli.2023, Brauns.2024, Parkavousi.2025,
Saha.2025} involving an asymmetric coupling matrix. Experimental studies
have also been performed with robotic systems~\cite{Meredith.2020,
Chen.2024}. These studies have revealed interesting nonreciprocity-induced
phenomena such as run-and-chase states~\cite{Mangeat.2025, Saha.2025},
traveling waves~\cite{Dinelli.2023, Brauns.2024},
clustering~\cite{Kreienkamp.2024}, phase separation~\cite{Saha.2020},
and extensive spatio-temporal chaos~\cite{Woo.2026}.

More recently, we have proposed a multi-species Vicsek model with a
nonreciprocal alignment interaction~\cite{Woo.20258m}. In contrast to nonreciprocal systems built upon an asymmetric relationship among constituent particles, the model is fully symmetric under any permutations of particle species. Such a nonreciprocal yet symmetric model is established by introducing a phase shift into the Vicsek-type alignment interaction as shown in Eqs.~\eqref{eq:model_rule} and~\eqref{eq:alpha_matrix}. The model with $Q$ species possesses the permutation~($S_Q$) symmetry, or equivalently Potts symmetry of the $Q$-state Potts model~\cite{Wu.1982}, as well as continuous rotation symmetry. It was shown that nonreciprocity and Potts symmetry result in intriguing collective behaviors. 

In this paper, we present a detailed analysis of the nonreciprocal
$Q$-species Vicsek model with Potts symmetry. In Sec.~\ref{sec:model}, we
introduce the model consisting of $Q$ species particles subject to a
Vicsek-type velocity alignment interaction. The model assumes a constant
phase shift in the inter-species velocity alignment interaction. We justify
the phase shift by assuming a time delay in signal transformation processes
between particles of different species. In Sec.~\ref{sec:ContinuumTheory},
we derive the Boltzmann equation and the hydrodynamic equation describing
the particle-based model in terms of continuum fields for species-dependent
particle density, polarization, and so on. A mean-field treatment, assuming
spatial homogeneity, predicts that the nonreciprocal phase shift gives rise
to  chirality. In Sec.~\ref{sec:phase_diagram}, we present extensive
numerical simulation results for $Q=2, \cdots, 6$. These numerical studies
reveal that the system displays a disordered phase, a chiral phase with
quasi-long-range-order, a species-separation phase, and a coexistence phase.
In the chiral phase, particles of all species are mixed and perform
counter-clockwise chiral motion. Surprisingly, a correlation function of the
local polarization follows a power-law decay indicating quasi-long-range
order.
We present an argument that the quasi-long-range order in the chiral phase has the same origin as the equilibrium $XY$ model in 2D. In the species-separation phase, Potts symmetry is broken spontaneously and particles of different species unmix. We demonstrate numerically that an effective repulsion between different species emerges from the nonreciprocal phase shift. We also present theoretical evidence for the repulsion from a perturbative analysis of the hydrodynamic equation derived in Sec.~\ref{sec:hydrodynamics}. Finally, we conclude the paper in Sec.~\ref{sec:discussion}.

\section{Nonreciprocal multi-species Vicsek model}\label{sec:model}

To analyze the emergence and consequences of nonreciprocity and $S_Q$ symmetry~(or Potts symmetry) in active  systems, we formulate a generic model of self-propelled particles with an internal phase degree of freedom. Particles tend to align with their neighbors based on {\em time-delayed} information about the state of these internal phases. Similar systems of agents that couple their motility with internal degrees of freedom have been considered in the past \cite{Tanaka.2007, OKeeffe.2017, Hong.2023}.

Concretely, we consider a $Q$-species ensemble of self-propelled particles in square boxes of size $L^2$ in two dimensions. Each species has the same population of $N_0 = \rho_0 L^2$ particles. The total number of particles is denoted as $N_{\rm tot.} = QN_0$ with $\rho_{\rm tot.} = Q\rho_0$. Each particle, indexed by $n=1, \cdots, N_{\rm tot.}$, is characterized by its position $\bm{r}_n=(x_n, y_n)$, its direction of motion $\hat{\bm{e}}(\theta_n) = (\cos\theta_n,\sin\theta_n)$ with polar angle $\theta_n\in(-\pi,\pi]$, a species index (or `spin') $s_n =1,\cdots, Q$, and a phase $\psi_n$ as an internal degree of freedom. This phase has an intrinsic eigen-frequency $\Wint$ and is subject to an alignment interaction.

{
Adopting alignment rules of the continuous-time Vicsek model, or the active Brownian particle~(ABP) model~\cite{Romanczuk.2012,Martin-Gomez.2018}, the particles update their phases based on time-delayed information about the phases of their neighbors and an additional noise:
\begin{equation}\label{eq:dot_psi}
\dot\psi_n(t) = \Wint
-J\sum_{m\in \mathcal{N}_n}
\sin[\psi_n(t)-\psi_m(t-\tau_{nm})]+ \xi_n(t) ,
\end{equation}
where $\mathcal{N}_n$ denotes the set of particles within a circle with radius $r_0$  around particle $n$, and $\xi_n(t)$ is Gaussian delta-correlated noise. This kind of dynamics is readily realizable with programmable micro-robots, c.f. \cite{fruchart2021non}. 
Particles are self-propelled to a direction set by their polar angles: $\dot{\bm{r}}_n(t) = v_0 \hat{\bm{e}}(\theta_n(t))$ in  continuous-time dynamics or $\bm{r}_n(t+\Delta t) = \bm{r}_n(t) + v_0 \Delta t \hat{\bm{e}}(\theta_n(t))$ in discrete-time dynamics with a self-propulsion speed $v_0$.
Our model assumes that the internal phase relates to the polar angle via
\begin{equation}\label{eq:psi_theta}
    \theta_n(t) = \psi_n(t)-\Wint t .
\end{equation}
Without time delays ($\tau_{nm}=0)$ and without an intrinsic eigen-frequency ($\Wint=0$), this is identical to the ordinary active Brownian particle dynamics.
}

The internal phase degree of freedom can be eliminated to derive the equations of motion for the polar angle $\theta_n$. First, we note the Taylor expansion for $\psi_m(t-\tau_{nm})$: 
\begin{equation}
\psi_m(t-\tau_{nm}) = \psi_m(t)-\dot\psi_m(t)\tau_{nm} + O(\tau_{nm}^2)
\end{equation}
with $\dot\psi_m(t)=\Wint+\dot\theta_m(t)$.
For a large eigen-frequency, $|\Wint|\gg|\dot\theta_m(t)|= O(J)$, we can approximate $\dot\psi_m(t) \approx \Wint$ to obtain 
\begin{equation} \label{eq:psi_tau_app}
\psi_m(t-\tau_{nm})\approx \psi_m(t) - \Wint \tau_{nm} .
\end{equation}
Plugging this into Eq.~\eqref{eq:dot_psi} and using $\psi_n(t) - \psi_m(t) = \theta_n(t)-\theta_m(t)$, we obtain
\begin{equation}\label{eq:model_rule_cont}
\begin{split}
    \dot\theta_n(t) &= -J\sum_{m\in \mathcal{N}_n}
    \sin[\theta_n(t)-\theta_m(t)-\alpha_{nm}]+  \xi_n(t)  \\
    \dot{\bm{r}_n}(t) &= v_0 \hat{\bm{e}}(\theta_n(t)) 
\end{split}
\end{equation}
with phase shifts
\begin{equation}\label{eq:alpha}
\alpha_{nm} = -\Wint \tau_{nm} \pmod{2\pi} .
\end{equation}
Reciprocity requires that $\sin(\theta_n-\theta_m-\alpha_{nm})=-\sin(\theta_m-\theta_n-\alpha_{mn})$. This is achieved only if the phase shifts are antisymmetric, i.e., $\alpha_{mn}=-\alpha_{nm}~\pmod{2\pi}$. Otherwise, the phase shift renders the alignment interaction nonreciprocal. 
{A phase-shifted interaction was studied in the context of the
synchronization problem in Ref.~\cite{Uchida.2009}.}

The discrete-time version of this dynamics, which we will consider throughout this paper for numerical works, is given by a conventional Vicsek model dynamics with phase shifts $\alpha_{nm}$:
\begin{equation}\label{eq:model_rule}
\begin{split}
    \theta_n(t+\Delta t) &= \Arg\left[ \sum_{m\in \mathcal{N}_n} e^{i(\theta_{m}(t)+\alpha_{nm})}\right] + \zeta_n(t)  ,\\
\bm{r}_n(t+\Delta t) &= \bm{r}_n(t)  + v_0 \Delta t \hat{\bm{e}}(\theta_n(t)),
\end{split}
\end{equation}
where $\zeta_n(t)$ is an independent random variable drawn from a uniform distribution on $[-\eta\pi, \eta\pi]$ with noise strength parameter $\eta$~\footnote{The discrete-time dynamics of Eq.~\eqref{eq:model_rule} can be derived from a discrete-time dynamics for the internal phase variable.}.
The interaction terms in Eqs.~\eqref{eq:model_rule_cont} and \eqref{eq:model_rule} enforce velocity alignment with a phase misfit $\alpha_{nm}$. We expect that  the continuous- and discrete-time models share the same qualitative characteristics, which we confirmed numerically.

The origin of time-delays $\tau_{nm}$ could be manifold. Here we assume that they are dominated by signal conversion/transformation processes rather than transmission speed (in which case the delays would become distance dependent~\cite{Tanaka.2007}). Moreover, here we focus on the case in which these processes are slower between particles of different species than between particles of the same species, but otherwise independent of the species index, which implies
\begin{equation}\label{eq:alpha_matrix}
    \alpha_{nm} = \alpha (1-\delta_{s_n s_m}) . 
\end{equation}
It suffices to consider the nonreciprocal phase shift in the range $0\leq \alpha \leq \pi$ due to global rotational symmetry in the polar angle and inversion symmetry under $(x,y)\to (x,-y)$.
It constitutes a minimal model for a multi-species flocking system with {\em nonreciprocal interaction} for $0<\alpha<\pi$. In this setting, the velocity alignment interaction among particles of different species is subject to a constant phase misfit $\alpha$: Particle $n$ tends to align with the {\em apparent} polar angle of neighboring particles, which is identical to the true angle for the same species, but shifted by $\alpha$ for different species. 
{We refer the readers to Fig.~1 of Ref.~\cite{Woo.20258m} for a
schematic illustration of the effect of the phase shift.}
This model will be called a $Q$-species nonreciprocal Vicsek model~($Q$-NRVM).

When $\alpha=0$, the model reduces to the original Vicsek model. When
$\alpha=\pi$, the dynamics favors alignment among particles of the same
species and anti-alignment among those of different species. The two-species
model with $\alpha=\pi$ was explored in Ref.~\cite{Chatterjee.2023}{:
t}he inter-species anti-alignment stabilizes a parallel flocking
state in which the two species self-organize into an alternating band
structure flowing in the same direction~{(parallel flocking~(PF) band state)}, 
and an anti-parallel flocking state in which the two species flow in the
opposite direction penetrating through each other~{(anti-parallel
flocking~(APF) state)}. 
{For $Q>2$, the system can resolve the anti-alignment frustration at
    $\alpha=\pi$ by
forming an APF state in which PF bands of $n$ species move in one
direction and PF bands of remaining $Q-n$ species move in the opposite
direction. In the present work, we focus on the nonreciprocal case with
$\alpha \neq 0, \pi$.}

The $Q$-NRVM possesses {\em permutation~($S_Q$) symmetry} or {\em Potts symmetry} of the $Q$-state Potts model~\cite{Wu.1982} since any permutation of the $Q$ particle species leaves the system invariant, meaning all particle species are equivalent. This permutation  symmetry is broken explicitly in other nonreciprocal systems based on prey-predator-type or pursuer-evader-type interactions~\cite{Saha.2020, fruchart2021non, Knezevic.2022, Bandini.2024, Dopierala.2025, Popli.2025, Ma.2025}. 

Finally, we note that the continuous-time dynamics for the polar angle, Eq.(\ref{eq:model_rule_cont}), resembles the equation of motion of the Kuramoto model~\cite{kuramoto1975lecture}. Flocking in the former corresponds to phase synchronization in the latter. The Kuramoto model includes quenched random noise instead of temporal noise, and the phase shift is absent. Sakaguchi and Kuramoto~\cite{Sakaguchi.1986} later introduced a constant phase shift term to the Kuramoto model. The phase shift term was shown to result in an asymmetric distribution of the angular velocity. Abrams {\it et al.}~\cite{Abrams.2008} and Pikovsky and Rosenblum~\cite{Pikovsky.2008} extended the Sakaguchi and Kuramoto model by introducing multi-species subpopulations of oscillators and found chimera states in which synchronized and desynchronized domains coexist~\cite{Kuramoto.2002}. In our model, the angle variables are coupled locally to the spatial degrees of freedom. A global coupling has also been studied recently~\cite{OKeeffe.2017, Hong.2023}.

\section{Continuum field theory}\label{sec:ContinuumTheory}

To explore potential collective states of our model, we derive a Boltzmann
equation and a hydrodynamic equation. Before proceeding, it is useful to
consider a naive mean-field limit where the alignment interaction is
infinite-ranged. Let $\Omega_\mu = \frac{1}{N_0} \sum_n e^{
i\theta_n}\delta_{\mu s_n} = A_\mu e^{i\Theta_\mu}$ be the polarization of
species $\mu$. When $A_1=\cdots=A_Q$, the discrete-time dynamics
Eq.~\eqref{eq:model_rule} yields an iterative map for the phase
{$\{\Theta_\mu\}$ at time $t$ to $\{\Theta_\mu'\}$ at time $t+\Delta
t$}:
\begin{equation}\label{eq:mean_field_map}
    \Theta'_\mu = \Arg \left[(1-e^{i\alpha})e^{i\Theta_\mu} + e^{i\alpha} \sum_{\nu=1}^Q e^{i\Theta_\nu}\right] .
\end{equation}
This map allows an {\em in-phase chiral} solution, in which all the phases are synchronized and advance by 
\begin{equation}
    \psi_{\rm in} = {\rm Arg}[1+(Q-1)e^{i\alpha}] > 0
\end{equation}
at each time step. In this state, particles perform {\em counter-clockwise chiral motion} with synchronized polar angles. 

The discrete map~\eqref{eq:mean_field_map} also allows an {\em out-of-phase chiral} solution with vanishing net polarization~($\sum_\nu e^{i\Theta_\nu} = 0$), in which the phase angles advance by 
\begin{equation}
    \psi_{\rm out} = {\rm Arg}(1-e^{i\alpha}) = -\frac{1}{2}(\pi-\alpha) < 0
\end{equation}
each time step. In this state, particles perform {\em clockwise chiral motion} with species-dependent polar angles. These mean-field considerations indicate that the nonreciprocal phase shift can generate chirality.

\subsection{Boltzmann equation}\label{app:Boltzmann}
We proceed further to derive a field theory based on the Boltzmann equation approach~\cite{bertinHydrodynamic2009, Peshkov.2014, mahault2018outstanding}. Let $f^{\mu}(\bm{r},\theta,t) := \left\langle \sum_n \delta(\bm{r}_n(t)-\bm{r}) \delta(\theta_n(t)-\theta) \delta_{\mu s_n}\right\rangle$ be the one-particle distribution function for species $\mu=1,\cdots, Q$. The Boltzmann equation approach assumes (i) a continuous-time dynamics incorporating  single particle diffusion at a rate $\lambda$, (ii)  an alignment interaction through binary collisions, and (iii) a factorization of multi-particles distribution function as a product of one-particle distribution functions, known as the molecular chaos assumption. 

The single-particle distribution function $f^\mu(\bm{r},\theta,t)$ is governed by the coupled equations
\begin{equation} \label{eq:Boltzmann}
\partial_t f^{\mu} = - v_0 \hat{\bm{e}}(\theta)\cdot \bm{\nabla} f^{\mu} + I_{\rm d}[f^\mu] + \sum_{\nu=1}^Q I_{\rm col}[f^{\mu}, f^\nu]. 
\end{equation}
The first term in the righthand side accounts for the self-propulsion. The operators in the second and the third terms, abbreviated to $I_{\rm d}^\mu=I_{\rm d}[f^\mu]$ and $I_{\rm col}^{\mu\nu} = I_{\rm col}[f^\mu,f^\nu]$, describe single-particle diffusions and  binary collisions, respectively. They are given by 
\begin{align}
    I_{\rm d}^\mu =& -\lambda f^\mu(\bm{r},\theta,t)  \nonumber \\
     & + \lambda \int_{-\pi}^\pi f^{\mu}(\bm{r},\theta',t)P_{\rm n}(\theta-\theta')d\theta'  \label{eq:kernel_dif} \\
    I_{\rm col}^{\mu\nu} =& - f^{\mu}(\bm{r},\theta,t) \int_{-\pi}^\pi K(\theta, \theta') f^{\nu}(\bm{r},\theta',t)d\theta' \nonumber \\
    &+\iint_{-\pi}^{\pi} \left[ f^\mu(\bm{r},\theta_1,t)f^\nu(\bm{r},\theta_2,t) K(\theta_1,\theta_2) \right. \nonumber \\
    & \left. \quad\quad\quad \times P_{\rm n}(\theta-\theta_1-\Theta_{\mu\nu}(\theta_1,\theta_2)) \right] d\theta_1 d\theta_2 . \label{eq:kernel_col} 
\end{align}
For simplicity, we adopt the Gaussian distribution
\begin{equation}\label{eq:noise_distribution}
P_{\rm n}(\phi) = \frac{1}{\sqrt{2\pi\eta^2}} \sum_{m=-\infty}^{\infty} e^{-(\phi-2\pi m)^2/(2\eta^2)}
\end{equation}
of mean zero and variance $\eta^2$ folded into the interval $[-\pi:\pi]$ for both the polar angle diffusion and the noise in the alignment interaction.
The scattering cross section $K(\theta_1,\theta_2)$ and the scattering angle $\Theta_{\mu\nu}(\theta_1,\theta_2)$ depend on the relative angle $\theta_2-\theta_1$. Hence we will regard them as functions of a single variable $\phi=\theta_2-\theta_1$. The scattering cross section is given by $K(\phi) = 4r_0 v_0 \left| \sin\frac{\phi}{2}\right|$~\cite{mahault2018outstanding}. The scattering angle for a reciprocal intra-species collision is given by $\Theta(\phi) = \phi/2$~\cite{mahault2018outstanding}. In contrast, upon a nonreciprocal inter-species collision, the scattering angle should be modified due to the phase shift $\alpha$. The scattering angle is written as $\Theta_{\mu\nu}(\phi) = \delta_{\mu\nu} \Theta_{\rm R}(\phi) + (1-\delta_{\mu\nu}) \Theta_{\rm NR}(\phi)$ with
\begin{equation}
\begin{split} \label{eq:scattering_angle}
    \Theta_{\rm R}(\phi) &:= \frac{\phi}{2} , \\
    \Theta_{\rm NR}(\phi) &:= \frac{(\phi+\alpha) \pmod{2\pi}}{2} .
\end{split}
\end{equation}

The Boltzmann equation can be cast into a dimensionless form by rescaling
$\bm{r} \rightarrow \frac{v_0}{\lambda}\bm{r}$, 
$t      \rightarrow \frac{1}{\lambda} t$, and
$f^\mu  \rightarrow  {\rho_0}f^\mu$.
In the dimensionless form, we can set $\lambda=v_0=\rho_0=1$ and the scattering cross section is replaced by 
\begin{equation}\label{eq:cross_section}
K(\phi) = 2\pi \kappa \left| \sin\frac{\phi}{2}\right| .
\end{equation}
with a dimensionless coupling constant
\begin{equation}\label{eq:couplingConstant}
    \kappa = \frac{2r_0 v_0 \rho_0}{\pi \lambda} .
\end{equation}

It is convenient to use the complex coordinates $z=x+iy$ and $z^*=x-iy$ instead of the Cartesian coordinates $x$ and $y$. In terms of the complex coordinates, the directional derivative becomes 
\begin{equation}\label{eq:complex_directional_deri}
\hat{\bm{e}}(\theta)\cdot \bm{\nabla} = (\cos\theta) \partial_x + (\sin\theta)\partial_y = (e^{i\theta}\partial_z + e^{-i\theta}\partial_{z^*})
\end{equation}
with $\partial_z = \frac{1}{2}(\partial_x - i \partial_y)$ and $\partial_{z*} = \frac{1}{2}(\partial_x + i \partial_y)$. 
We further expand the one-particle distribution function in terms of the Fourier or multipole modes
\begin{equation}\label{eq:Fourier_mode}
    f^{\mu}(\bm{r},\theta,t) = \frac{1}{2\pi} \sum_{k\in \mathbb{Z}} f^{\mu}_k(\bm{r},t) e^{-i k\theta} .
\end{equation}
These modes correspond to the local particle density $f^\mu_0(\bm{r},t) = \rho^{\mu}(\bm{r},t)$, the local complex polarization field $f^\mu_1(\bm{r},t) = \langle e^{i\theta}\rangle_{f^\mu} = \int d\theta e^{i\theta} f^\mu(\bm{r},\theta, t)$, and so on. The polarization field may be represented with a complex field $w^\mu({\bm r}, t) = m^\mu_x(\bm{r},t) + i m^\mu_y(\bm{r},t)$ or a vector field $\bm{m}^\mu(\bm{r},t) = (m^\mu_x(\bm{r},t), m^\mu_y(\bm{r},t))$ with $m^{\mu}_x({\bm r},t) = \langle \cos\theta\rangle_{f^\mu}$ and $m^\mu_y(\bm{r},t) = \langle \sin\theta \rangle_{f^{\mu}}$. 
Note that ${f}^\mu_{-1} = ({f}^\mu_1)^*$.

Plugging Eq.~\eqref{eq:Fourier_mode} into Eq.~\eqref{eq:Boltzmann},
one can derive the coupled equations for $f^{\mu}_k$. 
To handle the multi-species system with ease, we introduce a $Q$-dimensional {\em species space} and represent the multipole moments as a column vector
\begin{equation}
    \bm{f}_{k} = (f^1_k, \cdots, f^\mu_k, \cdots, f^{Q}_k)^T
\end{equation}
in the species space.
The interaction kernels between $k$th and $l$th modes can be written in a
compact form using  a $Q\times Q$ matrix $\bm{J}^{kl}$ whose matrix elements
are given by
\begin{equation}\label{eq:J_kl_mn}
    [\bm{J}^{kl}]_{\mu\nu} = \delta_{\mu\nu} J^{kl}_{\rm R}  + (1-\delta_{\mu\nu}) J^{kl}_{\rm NR}  .
\end{equation}
Here, $J^{kl}_{\rm R}$ and $J^{kl}_{\rm NR}$ denote intra-species reciprocal and inter-species nonreciprocal coupling constants, respectively~(see Sec.~\ref{sec:J_explicit} for their explicit expressions). We also introduce the Hadamard product $\circ$ which maps two species space vectors $\bm{a}$ and $\bm{b}$ to $\bm{c} = \bm{a}\circ \bm{b} = \bm{b} \circ \bm{a}$ where $c^\mu = a^\mu b^\mu$. The Boltzmann equation is then written in a compact form
\begin{equation}\label{eq:BoltzmannVector}
\begin{split}
    \partial_t \bm{f}_k =& - \partial_{z^*} \bm{f}_{k-1} - \partial_z
    \bm{f}_{k+1} \\
    & - (1-P_k) \bm{f}_k + \sum_{l\in\mathbb{Z}}
    {\bm{f}}_{k-l} \circ (\bm{J}^{kl} \bm{f}_l), 
    \end{split}
\end{equation}
where $P_k = e^{-k^2 \eta^2/2}$ is the $k$-th Fourier coefficient of the noise distribution in Eq.~\eqref{eq:noise_distribution}.
In particular, for $k=0$, Eq.~\eqref{eq:BoltzmannVector} reduces to a continuity equation
\begin{equation}\label{eq:continuum_eq}
    \partial_t \bm{f}_0 + \partial_{z^*} \bm{f}_{-1} + \partial_z \bm{f}_1 = 0, 
\end{equation}
or $\partial_t \rho^\mu + \bm{\nabla} \cdot \bm{m}^\mu = 0$. 

{It is instructive to summarize the symmetry property of the
    Boltzmann equation. Let $\bm{P}(P)$ be a $Q\times Q$ matrix corresponding to a
permutation $P:\{1,\cdots, Q\} \to \{1,\cdots, Q\}$. The species-species
interaction matrix $\bm{J}^{kl}$ in Eq.~\eqref{eq:J_kl_mn} has a symmetry
property 
$$ \bm{J}^{kl} = \bm{P}(P) \bm{J}^{kl} \bm{P}^\dagger(P) $$
for an arbitrary permutation $P$. Using this symmetry property, one can show that
$\bm{f}_k$ and $\tilde{\bm{f}}_k := \bm{P}\bm{f}_k$ satisfy the same
Boltzmann equation. Namely, the Boltzmann equation obeys Potts symmetry
of the microscopic model. 
The interaction matrix considered in Ref.~\cite{fruchart2021non}, however, includes
an antisymmetric part. Thus, the symmetry relation is broken explicitly. }

\subsection{{Mean field analysis}}\label{sec:lsa}
The Boltzmann equation has a trivial steady state solution
{$\bm{f}_k(\bm{r},t) = \bar{\bm{f}}_k := \delta_{k0} \bm{1}_Q$} with $\bm{1}_Q = (1,\cdots,1)^T$. This corresponds to a homogeneous disordered state. We inspect linear stability of this disordered state against a uniform temporal fluctuation $\bm{f}_k(t) = \bar{\bm{f}}_k + \bm{\epsilon}_k(t)$. Expanding the Boltzmann equation with respect to $\bm{\epsilon}_k$, we obtain
\begin{equation}
    \dot{\bm{\epsilon}}_k = \sum_{l} \bm{M}^{kl} \bm{\epsilon}_l
\end{equation}
with the stability matrix
\begin{equation}\label{eq:Mlst_app}
    \bm{M}^{kl} = \delta_{kl} ((P_k-1)\bm{E}+\bm{J}^{kk} + \bm{D}(\bm{J}^{k0}\bm{1}_Q)),
\end{equation}
{where $\bm{E}$ denotes the identity matrix with elements $E_{\mu\nu} =
\delta_{\mu\nu}$}.
Here, we introduce an operator $\bm{D}(\cdot)$ mapping a species space vector $\bm{u} = (u^1,\cdots, u^Q)^T$ to a diagonal matrix $\bm{D}(\bm{u}) = \diag(u^1,\cdots, u^Q)$~\footnote{The Hadamard product $\bm{a}\circ \bm{b}$ can be written as $\bm{a} \circ \bm{b} = \bm{D}(\bm{a}) \bm{b}$}.
Since different Fourier modes do not couple, it suffices to consider each
mode $k$ separately with the stability matrix $\bm{M}^{(k)} = \bm{M}^{kk}$.
Noting that $\bm{D}(\bm{J}^{k0}\bm{1}_Q) = [J^{k0}_{\rm R} +
(Q-1)J^{k0}_{\rm NR}]\bm{E}$, we obtain  $[\bm{M}^{(k)}]_{\mu\nu} = D^{(k)}\delta_{\mu\nu} + O^{(k)}(1-\delta_{\mu\nu})$ with 
\begin{equation}\label{eq:Mlst_element_app}
\begin{split}
    D^{(k)} =& P_k -1 + J^{kk}_{\rm R} + J^{k0}_{\rm R} + (Q-1)J^{k0}_{\rm NR} , \\
    O^{(k)} =& J^{kk}_{\rm NR} .
\end{split}
\end{equation}
The matrix $\bm{M}^{(k)}$ has a nondegenerate complex eigenvalue $\Lambda_{\rm in}^{(k)} = D^{(k)}+(Q-1)O^{(k)}$ with an eigenvector $\bm{\lambda}_{\rm in} = \bm{1}_Q$~[in-phase mode] and a $(Q-1)$-fold degenerate complex eigenvalue $\Lambda_{\rm out}^{(k)} = D^{(k)}-O^{(k)}$ with eigenvectors $\bm{\lambda}_{\rm out} \perp \bm{\lambda}_{\rm in}$~[out-of-phase mode]. We focus on the polarization field with $k=1$.

\begin{figure}
\includegraphics[width=\columnwidth]{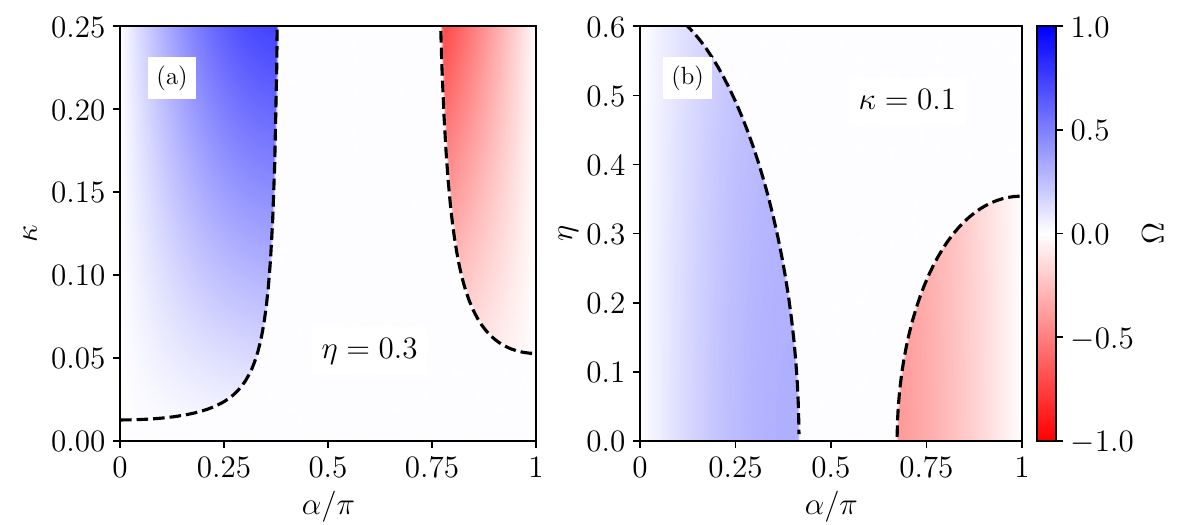}
    \caption{Mean-field phase diagram from the linear stability analysis at $Q=3$ in the $\alpha-\kappa$ plane with fixed $\eta=0.3$ in (a) and in the $\alpha-\eta$ plane with fixed $\kappa=0.1$ in (b). The chiral angular frequency $\Omega$ in the in-phase and out-of-phase chiral states are color-coded according to the bar chart.}
    \label{fig:lsa_pd}
\end{figure}

The disordered state is destabilized by the in-phase mode and the out-of-phase mode. The stability boundaries set by $\Re[\Lambda_{\rm in, out}^{(1)}]=0$ are given by
\begin{equation}\label{eq:kappa_in_out}
\begin{split} 
\kappa_{\rm in} &= \frac{1-e^{-\eta^2/2}}{4e^{-\eta^2/2}[1+(Q-1)\cos^3\frac{\alpha}{2}]-8Q/3},  \\
\kappa_{\rm out} &= \frac{1-e^{-\eta^2/2}}{4e^{-\eta^2/2}[1+\frac{Q}{2}\cos\frac{\alpha}{2}-\cos^3\frac{\alpha}{2}+\frac{Q\alpha}{4}\sin\frac{\alpha}{2}]-4Q} .
\end{split}
\end{equation}
In the in-phase chiral mode~($\kappa>\kappa_{\rm in}$), all particles perform a counter-clockwise chiral motion with identical phases~($\bm{\lambda}_{\rm in} = \bm{1}_Q$) with a positive angular frequency
\begin{equation}\label{eq:Win_MF}
\Omega_{\rm in} = \Im[\Lambda_{\rm in}^{(1)}] = 4\kappa (Q-1) e^{-\eta^2/2} \sin\frac{\alpha}{2} \cos^2\frac{\alpha}{2} > 0 .
\end{equation}
On the other hand, in the out-of-phase mode~($\kappa > \kappa_{\rm out})$,
all particles perform a clockwise chiral motion with a negative angular
frequency~\footnote{{The chiral frequency may deviate from the formulae
    in Eqs.~\eqref{eq:Win_MF} and \eqref{eq:Wout_MF} far from the stability
boundaries due to a nonlinear correction.}}
\begin{equation}\label{eq:Wout_MF}
    \Omega_{\rm out} = -\kappa  e^{-\eta^2/2} (Q\alpha + 2 \sin\alpha) \cos\frac{\alpha}{2} < 0 .
\end{equation}
The linear stability analysis yields the mean-field phase diagram as shown in Fig.~\ref{fig:lsa_pd}. It suggests that chiral states can emerge due to nonreciprocity.

\subsection{Hydrodynamic equation}\label{sec:hydrodynamics}
The Boltzmann equation is an infinite hierarchy of coupled equations for the multipole moments. For the long-time and large-distance behavior, it is useful to consider a hydrodynamic equation for the lowest order multipole moments: a density field~($k=0$) and a polarization field~($k=\pm 1$). 
Following Ref.~\cite{bertinHydrodynamic2009}, we will neglect higher order moments with $|k|\geq 3$ and take a quadrupole moment with $k=\pm 2$ as a fast variable enslaved by the density and polarization fields. The Boltzmann equation in Eq.~\eqref{eq:BoltzmannVector} with $k=2$ then yields
\begin{equation}
    \begin{split}
    \partial_{z^*} \bm{f}_1 \simeq & -(1-P_2) \bm{f}_2 + 
    \bm{f}_2\circ (\bm{J}^{20}\bm{f}_0) \\ 
    &+  \bm{f}_1\circ (\bm{J}^{21}\bm{f}_1) + 
    \bm{f}_0\circ (\bm{J}^{22}\bm{f}_2).  
    \end{split}
\end{equation}
Using $\bm{a}\circ \bm{b} = \bm{D}(\bm{a}) \bm{b} = \bm{D}(\bm{b})\bm{a}$, we can solve the above equation to obtain
\begin{equation}
    \bm{f}_2 = -\bm{X}(\bm{f}_0) \left\{ \partial_{z^*}\bm{f}_1 - \bm{f}_1 \circ (\bm{J}^{21} \bm{f}_1 ) \right\}
\end{equation}
with
\begin{equation}
    \bm{X}(\bm{f}_0) = \left\{ (1-P_2)\bm{E} - \bm{D}(\bm{f}_0) \bm{J}^{22} - \bm{D}(\bm{J}^{20} \bm{f}_0)\right\}^{-1}.
\end{equation}
Inserting $\bm{f}_2$ into Eq.~\eqref{eq:BoltzmannVector} with $k=1$, we finally obtain the hydrodynamic equation
\begin{equation}\begin{split}
    \partial_t \bm{f}_1  =& 
    - \partial_{z^*}\bm{f}_0 + \bm{Y}(\bm{f}_0) \bm{f}_1  \\
    & + \partial_{z}\left( \bm{X}(\bm{f}_0) \left( \partial_{z^*} \bm{f}_1 - \bm{D}(\bm{f}_1) \bm{J}^{21} \bm{f}_1 \right) \right) \\
    & + \bm{Z}(\bm{f}_{-1},\bm{f}_0) \left( \partial_{z^*} \bm{f}_1 - \bm{D}(\bm{f}_1) \bm{J}^{21} \bm{f}_1 \right) ,
    \label{eq:HydroEq}
\end{split}\end{equation}
where
\begin{equation}\begin{split}
        \bm{Y} &= -(1-P_1)\bm{E} + \bm{D}(\bm{J}^{10}\bm{f}_0) + \bm{D}(\bm{f}_0)\bm{J}^{11} \\
        \bm{Z} &= -\left[\bm{D}(\bm{J}^{1,-1} \bm{f}_{-1}) {+} \bm{D}(\bm{f}_{-1}) \bm{J}^{12} \right] \bm{X}(\bm{f}_0) .
\end{split}\end{equation}
Note that the $Q\times Q$ matrices $\bm{X}$ and $\bm{Y}$ depend only on $\bm{f}_0$ while  $\bm{Z}$ is linear in $\bm{f}_{-1}=\bm{f}_1^*$. The righthand side of Eq.~\eqref{eq:HydroEq} includes terms quadratic in $\bm{f}_{\pm 1}$, which manifests the broken time reversal symmetry. 
The hydrodynamic equation is highly nonlinear, which makes it improbable to obtain an analytic solution.  Nevertheless, one can gain useful physical insights considering limiting cases. 

We consider the hydrodynamic equation restricted within the species-symmetric subspace in which $\bm{f}_0(\bm{r},t) = \rho(\bm{r},t) \bm{1}_Q$ and $\bm{f}_1(\bm{r},t) = w(\bm{r},t) \bm{1}_Q$ with a constant vector $\bm{1}_Q = (1,\cdots, 1)^T$. 
Taking an inner product of Eqs.~\eqref{eq:continuum_eq} and \eqref{eq:HydroEq} with $\frac{1}{Q}\bm{1}_Q$, one can obtain the field equation  
\begin{align}
    \partial_t \rho =& -\partial_{z^*} w^* - \partial_{z} w \label{eq:continuumSym_app} \\
    \partial_t w =& - \partial_{z^*}\rho + \left(\mu(\rho) - \xi(\rho) |w|^2\right) w + 4 \partial_z \nu(\rho) \partial_{z^*} w \nonumber \\
    &+ 2 \zeta(\rho) w^* \partial_{z^*} w  + 2 \partial_z \eta(\rho) w^2 ,
    \label{eq:HydroSym_app}
\end{align}
where the coefficients are given by $\nu(\rho) := \frac{1}{4} \langle \bm{X} \rangle_Q$, $\mu(\rho) := \langle\bm{Y}\rangle_Q$, $\zeta(\rho) := \langle \bm{Z} \rangle_Q/(2w^*)$, $\eta (\rho) := \langle \bm{X}\bm{J}^{21}\rangle_Q/2$, and $\xi(\rho) := \langle \bm{Z}\bm{J}^{21}\rangle_Q/w^*$ with $\langle \bm{M}\rangle_Q := \bm{1}_Q^T \bm{M} \bm{1}_Q / Q$~\footnote{We remark that Eq.~\eqref{eq:HydroSym_app} reduces to a complex Ginzburg-Landau equation~\cite{Aranson.2002} when $\eta=\zeta=0$ and $\rho(\bm{r},t) = 1$.}.
These coefficients vary spatially through the $\bm{r}$ dependence of the density field. We emphasize that they are complex numbers in the presence of a nonreciprocal phase shift $\alpha \neq 0, \pi$. For instance, for $Q=2$, they are given by
\begin{align*}
    \nu(\rho) &= \frac{1}{4} \frac{1}{(1-P_2) + \frac{8}{15} \kappa \rho (14+5(1+e^{i\alpha})P_2)} \\
    \mu(\rho) &= -(1-P_1) + 4\kappa \rho \left( \left(1 +e^{i\alpha/2} \cos^2\frac{\alpha}{2} \right)P_1 - \frac{4}{3}\right) \\
    \zeta(\rho) &= 4 \kappa \nu(\rho) \left( P_1 e^{i\alpha/2}\cos\alpha(1+\cos\alpha) + 2P_1-\frac{8}{5}\right) \\
    \eta (\rho) &= 8\kappa \nu(\rho) \left( (1+e^{i\alpha}) P_2 + \frac{2}{3} \right) \\
    \xi(\rho) &= \frac{\zeta(\rho)\eta(\rho)}{\nu(\rho)} 
\end{align*}
with $P_k = e^{-k^2\eta^2/2}$.

If one further neglects spatial fluctuations such that $\rho(\bm{r},t) = 1$ and $w(\bm{r},t) = w(t)$, Eq.~\eqref{eq:HydroSym_app} reduces to a Stuart-Landau equation~\cite{Garcia-Morales.2012}
\begin{equation}\label{eq:StuartLandau}
    \dot w(t) = \mu_0 w(t) - \xi_0 |w(t)|^2 w(t)
\end{equation}
with complex parameters $\mu_0=\mu_r + i \mu_i$ and $\xi_0=\xi_r + i\xi_i$.  The Stuart-Landau equation undergoes a supercritical Hopf bifurcation~\cite{Garcia-Morales.2012, Strogatz.2024} at $\mu_r = 0$.  For $\mu_r >0$, it has a limit cycle solution $w(t) = A e^{i\Omega t}$ with a constant amplitude $A = \sqrt{\frac{\mu_r}{\xi_r}}$ and an angular velocity $\Omega = \mu_i - \frac{\mu_r \xi_i}{\xi_r}$. The limit cycle solution corresponds to the homogeneous in-phase chiral phase found in the linear stability analysis of the Boltzmann equation. 

In summary, we have derived the Boltzmann equation and the hydrodynamic
equation for the $Q$-NRVM. 
{The naive mean-field theory and the more
sophisticated {mean-field analysis based on the } continuum field
theory demonstrate {emergence of chirality due to the nonreciprocal
phase shift $\alpha$.} We will show, however, that the in-phase chiral
        states and the out-of-phase chiral state are not
    stable against spatial fluctuations. 
    The nature of emerging phases and
the phase diagram will be investigated in the next section.}

\section{Phase diagram}\label{sec:phase_diagram}

The continuum theory, {in the mean-field limit} assuming spatial homogeneity, predicted in-phase and out-of-phase chiral states. Here we investigate whether the chiral order survives temporal and spatial fluctuations. 

\begin{figure}
    \includegraphics[width=\columnwidth]{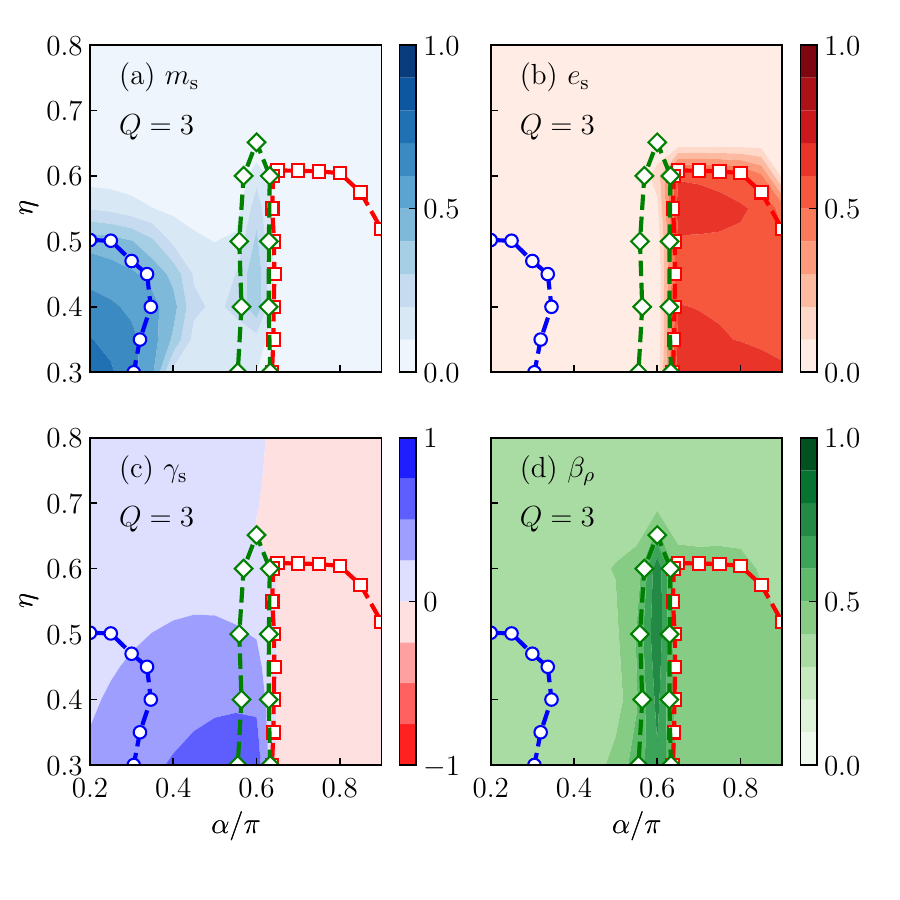}
    \caption{Color plot of the order parameters (a) $m_{\rm s}$, (b) $e_{\rm
    s}$, (c) $\gamma_{\rm s}$, and (d) the bimodality coefficient
$\beta_{\rho}$ for the three-species system of size $L=128$ with $\rho_0=2$.
Phase boundaries of a chiral phase~($\circ$), a species separation
phase~($\square$), and a coexistence phase~($\diamond$) 
{are determined using a finite-size scaling behavior $m_{\rm s}(L)
\sim L^{-\tilde\beta}$ with $\tilde\beta=\tilde\beta_{\rm BKT}=1/8$, $e_{\rm
s}=0.5$, and $\beta_\rho=5/9$, respectively~(see the main text).}}
    \label{fig:OrderParametersQ3}
\end{figure}

In {numerical simulations of the microscopic model},
we have measured a chirality 
\begin{equation}\label{eq:chirality_t}
    \gamma(t) = \frac{1}{N\Delta t} \sum_{n} \sin(\theta_n(t+\Delta t)-\theta_n(t)) 
\end{equation}
and a polarization
\begin{equation}\label{eq:polarization_t}
    m(t) = \left|\frac{1}{N}\sum_{n} \hat{\bm{e}}(\theta_n(t))\right|  .
\end{equation}
A positive~(negative) chirality indicates a collective
counter-clockwise~(clockwise) chiral motion. The polarization measures the degree of phase coherence. It is nonzero in the presence of long-range flocking order. It is also nonzero when particles exhibit chiral motion with synchronized polar angles. 
We have also measured an energy
\begin{equation}\label{eq:energy_t}
    e(t) = \frac{1}{(2\pi r_0^2 \rho_{\rm tot})N}\sum_{|\bm{r}_n-\bm{r}_m|<r_0} \left(\frac{Q \delta_{s_n s_m}-1}{Q-1} \right) .
\end{equation}
This quantity is analogous to the energy density of the $Q$-state Potts model~\cite{Wu.1982}. It equals zero if the particle species are perfectly mixed, and becomes positive when Potts symmetry is broken and particles of different species are spatially separated. The mean steady-state values $m_{\rm s} = \langle m(t)\rangle_{\rm s}$, $\gamma_{\rm s} = \langle \gamma(t)\rangle_{\rm s}$, and $e_{\rm s} = \langle e(t)\rangle_{\rm s}$ are used to characterize the macroscopic state of the system, where $\langle \dots\rangle_{\rm s}$ denotes a time average in the steady state. 

Figure~\ref{fig:OrderParametersQ3} presents an overall behavior of those
order parameters in the $\alpha-\eta$ plane for the three-species case,
along with the phase boundaries obtained in Ref.~\cite{Woo.20258m}.
Analyzing the order parameters quantitatively, we have identified four
distinct phases: a disordered phase, a chiral phase with quasi-long-range
order~(QLRO), a species-separation~(SS) phase with vortex cells, and a
coexistence phase~\cite{Woo.20258m}. We will characterize each phase in the subsequent subsections. The overall behavior of the order parameters, presented in Fig.~\ref{fig:MandEQ3}, indicates that the phase diagram has a similar structure for all values of $Q$. 
As $Q$ increases, the phase boundaries shift toward a higher $\alpha$ region.
{As we shall show in Sec.~\ref{sec:phase_separation}, an effective
repulsion between flocks of distinct species is responsible for the SS. This
effective repulsion works irrespective of $Q$. Thus, we expect that the SS
phase exists for any values of $Q$.}

\subsection{Chiral phase with quasi-long-range order}
\label{sec:chiral}

The chiral phase, occurring when $\alpha$ and $\eta$ are small, is characterized by species mixing~($e_{\rm s}\simeq 0$) and counter-clockwise chirality~($\gamma_{\rm s} > 0)$. The polarization order parameter $m_{\rm s}$ takes larger values than in the other regions. 

\begin{figure}[t]
    \includegraphics[width=\columnwidth]{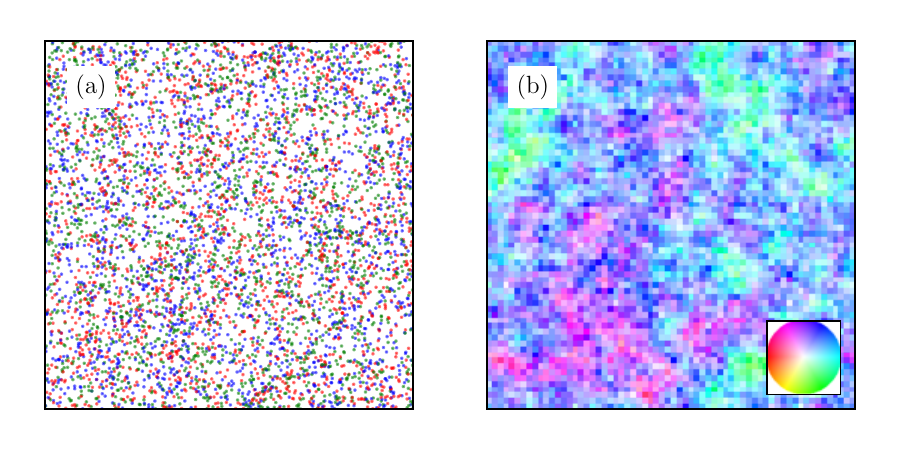}
    \caption{Snapshots of a particle configuration in (a) and a polarization field in (b) in the QLRO chiral phase. In (a), each dot representing a particle is color-coded according to particle species. In (b), the phase angle of a polarization field is color-coded according to the chart in the inset. Parameters: $Q=3$, $L=128$, $\rho_0=2$, $\alpha = 0.3\pi$, and $\eta=0.4$.}
    \label{fig:snapshot_chiral}
\end{figure}

Figure~\ref{fig:snapshot_chiral} presents representative snapshots of a particle configuration and a polarization field $\bm{m}(\bm{r},t) = \sum_n \hat{\bm{e}}(\theta_n(t)) \delta(\bm{r}-\bm{r}_n(t))$ obtained from a 
{numerical simulation of the microscopic model.}
The polarization field displays a long wavelength orientational fluctuation.
It turned out that polar order is not long-ranged but
quasi-long-ranged~\cite{Woo.20258m}. The steady-state correlation function $C_m(\bm{r}) := \langle \bm{m}(\bm{r}+\bm{r}_0,t) \cdot \bm{m}(\bm{r}_0,t)\rangle_{\bm{r}_0, \rm{s}}/\rho_{\rm tot.}$, averaged over $\bm{r}_0$ and $t$ in the steady state, decays algebraically as
\begin{equation}\label{eq:power_law_Cw}
    C_m(\bm{r}) \sim r^{-\tilde\eta},
\end{equation}
and the polarization order parameter decreases with increasing system size $L$ according to the power-law
\begin{equation}\label{eq:w_fss}
m_{\rm s}(L) \sim L^{-\tilde{\beta}} .
\end{equation}
The critical exponents $\tilde{\eta}$ and $\tilde{\beta}$ vary continuously
inside the chiral phase~\cite{Woo.20258m}.

\begin{figure}
    \includegraphics[width=\columnwidth]{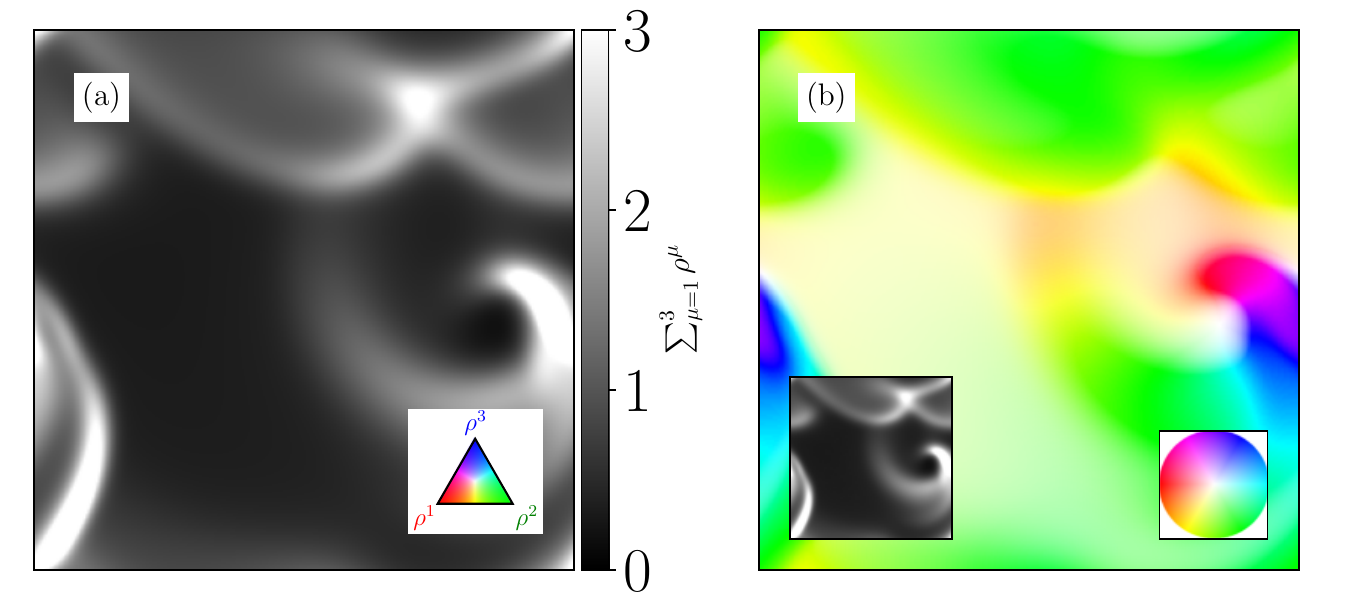}
    \caption{Snapshots of a density field in (a) and a polarization field in
    (b) computed with the  Boltzmann equation~\eqref{eq:BoltzmannVector} for
the three-species system~(see Mov.~1 in \cite{SM}). The Boltzmann equation,
truncated within $|k|\leq 4$~{($\bm{f}_k = 0$ for $|k|>4$)}, is
integrated numerically using the pseudo-spectral
method~\cite{fornberg1998practical, boyd2001chebyshev}. For numerical
convergence, the Boltzmann equation is regularized with a diffusion
term~\cite{mahault2018outstanding}. The snapshots are taken when the system,
starting from a random initial configuration, reaches a steady state. The
density field in (a) is color-coded using local densities $(\rho^1, \rho^2,
\rho^3)$ of the three species as an RGB code~{(see the inset). The total
density field $\sum_{\mu=1}^3\rho^\mu$ is drawn in the inset of (b).} These snapshots reveal that the three species particles are mixed with $\rho^1=\rho^2=\rho^3$ and that spatial fluctuations persist. Parameters: $L_x=L_y=64$, $\eta = 0.3$, $\alpha/\pi=0.1$ and $\kappa=0.053$.}
    \label{fig:snapshot_chiral_BE}
\end{figure}

The mean-field analysis for the Boltzmann equation predicted a homogeneous
in-phase chiral state for small $\alpha$ and $\eta$.
Figure~\ref{fig:snapshot_chiral_BE} presents typical snapshots of a density
field and a polarization field obtained by integrating numerically the
Boltzmann equation with a random initial
configuration~\footnote{{We discretized the Boltzmann equation using a
        uniform grid spacing $\Delta x = \Delta y = 1/16$ and a time step
        $\Delta t = 10^{-3}$ for a numerical study. 
        Time integration was performed using a fourth-order Runge-Kutta scheme. 
        The nonlinear terms were evaluated using a pseudo-spectral method,
        together with the standard $2/3$ dealiasing rule to reduce
        aliasing errors~\cite{fornberg1998practical, boyd2001chebyshev}. 
To avoid numerical instability, we included a diffusion term, $4 D_t
\partial_z\partial_{z^*} \bm{f}_k$ with
$D_t=0.1$~\cite{mahault2018outstanding}.}}. 
As these snapshots demonstrate, spatial fluctuations persist
in the in-phase chiral state. 
{We note that the Boltzmann equation displays stronger
density fluctuations than the microscopic system. 
A quantitative characterization of the fluctuation requires the
    analysis of the Boltzmann equation at larger sizes, which is
beyond our current numerical capacity.}

The QLRO chiral phase of the $Q$-NRVM reminds us of the QLRO phase of the equilibrium 2D $XY$ model~\cite{Berezinskii.1971, Kosterlitz.1973, Jose.1977, Izyumov.1988}. This model describes ferromagnetic ordering of continuous planar spins $\bm{s}_n = (\cos\theta_n, \sin\theta_n)$ on a 2D lattice. The spins interact via the $XY$ Hamiltonian 
\begin{equation}\label{eq:Hxy}
H_{XY} = -J \sum_{\langle n, m\rangle} \bm{s}_n \cdot \bm{s}_m = -J \sum_{\langle n, m\rangle} \cos(\theta_n-\theta_m),
\end{equation}
where $J>0$ and the sum is over nearest-neighbor pairs. 
The Hamiltonian has continuous rotational symmetry. Consequently, according to the Mermin-Wagner theorem~\cite{Mermin.1966}, the equilibrium $XY$ system in 2D cannot maintain long-range order. Instead, below a Berezinskii-Kosterlitz-Thouless~(BKT) transition temperature $T_{\rm BKT}$, the system displays QLRO, which is characterized by the power-law scaling of the correlation function as in Eq.~\eqref{eq:power_law_Cw} and the order parameter as in Eq.~\eqref{eq:w_fss} with the exponents $\tilde{\eta}$ and $\tilde\beta$ related by a scaling relation $\tilde\beta = \tilde{\eta}/2$~\cite{Jose.1977}. The exponents vary continuously in the QLRO phase and take the universal values $\tilde\beta_{\rm BKT} = 1/8$ and $\tilde\eta_{\rm BKT}=1/4$ at the transition temperature $T_{\rm BKT}$~\cite{Jose.1977}. The BKT transition is driven by topological excitations, vortices and antivortices~\cite{Kosterlitz.1973}. 

The original 2D Vicsek model has continuous rotational symmetry but achieves long-range order because the self-propulsion drives the system strongly out of equilibrium~\cite{Toner.1995}. Variants of the $XY$ model were recently studied, where $XY$ spins diffuse independently while interacting via a ferromagnetic coupling with local neighbors~\cite{Woo.2024, Rouzaire.2025}. These models are a passive version of the Vicsek model. Although passive diffusion drives the system out of equilibrium, the numerical study showed that the system undergoes a BKT transition between a QLRO phase and a disordered phase~\cite{Woo.2024}. The QLRO was also observed in a Vicsek-type model in which self-propelled particles reverse its velocity stochastically~\cite{Mahault.2018z9b}. 

Given that the $Q$-NRVM consists of self-propelled particles, it is an intriguing question whether the QLRO in our model 
{can be characterized as in } 
the 2D equilibrium $XY$ model. The numerical finding makes this scenario plausible, and we will elaborate on this possibility using an analytic argument. In essence, we will show that, in the chiral phase, the angles behave like motile $XY$ spins, and the spatial fluctuations of the particles around their circular orbits are diffusive. This result, combined with the findings in Ref.~\cite{Woo.2024}, implies QLRO.

We can simplify the continuous-time dynamics given in Eq.~\eqref{eq:model_rule_cont} in the QLRO chiral phase.
The mean angular velocity of particle $n$ is approximately given by $\Omega_0 = \langle \dot\theta_n\rangle \approx J \sum_{m\in \mathcal{N}_n}\sin\alpha_{nm} \approx \pi r_0^2 \rho_0 (Q-1)J \sin\alpha$. Then, the equation of motion for a phase angle $\phi_n := \theta_n - \Omega_0 t$ in the {\em co-rotating frame} becomes
\begin{equation}\label{eq:dot_phi1}
    \dot{\phi}_n \approx -J\sum_{m\in \mathcal{N}_n} V(\phi_n-\phi_m, \alpha_{nm})  + \xi_n(t) 
\end{equation}
with 
\begin{equation}\label{eq:dot_phi2}
V(\Delta \phi, \varphi) := \sin(\Delta \phi-\varphi)+\sin\varphi
\approx \cos\varphi \sin\Delta\phi . 
\end{equation}
To obtain the last expression, we neglected the higher-order term $(1-\cos\Delta \phi) = O(\Delta \phi^2)$. Inserting Eq.~\eqref{eq:dot_phi2} into Eq.~\eqref{eq:dot_phi1}, we obtain 
\begin{equation}\label{eq:dot_phi3}
\dot{\phi}_n \approx -J_{\rm eff}  \sum_{m\in\mathcal{N}_n}\sin(\phi_n-\phi_m) + \xi_n(t)
\end{equation}
with $J_{\rm eff} = \frac{(Q-1)J \cos\alpha}{Q}$. The resulting equation has the form of a Langevin equation for $XY$ spins. 

The key difference from the $XY$ model lies in the particle motility. A particle displacement is given by $\Delta z = \int_{t_0}^{t_0+ t} dt' v_0 e^{i(\Omega_0 t'+\phi(t'))}$ with complex coordinate $z=x+iy$. The mean square displacement is given by 
\begin{equation}\label{eq:msd}
    \langle|\Delta z|^2\rangle = v_0^2 \iint_{t_0}^{t_0+t} dt' dt''~ e^{i\Omega_0(t'-t'')}\left\langle e^{i(\phi(t')-\phi(t''))}\right\rangle .
\end{equation}
It is reasonable to assume that $\left\langle e^{i(\phi(t')-\phi(t''))}\right\rangle = a e^{-|t'-t''|/\tau_{\rm p}}$ with a constant $a>0$ and a persistent time $\tau_{\rm p}$. The integral can then be evaluated exactly.

When $\alpha=0$~(Vicsek model case), $\Omega_0=0$ and we get 
\begin{equation}\label{eq:msd_zerochirality}
    \langle|\Delta z|^2\rangle \simeq \begin{cases}
        av_0^2 t^2,& (t \ll \tau_{\rm p}) \\
        2av_0^2 \tau_{\rm p} t, & (t \gg \tau_{\rm p}) .
    \end{cases}
\end{equation}
This shows a crossover from ballistic motion to diffusive motion. This ballistic motion, occurring over the persistence time scale, causes strong temporal fluctuations in the interaction network, which we believe to be the reason why the Vicsek model is not constrained by the Mermin-Wagner theorem.
In contrast, when $\alpha\neq 0$ and $\tau_{\rm p}\Omega_0 \gg 1$ (meaning particles can complete multiple revolutions within a persistent time scale), the mean square displacement is given by
\begin{equation}
   \frac{\langle|\Delta z|^2\rangle}{(4av_0^2/\Omega_0^2)} \simeq \begin{cases}
       \frac{t}{\tau_{\rm p}} + (1-\frac{t}{\tau_{\rm p}}) \sin^2\frac{\Omega_0 t}{2},& (t \ll \tau_{\rm p})\\
       \frac{t}{2\tau_{\rm p}},& (t \gg \tau_{\rm p}) .
   \end{cases}
\end{equation}
Interestingly, chirality makes the particles' motion diffusive at all time scales. Since particles disperse slowly, the temporal fluctuation of the mutual interaction network is weaker than in the original Vicsek model. 
{This argument explains the origin of the QLRO chiral phase in our
model.

The QLRO phase is separated from a disordered phase. The nature of the phase
transition remains an open question for diffusing $XY$ spin systems. It is
reported that the $p$-state clock model, consisting of passive Brownian
particles, undergoes the BKT transition from the QLRO phase to the
disordered phase for $p>4$~\cite{Woo.2024}. On the other hand,
Ref.~\cite{Mahault.2018z9b} reports that a variant of the Vicsek
model with random velocity reversal undergoes a non-BKT transition from the
QLRO phase with continuously varying scaling exponent. Interestingly, in
Ref.~\cite{Mahault.2018z9b}, the finite-size-scaling exponent $\tilde\beta$ 
for the order parameter at the transition takes the value of $1/8$, which
coincides with the universal value $\tilde\beta_{\rm BKT}=1/8$ associated
with the BKT transition. We used this criterion $\tilde\beta=1/8$ to draw the
phase boundary of the QLRO phase in Fig.~\ref{fig:OrderParametersQ3}.}

{Topological excitations, such as vortices and antivorticies, can
destroy QLRO.} In the $Q$-NRVM, these topological excitations
destroy the phase coherence and also generate inhomogeneities in the density
field: we find that the density field develops an oscillating {\em monopole
moment} at a vortex core while a rotating {\em quadrupole moment} at an
antivortex core~(see App.~\ref{app:vortex_antivortex}). Thus, the disordered
phase is characterized by phase incoherence and {enhanced} density fluctuations.

In summary, we have shown in this subsection that the $Q$-NRVM is in a QLRO
chiral phase when $\alpha$ and $\eta$ are small. In this phase, the
polarization correlation function decays algebraically and the polarization
order parameter obeys a critical {finite-size scaling} behavior. 

\subsection{Species separation}\label{sec:phase_separation}

\begin{figure}
    \includegraphics[width=\columnwidth]{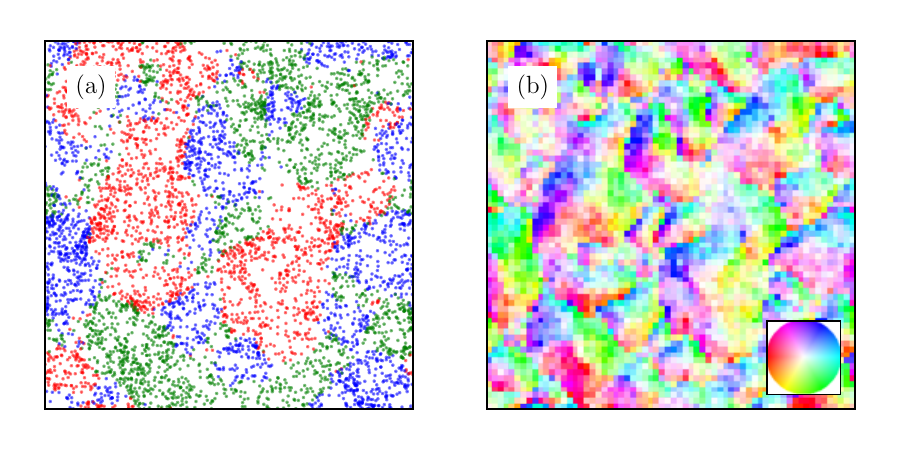}
    \caption{Snapshots of a particle configuration in (a) and a polarization field in (b) in the SS phase. The same color scheme is used as in Fig.~\ref{fig:snapshot_chiral}. Snapshots for different values of $Q$ are found in Fig.~\ref{fig:SS245}. Parameters: $Q=3$, $L=128$, $\rho_0=2$, $\alpha=0.8\pi$, and $\eta=0.35$.}
    \label{fig:snapshot_ps}
\end{figure}

For large $\alpha$ and small $\eta$, the $Q$-NRVM exhibits species separation. 
If particles of different species were mixed well, they would form an out-of-phase chiral state as predicted by the mean-field theory. We find that the nonreciprocal phase shift generates repulsion between counter-propagating flocks of different species, which destabilizes a species-mixed out-of-phase chiral state. 
In this section, we will characterize the SS phase, and present an analytic argument for its existence based on the hydrodynamic equation.

Figure~\ref{fig:snapshot_ps} presents representative snapshots of a particle
configuration and a polarization field in the SS phase. Particles are
species-separated and self-organize into a vortex cell~(VC) structure. Each
cell is occupied predominantly by particles of a single species flowing
clockwise along a boundary. Accordingly, the SS phase is characterized by a
positive energy order parameter $e_{\rm s}$ and a negative chirality
$\gamma_{\rm s}$~(see Figs.~\ref{fig:OrderParametersQ3} and~\ref{fig:MandEQ3}). 

\begin{figure}
    \includegraphics[width=\columnwidth]{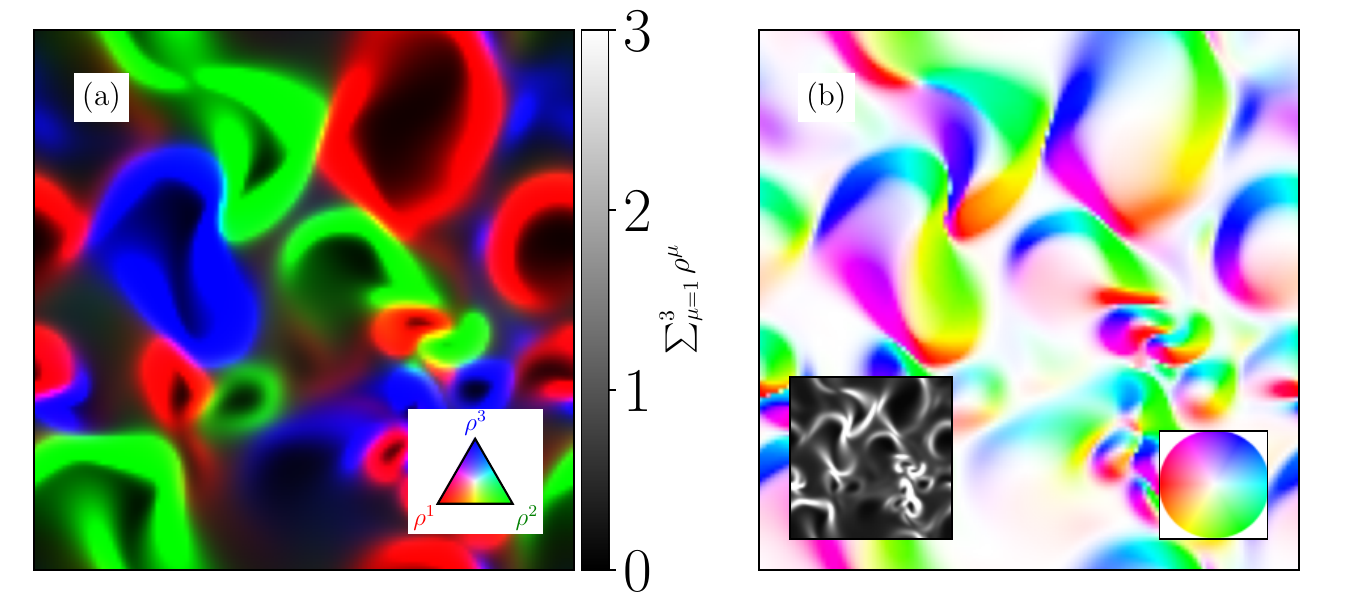}
    \caption{Snapshots of a density field in (a) and a polarization field in
        (b) computed with the  Boltzmann equation~\eqref{eq:BoltzmannVector}
        for the three-species system~(see Mov.~2 in \cite{SM}). {We
            adopt the same numerical integration method as specified in the
        caption of Fig.~\ref{fig:snapshot_chiral_BE},} and adopt the same color-coding scheme as in Fig.~\ref{fig:snapshot_chiral_BE}.
    Parameters: $L_x=L_y=64$, $\eta = 0.3$, $\alpha/\pi=0.9$ and $\kappa=0.212$.}
    \label{fig:VC_Boltzmann}
\end{figure}

The continuum Boltzmann equation~\eqref{eq:BoltzmannVector} also confirms the SS phase solution~(see Fig.~\ref{fig:VC_Boltzmann}). This assures that the continuum field theory captures the physics of species separation.

\begin{figure}
    \includegraphics[width=\columnwidth]{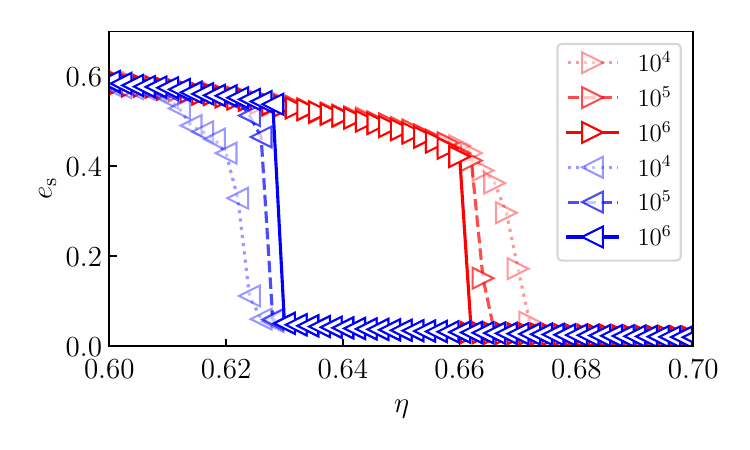}
    \caption{{Hysteresis loop for $e_{\rm s}$ at a fixed $\alpha=0.7\pi$.
    The order parameter is measured as $\eta$ increases~($\triangleright$)
or decreases~($\triangleleft$) at a rate of $|\Delta \eta| = 0.002$ per 
$\Delta T$ time steps, specified in the legend. Parameters: $Q=3$, $L=128$,
and $\rho_0 = 2$.}}
    \label{fig:hysteresis}
\end{figure}

The phase transition into the SS phase is signified
by a discontinuous jump in $e_{\rm s}$.
{The order parameter exhibits a hysteresis phenomenon, as shown in
    Fig.~\ref{fig:hysteresis}. We compare the order parameter values
measured using two different protocols: we increase~(decrease) the value of
$\eta$ by $|\Delta \eta| = 0.002$ every $\Delta T~( = 10^4, 10^5, 10^6)$
time steps while measuring $e_{\rm s}$ during the second half of each
interval. The curves from the two protocols form a hysteresis loop whose
area decreases as $\Delta T$ increases. Therefore, we conclude that the
transition between the SS phase and the disordered phase is of first order.}

\begin{figure}
    \includegraphics[width=\columnwidth]{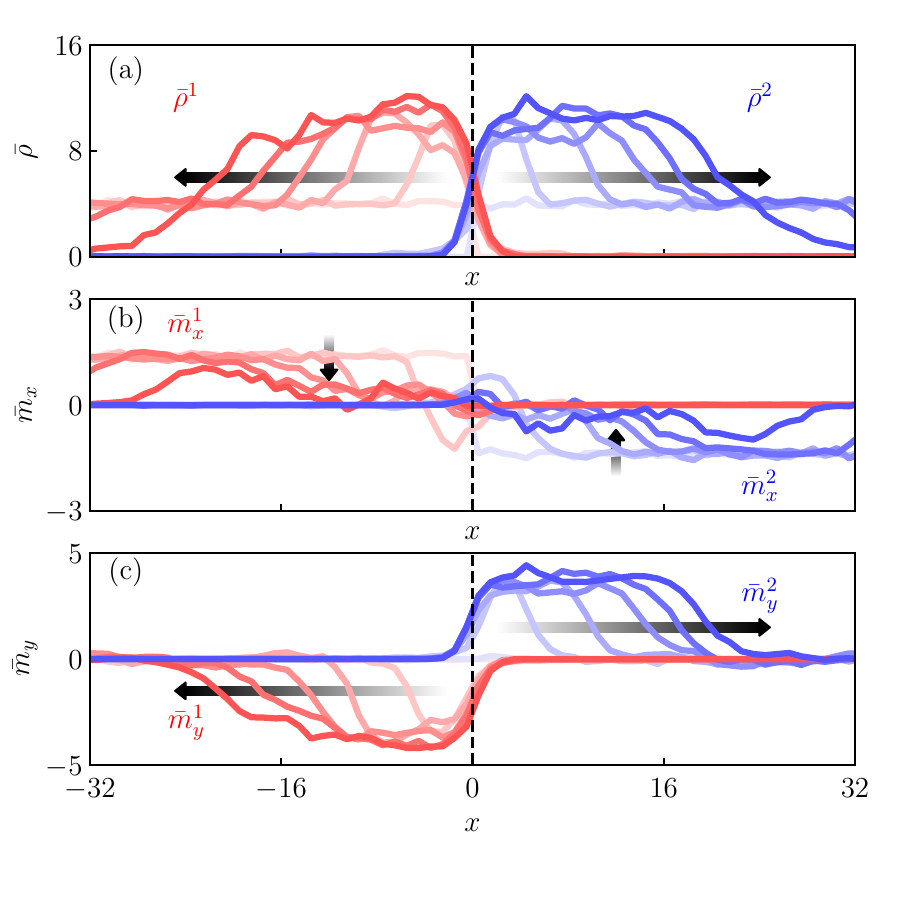}
    \caption{{Numerical simulations of the microscopic model showing} effective repulsion between counter-propagating flocks of
    species $\mu=1,2$. Each panel presents columnar profiles of (a) particle
density, (b) $x$ component of polarization, and (c) $y$ component of
polarization at successive time steps $t_n = 20\times n$ with $n=0, \cdots,
5$. Curves for species $\mu=1~(2)$ are drawn in red~(blue) color getting
darker as $t_n$ increases {as indicated by arrows}. Parameters:
$L_x=L_y=128$, $Q=2$, $\eta=0.4$, $\alpha=0.85\pi$, and {$\rho_0=2$}.}
    \label{fig:Repulsion}
\end{figure}

In the SS phase, particles of different species do not mix with each other as if there were an inter-species repulsive interaction. We demonstrate this effective repulsion using a setting in which particles of species $\mu=1$~($\mu=2$) are distributed uniformly in the $x<0$~($x>0$) region and flow collectively to the positive~(negative)-$x$ direction. These counter-propagating flocks collide at time $t=0$ at $x=0$. We have measured density and polarization profiles $\bar{\rho}^\mu(x)$ and $(\bar{m}_x^\mu(x), \bar{m}_y^\mu(x))$ at successive time steps after the collision~(see Fig.~\ref{fig:Repulsion}). The numerical data clearly show that particles of different species cannot penetrate into each other beyond a narrow collision band. The inter-species nonreciprocal interaction within the band causes the polarization field of each species to turn clockwise from the initial longitudinal direction to the perpendicular transverse direction~\cite{Banerjee.2017}. This reorientation prevents mixing and leads to SS.

We present an analytic theory for the scattering of counter-propagating flocks using the hydrodynamic equation~\eqref{eq:HydroEq} with $Q=2$. Our theory relies on a perturbative description for the hydrodynamic fields within the narrow collision band for small $\epsilon_\alpha := \pi - \alpha$. Let $w^{\mu}(t)$ be the complex polarization field of species $\mu=1,2$ near $x\simeq 0$. A spatial variation of the polarization field is assumed to be negligible at $x\simeq 0$.  

When $\alpha=\pi$, the counter-propagating flocks form an anti-parallel flocking~(APF) state~\cite{Chatterjee.2023}, in which they counter-propagate through each other with opposite polarizations $w^1(t) = -w^2(t) = w(t)$. The hydrodynamic equation~\eqref{eq:HydroEq} at $\alpha=\pi$ for the APF state becomes
\begin{equation}
    \partial_t w(t) = \left(\mu_\pi - \xi_\pi |w|^2 \right) w ,
\end{equation}
where
\begin{align}
    \mu_\pi &= (1+2(\pi+2)\kappa )P_1 - (1+8\kappa) \\
    \xi_\pi &= 32\kappa^2 \frac{P_2(15 P_1-2)}{15+112\kappa-15P_2}
\end{align}
with $P_k = e^{-k^2\eta^2/2}$. The steady-state polarizations in the APF state are given by $w^1 = -w^2 = w_\pi$ with 
\begin{equation}
    w_{\pi} = \sqrt{\frac{\mu_\pi}{\xi_\pi}}.
\end{equation}

When $\alpha < \pi$, the counter-propagating flocks deviate from the APF state. We investigate the deviation using the hydrodynamic equation linearized with respect to $\delta w_1(t) := w^1(t) - w_\pi$ and $\delta w_2(t) := w^2(t) + w_\pi$. After lengthy but straightforward algebra, we have derived the linearized equations of motion
\begin{align}
    \partial_t \delta w_S =& a_1 \delta w_S + a_2 \delta w_S^* \label{eq:dwSdt}\\
    \partial_t \delta w_A =& 
    b_1  + 
    (b_1+b_2 )\delta w_A + b_2\delta w_A^* , \label{eq:dwAdt}
\end{align}
for a symmetric part $\delta w_S := \delta w^1 + \delta w^2$ and an anti-symmetric part $\delta w_A := (\delta w^1-\delta w^2)/(2w_\pi)$. 

It is instructive to consider a limiting case with $\eta=0$.
In App.~\ref{app:theory_for_sp}, we present explicit expressions for the coefficients $a_{1,2}$ and $b_{1,2}$, and four normal mode solutions. It turns out that $\delta w_A$ has an unstable normal mode $e^{i\Theta_{A,+}}e^{\Lambda_{A,+}t}$ whose Lyapunov exponent $\Lambda_{A,+}$ is real and positive and eigen-direction $e^{i\Theta_{A,+}}\simeq -i$ points toward the negative $y$ direction for sufficiently small $\epsilon_\alpha$~(see Eqs.~\eqref{eq:lambda_A2_app} and~\eqref{eq:Lambda_A2_phase}). 
Due to this unstable normal mode, the two flocks turn their polarization by $-\pi/2$ upon collision, and flow eventually along the boundary between them in the opposite direction. Consequently, the two species cannot mix.

\begin{figure}
    \includegraphics[width=\columnwidth]{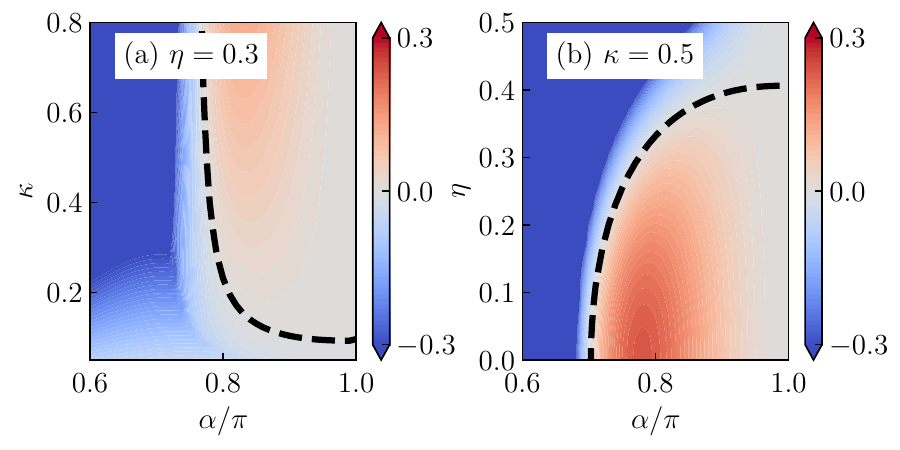}
    \caption{Real part of the Lyapunov exponent $\Lambda_{A,+}$ for the anti-symmetric part $\delta w_A$ in the $\alpha-\kappa$ plane with fixed $\eta = 0.3$ in (a) and in the $\alpha-\eta$ plane with fixed $\kappa = 0.5$ in (b). The thick dashed lines represent the stability boundary $\Re[\Lambda_{A,+}]=0$.}
    \label{fig:Lyapunov}
\end{figure}

For general $\eta$, we can evaluate the Lyapunov exponents numerically.
Lyapunov exponents for the symmetric part remain negative. On the other
hand, one normal mode for the antisymmetric part has a real positive Lyapunov exponent in a broad parameter range. We present the real part of the most relevant Lyapunov exponent $\Lambda_{A,+}$ in Fig.~\ref{fig:Lyapunov}. The stability boundary, $\Re[\Lambda_{A,+}]=0$, has a similar shape as the phase boundary of the SS phase. We conclude that the instability is responsible for the SS and the VC pattern. 

Species separation, such as cell sorting during embryogenesis, is an important phenomenon in biological processes. It has been shown that a mixture of two active matter systems with different mechanical properties can phase separate~\cite{Rozman.2024,Graham.2024}.  Our model reveals that nonreciprocity gives rise to species separation of multiple species that are {\em equivalent} to each other.  Vortex cell patterns have been reported in various experimental systems~\cite{Riedel.2005, Sumino.2012, Han.2020}. Some studies show that geometric confinement creates vortex cells~\cite{Wioland.2013, Opathalage.2019, Nishiguchi.2025}.  Other studies show that particles with intrinsic chirality can self-organize into a vortex cell pattern~\cite{Sumino.2012, Denk.2015, Han.2020, Faluweki.2023, Cammann.2024}. Our model can serve as a minimal model for vortex cell patterns in active matter systems. 

\subsection{Coexistence phase}\label{sec:coexist}

\begin{figure}
    \includegraphics[width=\columnwidth]{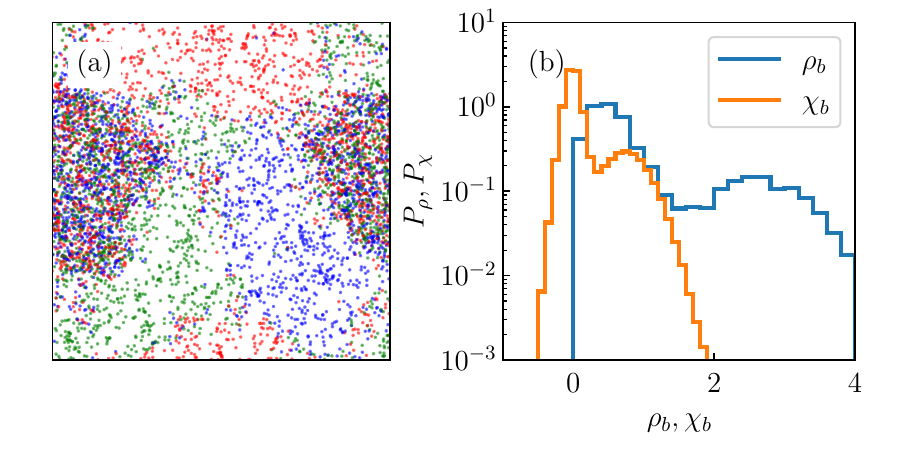}
    \caption{(a) Snapshot of a particle configuration for the three-species system in the coexistence phase. (b) Probability distribution of a local particle density $\rho_b$ and a local chirality $\chi_b$ at square meshes of size $b\times b$ with $b=2$. These quantities are normalized by $\rho_{\rm tot.}b^2$. The double-peak structure manifests the coexistence. Parameters: $Q=3$, $L= 128$, $\rho_0=2.0$, $\eta=0.55$, $\alpha=0.6\pi$.}
    \label{fig:snapshot_coexist}
\end{figure}

The species-mixed chiral phase and the species-separated vortex cell phase can coexist for intermediate values of $\alpha$. Figure~\ref{fig:snapshot_coexist} demonstrates this coexistence in the three-species system. The probability distribution functions of the local density and the local chirality density are characterized by double peaks. These peaks correspond to the species-mixed chiral phase~(high density and positive chirality) and the SS phase~(low density and negative chirality), respectively.
{The low particle density in the SS domain renders the
species-separation order parameter $e_{\rm s}$ smaller than in the SS phase
in Fig.~\ref{fig:OrderParametersQ3}.}

The system is in a {\em dynamical equilibrium} in the coexistence phase. Inside a high-density species-mixed chiral cluster, particles perform counter-clockwise chiral motion with almost identical but fluctuating phases. Phase coherence is reinforced by the nonreciprocal inter-species interaction. Such an interaction becomes weaker near a cluster boundary. Thus, a group of one species losing phase coherence with the other species can leak from the cluster, which leads to a nucleation of a vortex cell. A vortex cell can also be absorbed into a chiral cluster. These nucleation and absorption processes are balanced in the steady state.

We quantify this coexistence with the {\em bimodality coefficient} $\beta_\rho$ of the local density distribution function~\footnote{The bimodality coefficient $\beta_x$ of a random variable $x$ is defined as $(1+m_3^2(x))/m_4(x)$ where $m_3$ and $m_4$ are the skewness and the kurtosis~\cite{Knapp.2007}.}. The bimodality coefficient takes the minimum value of $0$ for a unimodal Gaussian distribution and the maximum value of $1$ for a distribution with two distinct delta peaks. It equals $5/9$ for a uniform distribution, which is taken as a bimodality threshold phenomenologically~\cite{Knapp.2007}. Using this threshold value, we construct the phase boundary of the coexistence phase in Fig.~\ref{fig:OrderParametersQ3}. 

\begin{figure}
    \includegraphics*[width=\columnwidth]{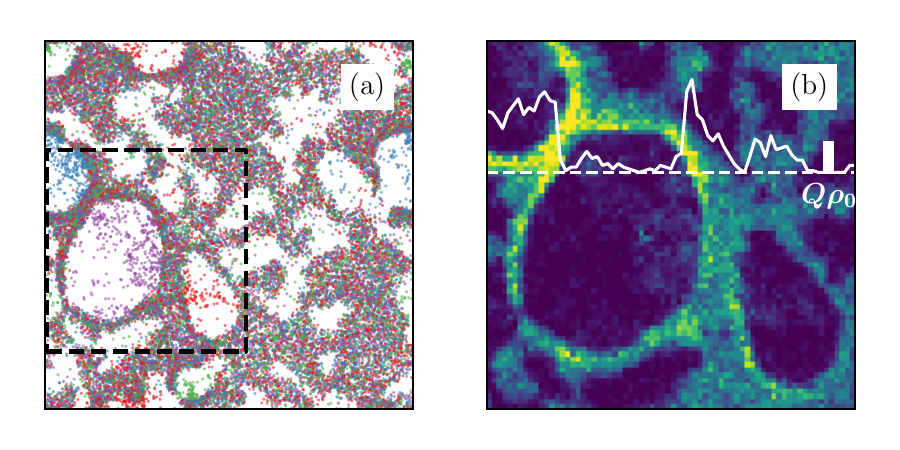}
    \caption{Bubble state in the coexistence phase for the four-species system with $L=128$, $\rho_0=3$, $\eta=0.3$, and $\alpha = 0.72\pi$. (a) Snapshot of individual particles colored according to their species. Each bubble is inhabited by particles of single species. The local density field within the dashed square region in (a) is drawn in (b). The brighter the image is, the higher the local density is. The density field across the horizontal section, represented by a dashed line, is plotted with solid line. The thick vertical segment represents the scale for the overall particle density $\rho_{\rm tot.} = \rho_0 Q$. The bubble boundary is densely populated.}
    \label{fig:snapshot_bubble}
\end{figure}

\begin{figure}
    \includegraphics[width=\columnwidth]{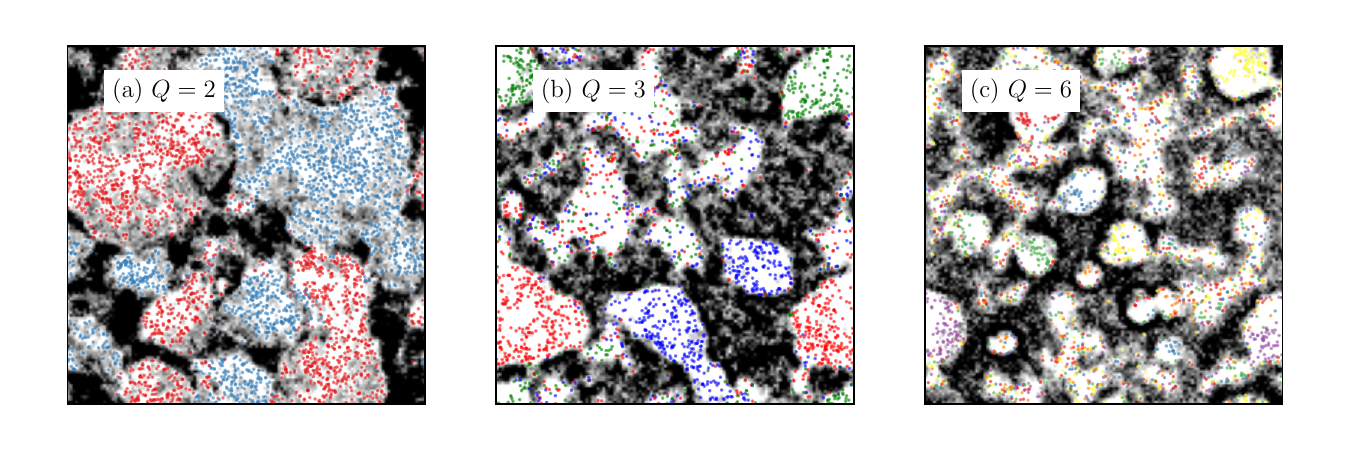}
    \caption{Bubble states for $Q=2, 3, 6$ with $(\alpha/\pi, \eta) = (0.40, 0.37)$, $(0.57, 0.30)$, and $(0.86, 0.30)$, respectively. Particles only in a low density region with $\rho<\rho_{\rm tot.}/2$ are plotted with dots colored according to their species. A high density region with $\rho>\rho_{\rm tot.}/2$ is filled in gray scale. The darker, the local density is higher. Bubbles inhabited by a single species are bounded by a sharp boundary. On the other hand, bubbles inhabited by multiple species have an indistinct boundary. These are annihilating bubbles. Parameters: $L=128$ and $\rho_{\rm tot.}=12$.}
    \label{fig:bubbleQ236}
\end{figure}

The coexistence phase features intriguing dynamical patterns depending on
relative areal fractions of the chiral and the SS phases. For instance, we
observe a {\em bubble state}, in which a species-mixed chiral background is
punctured by vortex cells~(see Fig.~\ref{fig:snapshot_bubble} for $Q=4$ and
Fig.~\ref{fig:bubbleQ236} for other values of $Q$). A bubble is nucleated
spontaneously when a coherent group of single species is separated from the
others due to statistical fluctuations. These particles interact
nonreciprocally with surrounding particles acquiring clockwise chirality. As
they flow and push other species particles away, a bubble grows with a sharp
boundary.  However, these bubbles are not robust. As a bubble grows, an
inhabiting chiral gas becomes dilute and cannot exert enough pressure
against background particles. Consequently, such a bubble dissolves
eventually. Bubbles can also be annihilated in pairs. When two bubbles of
different species collide with each other and merge into one, particles are
mixed. Then, chiral motion along the bubble boundary weakens, which
destabilizes both bubbles~(see Fig.~\ref{fig:bubbleQ236}). The emergent
dynamical pattern of the bubble as well as the vortex cell array can be
useful in control and microfabrication of active mixture. 
Quantitative and
theoretical understanding of these emergent dynamical pattern is in order. 
{It would also be interesting to investigate whether the coexistence
phase could be further separated into sub-phases, which is not pursued in
this work.} 

In the coexistence phase, the system is phase-separated into high- and
low-density regions. Phase separation is common in active matter systems,
and various field theoretical models have been proposed to explain the
mechanism. The active Cahn-Hilliard equation~\cite{Speck.2014} and the
$\phi^4$-type scalar field theory~\cite{Wittkowski.2014,Tjhung.2018}
successfully describe phase separation in single-component active matter.
The nonreciprocal multi-component Cahn-Hilliard equation  displays phase
separation and pattern formation~\cite{Saha.2020}. It will be interesting to
establish a theoretical framework for species separation in a nonreciprocal
multi-species system with Potts symmetry, like the $Q$-NRVM. The coexistence
phase was not captured by the linear stability analysis. A theoretical
framework for species separation in the $Q$-NRVM, possibly by extending the
linear stability analysis to include $\bm{r}$-dependent perturbations, is warranted.

\section{Discussion and outlook}
\label{sec:discussion}

This work proposes a $Q$-species Vicsek model and a corresponding continuum hydrodynamic equation as a minimal model for nonreciprocal active matter systems with Potts symmetry ($S_Q$ permutation symmetry). The unique feature of our model is the introduction of a constant phase shift in the velocity alignment interaction. This phase shift makes the mutual interaction nonreciprocal while maintaining symmetric coupling amplitudes between all species. In contrast, most prior studies incorporate nonreciprocity through asymmetric coupling amplitudes among non-equivalent agents. 

The nonreciprocal phase shift $\alpha$ is the origin of the emergent
collective chiral motion, counter-clockwise or clockwise. For small
$\alpha$, particles are well mixed and perform counterclockwise chiral
motion characterized by QLRO.  
{We argue that the emergent chirality impedes particle motility,
    rendering the motion of chiral particles diffusive.
Thus, the chiral particles can display QLRO as diffusing $XY$ spins.}

{A recent work reports that two-dimensional single species chiral Malthusian flocks
    belong to the universality class of compact Kardar-Parisi-Zhang~(KPZ)
equation~\cite{Chen.2026}. In this class, a QLRO phase is unstable on 
asymptotically large length scales~\cite{Lauter.2017, Daviet.2025}. 
We remark that the density field is not slaved to the polarization field in
our model. Thus, the theory of Ref.~\cite{Chen.2026} cannot be applied to
our model.} 

For large $\alpha$, species separation occurs and a vortex cell pattern emerges. Each vortex cell is predominantly inhabited by a single species. Particles in adjacent cells interact along the cell boundary to generate clockwise chiral motion. The perturbative linear stability analysis in Sec.~\ref{sec:phase_separation} reveals an instability that the polarizations of two distinct species flocks should turn by $\pi/2$ upon head-on collision. This instability generates an effective repulsion among different species particles and is responsible for species separation. We note that this rotation of polarization can be regarded as an odd viscosity phenomenon~\cite{Banerjee.2017}.

We found a coexistence phase in which the QLRO chiral phase and the SS phase coexist in space. Phase separation is common in active matter systems~\cite{Speck.2014, Wittkowski.2014, Tjhung.2018, Saha.2020}.
The coexistence phase in our model cannot be fully accounted for by
existing active matter theories for phase separation: First, our model
possesses permutation symmetry, and second, the density field is coupled to
the chirality field emerging from the nonreciprocal interactions. Therefore,
the coexistence phase calls for an effective field theory. 

{We demonstrated numerically that the Boltzmann equation system captures
the species-mixed chiral state and the species-separated state with vortex
cells. However, whether the continuum equation also accommodates the highly
heterogeneous coexistence state remains an open question.
Investigating the complete phase diagram of the continuum equation is
another important avenue for future research. A comprehensive
understanding of the continuum system requires substantial computational
resources and will be pursued in future work.}

In summary, we have established a minimal model for multi-species active chiral fluids with Potts symmetry. One may consider species-dependent phase shifts in the velocity-alignment interaction to explore systems with broken $S_Q$ symmetry. It will also be interesting to investigate its transport properties, particularly in view of the odd viscosity and odd diffusivity. We leave these for future work.

The data that support the findings of this article are openly available~\cite{github}.

\begin{acknowledgments}
We acknowledge useful discussions with Masaki Sano and Euijoon Kwon. 
\end{acknowledgments}

\appendix

\section{Explicit expressions for interaction kernels}\label{sec:J_explicit}
The interaction kernels $J^{kl}_{\rm R}$ and $J^{kl}_{\rm NR}$ represent the coupling amplitudes between the $k$th and $l$th Fourier modes of the same and different species, respectively. They are given by
\begin{equation}\label{eq:Js}
\begin{split}
    J^{kl}_{\rm R} =& P_k L^{kl}_{\rm R} - I_l,  \\
    J^{kl}_{\rm NR} =& P_k L^{kl}_{\rm NR} - I_l , 
\end{split}
\end{equation}
where
\begin{equation}\label{eq:PILs}
\begin{split}
    P_k =& \int_{-\pi}^\pi d\phi \ P_{\rm n}(\phi) e^{ik\phi} = e^{-k^2 \eta^2/2} \\
    I_k =& \int_{-\pi}^\pi \frac{d\phi}{2\pi} K(\phi)e^{ik\phi} \\
    L^{kl}_{\rm R} =& \int_{-\pi}^\pi \frac{d\phi}{2\pi} K(\phi) e^{ik \Theta_{\rm R}(\phi) - il \phi}  \\
    L^{kl}_{\rm NR} =& \int_{-\pi}^\pi \frac{d\phi}{2\pi} K(\phi) e^{ik \Theta_{\rm NR}(\phi) - il \phi}
\end{split}
\end{equation}
with the noise distribution function $P_{\rm n}(\phi)$ in Eq.~\eqref{eq:noise_distribution}, the scattering angle $\Theta_{\rm R, NR}$ in Eq.~\eqref{eq:scattering_angle}, and  the scattering cross section $K(\phi)$ in Eq.~\eqref{eq:cross_section}.

We note that $I_k$ and $L^{kl}_{\rm R, NR}$ involve Fourier coefficients of $K(\phi)$. Thus, it is convenient to introduce an auxiliary function
\begin{equation}\label{eq:aux_func_app}
F(c,\beta) := \int_{\beta}^\pi d\theta~ \left|\sin\frac{\theta}{2}\right| e^{i c\theta} 
\end{equation}
for an integer or a half-integer $c$. Then, one obtains that
\begin{equation}\label{eq:coupling_app}
\begin{split}
    I_k =& \kappa F(k,-\pi) = \frac{4\kappa}{1-4k^2},  \\
    L^{kl}_{\rm R} =& \kappa F\left(\frac{k}{2}-l,-\pi\right),\\
    L^{kl}_{\rm NR} =& \kappa e^{\frac{i}{2} k \alpha } \left[ F\left(\frac{k}{2}-l,-\pi\right) \right. \\
    & \quad \quad \quad \left. - (1-e^{-i\pi k}) F\left(\frac{k}{2}-l,\pi-\alpha\right)\right] .
\end{split}
\end{equation}
The nonreciprocity parameter $\alpha$ appears in $L^{kl}_{\rm NR}$, hence in $J^{kl}_{\rm NR}$.


\section{Order parameters and snapshots}\label{app:numerics_app}
 
In the main text, we presented the numerical result primarily for the three-species case. Here, we present supplemental numerical data for other values of $Q$.

\begin{figure*}
    \includegraphics[width=1.8\columnwidth]{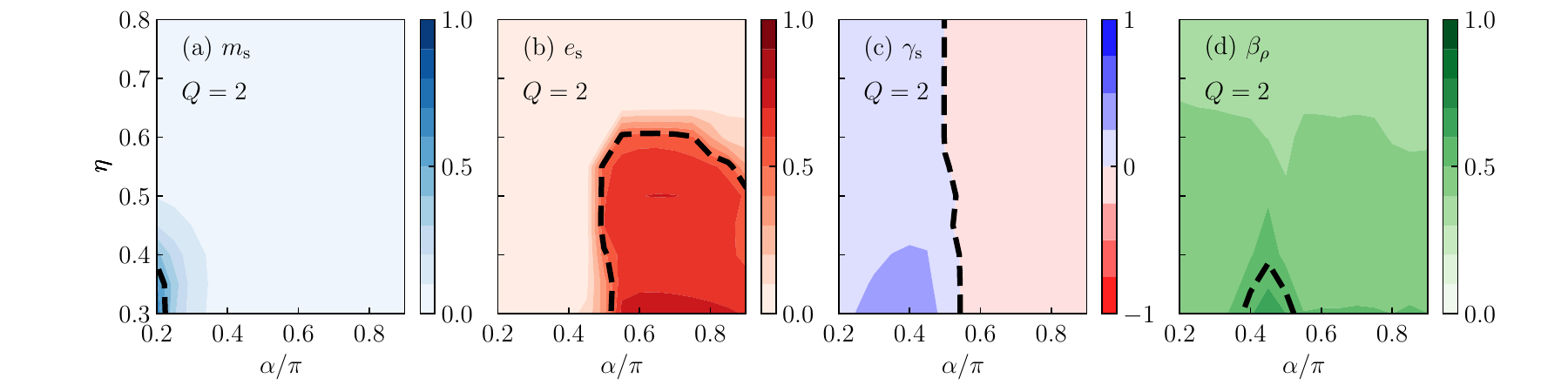} \\
    \includegraphics[width=1.8\columnwidth]{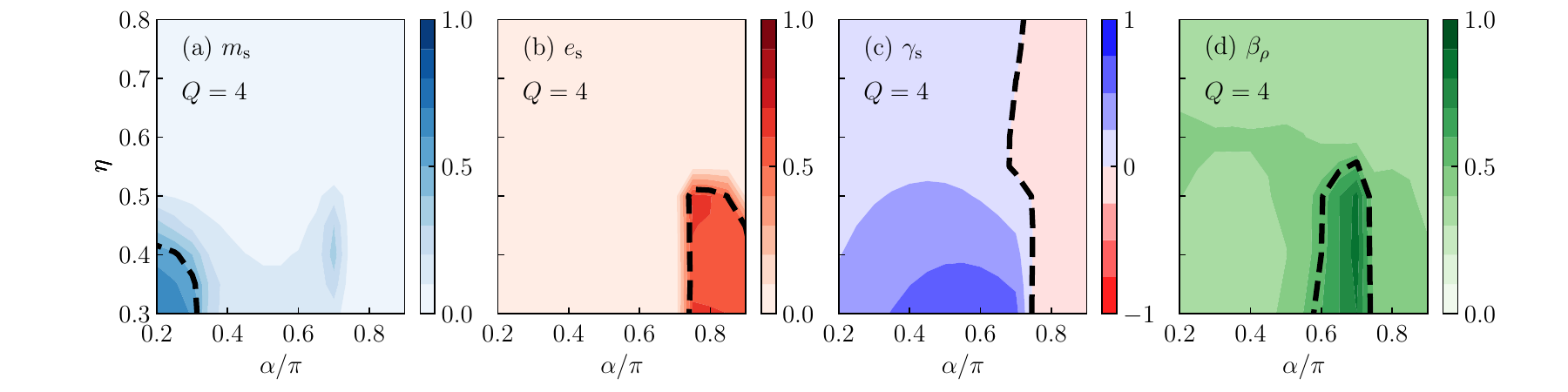}
    \includegraphics[width=1.8\columnwidth]{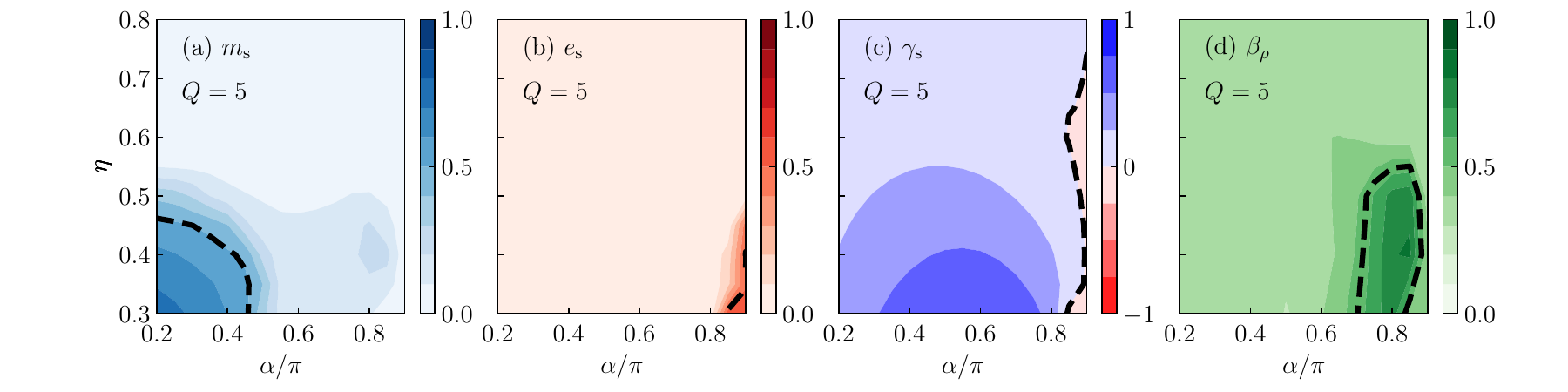}
    \includegraphics[width=1.8\columnwidth]{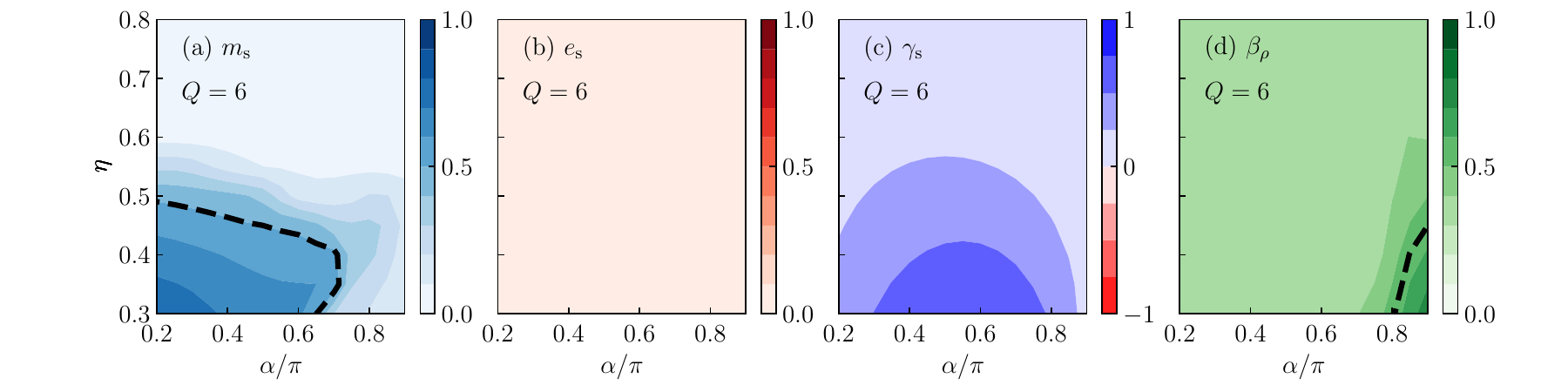}
    \caption{Density plot of $m_{\rm s}$,  $e_{\rm s}$,  $\gamma_{\rm s}$, and $\beta_\rho$ in the $\alpha-\eta$ plane for $Q=2$ with $\rho_0=2$ and $Q=4, 5, 6$ with  $\rho_0=1$. Numerical simulations were performed on a system of size $128\times 128$. The contour lines drawn at $m_{\rm s}=0.5$,  $e_{\rm s}=0.5$, $\gamma_{\rm s}=0$, and  $\beta_\rho=5/9$, are guides for the eye.}
    \label{fig:MandEQ3}
\end{figure*}

Figure~\ref{fig:MandEQ3} compares the order parameters in the $\alpha-\eta$
plane for $Q=2, 4, 5, 6$. These results suggest that the $Q$-NRVM displays
the four distinct phases irrespective of the $Q$ values. On the other hand,
the phase boundaries shift toward the large $\alpha$ side as $Q$ increases.
The SS phase is invisible for $Q=5$ and $6$ in these plots. We have confirmed the SS phase using higher values of $\rho_0$.

\begin{figure}
    \includegraphics[width=\columnwidth]{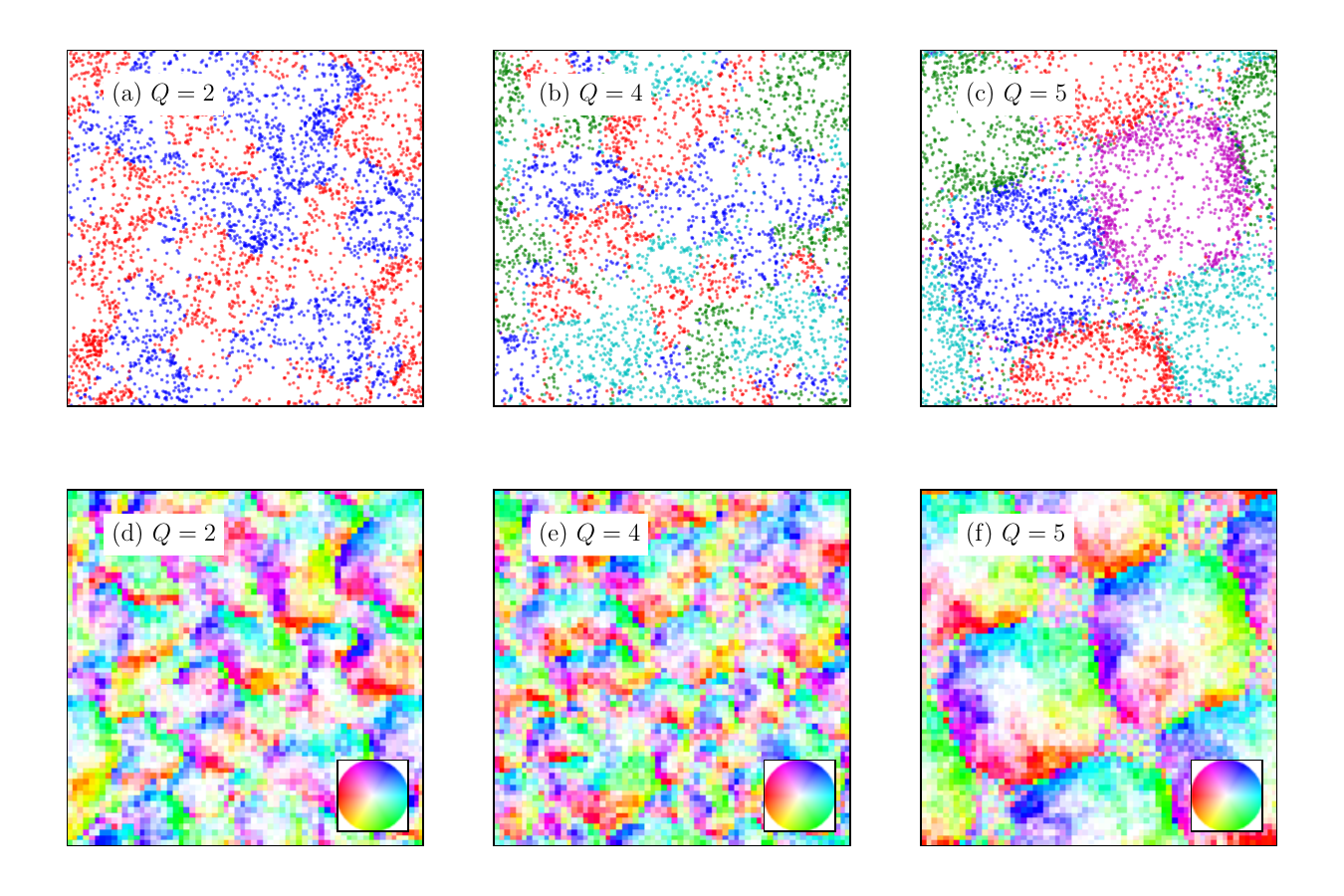}
    \caption{Snapshots of a particle configuration and a polarization field
    in the SS phase for $Q=2$ in (a, d), $Q=4$ in (b,e), $Q=5$ in (c,f).
Parameters: $(\rho_0, \alpha/\pi, \eta) = (2.0, 0.8, 0.3)$ for $Q=2$, $(1.0,
0.8, 0.35)$ for $Q=4$, and $(1.0, 0.95, 0.3)$ for $Q=5$.}
    \label{fig:SS245}
\end{figure}

Figure~\ref{fig:SS245} presents snapshots in the SS phase for $Q=2$, $4$, and $5$. Species separation is manifest in the particle configurations. The polarization fields clearly demonstrate vortex cells with clockwise chirality. 

\section{Density field near vortex and antivortex cores}
\label{app:vortex_antivortex}

Vortex and antivortex excitations generate density fluctuations.
Consider an idealized situation in which a perfect in-phase chiral state is perturbed by a single vortex-antivortex pair:  Particles revolve on circular orbits whose centers are located regularly at positions $u_n = u_{n,x} + i u_{n,y}$ with $u_{n,x}, u_{n,y} \in \mathbb{Z}$ forming a square lattice. Particle $n$ moves along a trajectory $z_n(t) = u_n + \frac{v_0}{i\Omega_0} e^{i(\Omega_0 t + \phi_n)}$ with velocity $v_n = v_0 e^{i(\Omega_0 t + \phi_n)}$, regardless of its species, with different phase $\phi_n$ and identical angular frequency $\Omega_0$. 

\begin{figure}[t]
    \includegraphics[width=\columnwidth]{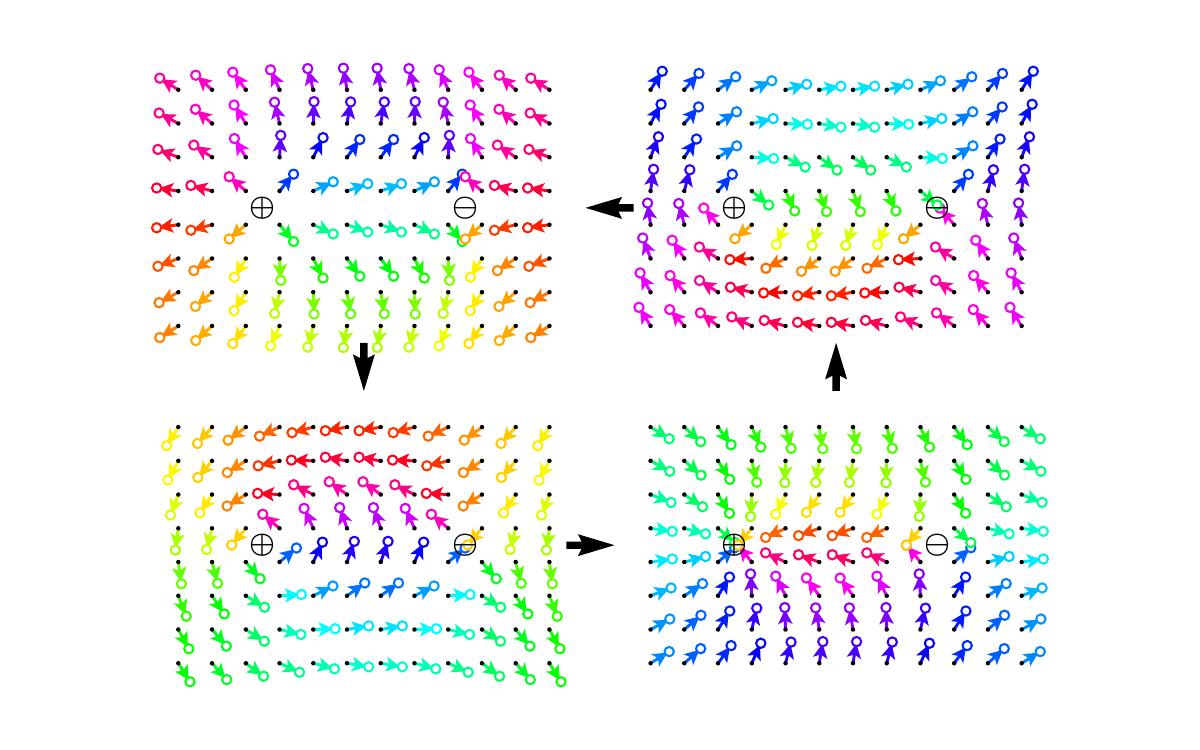}
    \caption{Illustration of a series of particle configurations in the presence of a vortex at position marked with $\oplus$ and an antivortex at position marked with $\ominus$. Particles are marked with an empty circle and orbit centers with a dot. The displacement from the orbit center is represented with an arrow color-coded according to the direction.}
    \label{fig:vapair}
\end{figure}

A vortex at $z=z_{\oplus}$ and an antivortex at $z=z_{\ominus}$ are introduced by choosing
\begin{equation}
    \phi_n = \Arg~[u_n-z_\oplus] + \Arg~[(u_n-z_\ominus)^*] 
\end{equation}
We illustrate four snapshots taken consecutively in a time step of $\Delta t = \frac{1}{4}\frac{2\pi}{\psi}$ in Fig.~\ref{fig:vapair}. The snapshots demonstrate that these topological excitations generate a time-periodic modulation in the particle density field. An oscillating monopole moment develops near the vortex core, while a rotating quadrupole moment develops near the antivortex core. 

We derive the exact density modulation due to a vortex or an antivortex excitation in the otherwise in-phase chiral state in the continuum limit. Let $\bm{r}(\bm{u},t)= (x(\bm{u},t), y(\bm{u},t))$ denote the position vector of a particle revolving on a circular orbit centered at $\bm{u}=(u_x, u_y)$ with angular frequency $\Omega_0$.
The orbit centers are distributed uniformly with a constant density function $\rho_c(\bm{u}) = \rho$ with $\rho = \rho_0 Q$.
We use the complex coordinate $z=x+iy$ and its complex conjugate $z^*$, the polar coordinate $r=\sqrt{x^2+y^2}$ and $\theta = \tan^{-1}(y/x)$, as well as the Cartesian coordinate. We will use these systems interchangeably.

In the chiral state, the particle position $z = x + i y$ is given by the mapping
\begin{equation}
    z = f(u,u^*, t) = u + r_0 e^{i(\Omega_0 t+\phi(u, u^*))}
\end{equation}
from $u=u_x+iu_y$ and its complex conjugate $u^*$, where $r_0$ is the radius of the orbit, $\Omega_0$ is the angular frequency, and the phase field $\phi(u,u^*)$ describes an excitation from the in-phase chiral state. Assuming particles are well mixed, we do not distinguish the particle species. 

In the presence of a vortex excitation with its core at the origin $u=u^*=0$,  the phase field $\phi(u,u^*)$ becomes
\begin{equation}
    \phi_\oplus(u, u^*) = \Arg[u] = \tan^{-1}\left(\frac{u_y}{u_x}\right),
\end{equation}
and the mapping is given by   
\begin{equation}\label{eq:fvortex_app}
    f_\oplus(u,u^*, t) = u + r_0 \frac{u}{\sqrt{u u^*}} e^{i\Omega_0 t} .
\end{equation}
The particle density is then given by
\begin{equation}
    \rho_\oplus(z,z^*,t) = \rho_c(u, u^*)  \left| \frac{\partial (f_\oplus, f_\oplus^*)}{\partial (u,u^*)}\right|^{-1} .
\end{equation}
The last term is the Jacobian of the transformation~\eqref{eq:fvortex_app}, which equals
\begin{equation}
    \left| \frac{\partial (f_\oplus, f_\oplus^*)}{\partial (u,u^*)}\right| = 1 + \frac{r_0}{\sqrt{u u^*}} \cos\Omega_0 t .
\end{equation}
The transformation~\eqref{eq:fvortex_app} is inverted to yield that $z = f_\oplus \simeq u$ for $|u|, |z| \gg r_0$. Thus, we  obtain the density function
\begin{equation}\label{eq:rho_vortex_app}
    \rho_\oplus(\bm{r},t) \simeq \rho \left(1 - \frac{r_0}{r} \cos\Omega_0 t \right)
\end{equation}
away from the vortex core~($r \gg r_0$).
The vortex excitation induces an isotropic {\em monopole} moment in the particle density. Its amplitude decays algebraically with the distance from the core as $1/r$ and oscillates with the angular frequency $\Omega_0$.

In the presence of an antivortex with its core at the origin $u=u^*=0$, the phase field becomes
\begin{equation}
    \phi_\ominus(u, u^*) = \Arg[u^*] = -\tan^{-1}\left(\frac{u_y}{u_x}\right),
\end{equation}
which yields that  
\begin{equation}\label{eq:fantivortex_app}
    f_\ominus(u,u^*, t) = u + r_0 \frac{u^*}{\sqrt{u u^*}} e^{i\Omega_0 t} .
\end{equation}
It is straightforward to derive that 
\begin{equation}
    \left| \frac{\partial (f_\ominus, f_\ominus^*)}{\partial (u,u^*)}\right| = 1 - \frac{r_0}{\sqrt{u u^*}} \cos(\Omega_0 t + 2 \phi_\ominus(u,u^*))  .
\end{equation}
The transformation~\eqref{eq:fantivortex_app} is inverted to yield $z \simeq u$ for $|z|, |u| \gg r_0$. Thus, we finally obtain that
\begin{equation}\label{eq:rho_antivortex_app}
    \rho_\ominus(\bm{r},t) \simeq \rho \left(1 + \frac{r_0}{r} \cos(\Omega_0 t-2\theta) \right)
\end{equation}
away from the vortex core, $r \gg r_0$.
The anti-vortex excitation induces a {\em quadrupole} moment in the particle density. The amplitude decays algebraically with the distance from the core as $1/r$. The principal direction rotates with the angular frequency $\Omega_0$. 

\section{Perturbative analysis of the hydrodynamic equation for the two-species system}
\label{app:theory_for_sp}

We present a normal-mode analysis for the linearized hydrodynamic equations~\eqref{eq:dwSdt} and~\eqref{eq:dwAdt} in a limiting case with $\eta=0$. Before addressing the problem, we first document the normal mode solution for a general linear equation for a complex variable $z(t)$
\begin{equation}\label{eq:generalLinear}
\partial_t z = p z + q z^*
\end{equation}
with complex coefficients $p$ and $q$. It can be recast into a coupled linear system for $z$ and $z^*$:
\begin{equation}
    \begin{pmatrix} \partial_t z \\ \partial_t z^* \end{pmatrix} = \begin{pmatrix} p & q \\ q^* & p^* \end{pmatrix} 
    \begin{pmatrix} z \\ z^* \end{pmatrix} ,
\end{equation}
whose normal modes can be found from eigenvectors and eigenvalues of the $2\times 2$ matrix. Explicitly, the eigenvalues are given by
\begin{equation}\label{eq:eigenvalue}
\Lambda_\pm = \Re[p] \pm \sqrt{ |q|^2 - \Im[p]^2},
\end{equation}
with the corresponding eigenvectors $(1,(\Lambda_{\pm}-p)/q)^T$.
For $|q|^2 \geq \Im[p]^2$, in particular, the two eigenvalues are real and the normal modes are given by
\begin{equation}
    z_\pm(t) = e^{i\Theta_\pm} e^{\Lambda_{\pm} t}
\end{equation}
with the phase angle
\begin{equation}\label{eq:eigenphase}
    \Theta_{\pm} = \frac{1}{2}\Arg \left[ \frac{q}{\Lambda_{\pm}-p}  \right] .
\end{equation}

At $\eta=0$, the symmetric mode is governed by $\partial_t \delta w_S = a_1 \delta w_S + a_2 \delta w_S^*$ with
\begin{equation}\label{eq:coeffS_app}
    \begin{split}
        a_1 &= -\frac{4\kappa}{39} \left(21\pi- 29\right) + O(\epsilon_\alpha^2), \\
        a_2 &= -\frac{6(\pi-2)\kappa}{91} \left( 7 - 6 i \epsilon_\alpha \right)+O(\epsilon_\alpha^2). 
    \end{split}
\end{equation}
Using Eq.~\eqref{eq:eigenvalue}, we find that two normal modes for $\delta w_S$ have negative Lyapunov exponents $\Lambda_{S,+} \approx -\frac{2}{39}(33\pi-40)\kappa$ and $\Lambda_{S,-} \approx -\frac{2}{39}(51\pi-76)\kappa$. Therefore, the two flocks flow in the opposite direction for small $\epsilon_\alpha = \pi-\alpha$ and at $\eta=0$.

At $\eta=0$, the anti-symmetric mode is governed by $\partial_t \delta w_A = b_1 + (b_1 + b_2) \delta w_A + b_2 \delta w_A^*$ with 
\begin{equation}\label{eq:coeffA_app}
    \begin{split}
        b_1 &= -\frac{i\kappa}{7}\left(24- 5\pi\right)\epsilon_\alpha+O(\epsilon_\alpha^2), \\
        b_2 &=-\frac{2(\pi-2)\kappa}{7} \left( 7 - 6i\epsilon_\alpha\right)+O(\epsilon_\alpha^2) .
    \end{split}
\end{equation}
The inhomogeneous term $b_1=O(\epsilon_\alpha)$ with negative imaginary part accounts for a clockwise bending of $\delta w_A$. The inhomogeneous term also renders a fixed point at 
\begin{equation}
    \delta w_{A,f} = i \frac{28(\pi-2)}{(29\pi-72)\epsilon_\alpha} + O(\epsilon_\alpha^0).
\end{equation}
Stability is determined by the Lyapunov exponent. Using Eq.~\eqref{eq:eigenvalue}, we obtain that
\begin{align}
    \Lambda_{A,-} &= -3(\pi-2)\kappa + O(\epsilon_\alpha) \\
    \Lambda_{A,+} &= \frac{47\pi^2 + \pi-288}{98(\pi-2)} \kappa \epsilon_\alpha^2 + O(\epsilon_\alpha^3) . \label{eq:lambda_A2_app}
\end{align}
The first Lyapunov exponent $\lambda_{A,-}$ is negative. Thus, the normal mode associated with it is irrelevant. However, the second Lyapunov exponent $\Lambda_{A,+}$ is real and positive  with the phase angle 
\begin{equation}\label{eq:Lambda_A2_phase}
\Theta_{A,+} \simeq -\pi/2 + O(\epsilon_\alpha).
\end{equation}
Thus, the anti-symmetric mode grows indefinitely as $\delta w_A(t) \sim e^{i \Theta_{A,+} \Lambda_{A,+}t} = -i e^{\Lambda_{A,+}t}$. Recalling that $w^1 = w_\pi + \delta w_S/2 + w_\pi \delta w_A$ and $w^2 = -w_\pi + \delta w_S/2 - w_\pi \delta w_A$, we conclude that the unstable mode  $e^{i\Theta_{A,+} \Lambda_{A,+}t}$ is responsible for the effective repulsion between different species.

\bibliography{preprint_rev}

\begin{thebibliography}{92}%
\makeatletter
\providecommand \@ifxundefined [1]{%
 \@ifx{#1\undefined}
}%
\providecommand \@ifnum [1]{%
 \ifnum #1\expandafter \@firstoftwo
 \else \expandafter \@secondoftwo
 \fi
}%
\providecommand \@ifx [1]{%
 \ifx #1\expandafter \@firstoftwo
 \else \expandafter \@secondoftwo
 \fi
}%
\providecommand \natexlab [1]{#1}%
\providecommand \enquote  [1]{``#1''}%
\providecommand \bibnamefont  [1]{#1}%
\providecommand \bibfnamefont [1]{#1}%
\providecommand \citenamefont [1]{#1}%
\providecommand \href@noop [0]{\@secondoftwo}%
\providecommand \href [0]{\begingroup \@sanitize@url \@href}%
\providecommand \@href[1]{\@@startlink{#1}\@@href}%
\providecommand \@@href[1]{\endgroup#1\@@endlink}%
\providecommand \@sanitize@url [0]{\catcode `\\12\catcode `\$12\catcode
  `\&12\catcode `\#12\catcode `\^12\catcode `\_12\catcode `\%12\relax}%
\providecommand \@@startlink[1]{}%
\providecommand \@@endlink[0]{}%
\providecommand \url  [0]{\begingroup\@sanitize@url \@url }%
\providecommand \@url [1]{\endgroup\@href {#1}{\urlprefix }}%
\providecommand \urlprefix  [0]{URL }%
\providecommand \Eprint [0]{\href }%
\providecommand \doibase [0]{https://doi.org/}%
\providecommand \selectlanguage [0]{\@gobble}%
\providecommand \bibinfo  [0]{\@secondoftwo}%
\providecommand \bibfield  [0]{\@secondoftwo}%
\providecommand \translation [1]{[#1]}%
\providecommand \BibitemOpen [0]{}%
\providecommand \bibitemStop [0]{}%
\providecommand \bibitemNoStop [0]{.\EOS\space}%
\providecommand \EOS [0]{\spacefactor3000\relax}%
\providecommand \BibitemShut  [1]{\csname bibitem#1\endcsname}%
\let\auto@bib@innerbib\@empty
\bibitem [{\citenamefont {Marchetti}\ \emph {et~al.}(2013)\citenamefont
  {Marchetti}, \citenamefont {Joanny}, \citenamefont {Ramaswamy}, \citenamefont
  {Liverpool}, \citenamefont {Prost}, \citenamefont {Rao},\ and\ \citenamefont
  {Simha}}]{Marchetti.2013}%
  \BibitemOpen
  \bibfield  {author} {\bibinfo {author} {\bibfnamefont {M.~C.}\ \bibnamefont
  {Marchetti}}, \bibinfo {author} {\bibfnamefont {J.~F.}\ \bibnamefont
  {Joanny}}, \bibinfo {author} {\bibfnamefont {S.}~\bibnamefont {Ramaswamy}},
  \bibinfo {author} {\bibfnamefont {T.~B.}\ \bibnamefont {Liverpool}}, \bibinfo
  {author} {\bibfnamefont {J.}~\bibnamefont {Prost}}, \bibinfo {author}
  {\bibfnamefont {M.}~\bibnamefont {Rao}},\ and\ \bibinfo {author}
  {\bibfnamefont {R.~A.}\ \bibnamefont {Simha}},\ }\bibfield  {title} {\bibinfo
  {title} {{Hydrodynamics of soft active matter}},\ }\href
  {https://doi.org/10.1103/revmodphys.85.1143} {\bibfield  {journal} {\bibinfo
  {journal} {Rev. Mod. Phys.}\ }\textbf {\bibinfo {volume} {85}},\ \bibinfo
  {pages} {1143 } (\bibinfo {year} {2013})}\BibitemShut {NoStop}%
\bibitem [{\citenamefont {Cates}\ and\ \citenamefont
  {Tailleur}(2015)}]{Cates.2014}%
  \BibitemOpen
  \bibfield  {author} {\bibinfo {author} {\bibfnamefont {M.~E.}\ \bibnamefont
  {Cates}}\ and\ \bibinfo {author} {\bibfnamefont {J.}~\bibnamefont
  {Tailleur}},\ }\bibfield  {title} {\bibinfo {title} {{Motility-Induced Phase
  Separation}},\ }\href
  {https://doi.org/10.1146/annurev-conmatphys-031214-014710} {\bibfield
  {journal} {\bibinfo  {journal} {Annu. Rev. Condens. Matter Phys.}\ }\textbf
  {\bibinfo {volume} {6}},\ \bibinfo {pages} {219} (\bibinfo {year}
  {2015})}\BibitemShut {NoStop}%
\bibitem [{\citenamefont {Bechinger}\ \emph {et~al.}(2016)\citenamefont
  {Bechinger}, \citenamefont {Leonardo}, \citenamefont {Löwen}, \citenamefont
  {Reichhardt}, \citenamefont {Volpe},\ and\ \citenamefont
  {Volpe}}]{Bechinger.2016}%
  \BibitemOpen
  \bibfield  {author} {\bibinfo {author} {\bibfnamefont {C.}~\bibnamefont
  {Bechinger}}, \bibinfo {author} {\bibfnamefont {R.~D.}\ \bibnamefont
  {Leonardo}}, \bibinfo {author} {\bibfnamefont {H.}~\bibnamefont {Löwen}},
  \bibinfo {author} {\bibfnamefont {C.}~\bibnamefont {Reichhardt}}, \bibinfo
  {author} {\bibfnamefont {G.}~\bibnamefont {Volpe}},\ and\ \bibinfo {author}
  {\bibfnamefont {G.}~\bibnamefont {Volpe}},\ }\bibfield  {title} {\bibinfo
  {title} {{Active particles in complex and crowded environments}},\ }\href
  {https://doi.org/10.1103/revmodphys.88.045006} {\bibfield  {journal}
  {\bibinfo  {journal} {Rev. Mod. Phys.}\ }\textbf {\bibinfo {volume} {88}},\
  \bibinfo {pages} {045006} (\bibinfo {year} {2016})}\BibitemShut {NoStop}%
\bibitem [{\citenamefont {Bowick}\ \emph {et~al.}(2022)\citenamefont {Bowick},
  \citenamefont {Fakhri}, \citenamefont {Marchetti},\ and\ \citenamefont
  {Ramaswamy}}]{Bowick.2022}%
  \BibitemOpen
  \bibfield  {author} {\bibinfo {author} {\bibfnamefont {M.~J.}\ \bibnamefont
  {Bowick}}, \bibinfo {author} {\bibfnamefont {N.}~\bibnamefont {Fakhri}},
  \bibinfo {author} {\bibfnamefont {M.~C.}\ \bibnamefont {Marchetti}},\ and\
  \bibinfo {author} {\bibfnamefont {S.}~\bibnamefont {Ramaswamy}},\ }\bibfield
  {title} {\bibinfo {title} {{Symmetry, Thermodynamics, and Topology in Active
  Matter}},\ }\href {https://doi.org/10.1103/physrevx.12.010501} {\bibfield
  {journal} {\bibinfo  {journal} {Phys. Rev. X}\ }\textbf {\bibinfo {volume}
  {12}},\ \bibinfo {pages} {010501} (\bibinfo {year} {2022})}\BibitemShut
  {NoStop}%
\bibitem [{\citenamefont {Shaebani}\ \emph {et~al.}(2020)\citenamefont
  {Shaebani}, \citenamefont {Wysocki}, \citenamefont {Winkler}, \citenamefont
  {Gompper},\ and\ \citenamefont {Rieger}}]{Shaebani.2020}%
  \BibitemOpen
  \bibfield  {author} {\bibinfo {author} {\bibfnamefont {M.~R.}\ \bibnamefont
  {Shaebani}}, \bibinfo {author} {\bibfnamefont {A.}~\bibnamefont {Wysocki}},
  \bibinfo {author} {\bibfnamefont {R.~G.}\ \bibnamefont {Winkler}}, \bibinfo
  {author} {\bibfnamefont {G.}~\bibnamefont {Gompper}},\ and\ \bibinfo {author}
  {\bibfnamefont {H.}~\bibnamefont {Rieger}},\ }\bibfield  {title} {\bibinfo
  {title} {{Computational models for active matter}},\ }\href
  {https://doi.org/10.1038/s42254-020-0152-1} {\bibfield  {journal} {\bibinfo
  {journal} {Nat. Rev. Phys.}\ }\textbf {\bibinfo {volume} {2}},\ \bibinfo
  {pages} {181} (\bibinfo {year} {2020})}\BibitemShut {NoStop}%
\bibitem [{\citenamefont {Chaté}(2020)}]{Chate.2020}%
  \BibitemOpen
  \bibfield  {author} {\bibinfo {author} {\bibfnamefont {H.}~\bibnamefont
  {Chaté}},\ }\bibfield  {title} {\bibinfo {title} {{Dry Aligning Dilute
  Active Matter}},\ }\href
  {https://doi.org/10.1146/annurev-conmatphys-031119-050752} {\bibfield
  {journal} {\bibinfo  {journal} {Annu. Rev. Condens. Matter Phys.}\ }\textbf
  {\bibinfo {volume} {11}},\ \bibinfo {pages} {189} (\bibinfo {year}
  {2020})}\BibitemShut {NoStop}%
\bibitem [{\citenamefont {Kosterlitz}\ and\ \citenamefont
  {Thouless}(1973)}]{Kosterlitz.1973}%
  \BibitemOpen
  \bibfield  {author} {\bibinfo {author} {\bibfnamefont {J.~M.}\ \bibnamefont
  {Kosterlitz}}\ and\ \bibinfo {author} {\bibfnamefont {D.~J.}\ \bibnamefont
  {Thouless}},\ }\bibfield  {title} {\bibinfo {title} {{Ordering, metastability
  and phase transitions in two-dimensional systems}},\ }\href
  {https://doi.org/10.1088/0022-3719/6/7/010} {\bibfield  {journal} {\bibinfo
  {journal} {J. Phys. C}\ }\textbf {\bibinfo {volume} {6}},\ \bibinfo {pages}
  {1181} (\bibinfo {year} {1973})}\BibitemShut {NoStop}%
\bibitem [{\citenamefont {Vicsek}\ \emph {et~al.}(1995)\citenamefont {Vicsek},
  \citenamefont {Czirók}, \citenamefont {Jacob}, \citenamefont {Cohen},\ and\
  \citenamefont {Shochet}}]{Vicsek.1995}%
  \BibitemOpen
  \bibfield  {author} {\bibinfo {author} {\bibfnamefont {T.}~\bibnamefont
  {Vicsek}}, \bibinfo {author} {\bibfnamefont {A.}~\bibnamefont {Czirók}},
  \bibinfo {author} {\bibfnamefont {E.~B.}\ \bibnamefont {Jacob}}, \bibinfo
  {author} {\bibfnamefont {I.}~\bibnamefont {Cohen}},\ and\ \bibinfo {author}
  {\bibfnamefont {O.}~\bibnamefont {Shochet}},\ }\bibfield  {title} {\bibinfo
  {title} {{Novel Type of Phase Transition in a System of Self-Driven
  Particles}},\ }\href {https://doi.org/10.1103/physrevlett.75.1226} {\bibfield
   {journal} {\bibinfo  {journal} {Phys. Rev. Lett.}\ }\textbf {\bibinfo
  {volume} {75}},\ \bibinfo {pages} {1226 } (\bibinfo {year}
  {1995})}\BibitemShut {NoStop}%
\bibitem [{\citenamefont {Mermin}\ and\ \citenamefont
  {Wagner}(1966)}]{Mermin.1966}%
  \BibitemOpen
  \bibfield  {author} {\bibinfo {author} {\bibfnamefont {N.~D.}\ \bibnamefont
  {Mermin}}\ and\ \bibinfo {author} {\bibfnamefont {H.}~\bibnamefont
  {Wagner}},\ }\bibfield  {title} {\bibinfo {title} {{Absence of Ferromagnetism
  or Antiferromagnetism in One- or Two-Dimensional Isotropic Heisenberg
  Models}},\ }\href {https://doi.org/10.1103/physrevlett.17.1133} {\bibfield
  {journal} {\bibinfo  {journal} {Phys. Rev. Lett.}\ }\textbf {\bibinfo
  {volume} {17}},\ \bibinfo {pages} {1133} (\bibinfo {year}
  {1966})}\BibitemShut {NoStop}%
\bibitem [{\citenamefont {Nagy}\ \emph {et~al.}(2010)\citenamefont {Nagy},
  \citenamefont {Ákos}, \citenamefont {Biro},\ and\ \citenamefont
  {Vicsek}}]{Nagy.2010}%
  \BibitemOpen
  \bibfield  {author} {\bibinfo {author} {\bibfnamefont {M.}~\bibnamefont
  {Nagy}}, \bibinfo {author} {\bibfnamefont {Z.}~\bibnamefont {Ákos}},
  \bibinfo {author} {\bibfnamefont {D.}~\bibnamefont {Biro}},\ and\ \bibinfo
  {author} {\bibfnamefont {T.}~\bibnamefont {Vicsek}},\ }\bibfield  {title}
  {\bibinfo {title} {{Hierarchical group dynamics in pigeon flocks}},\ }\href
  {https://doi.org/10.1038/nature08891} {\bibfield  {journal} {\bibinfo
  {journal} {Nature}\ }\textbf {\bibinfo {volume} {464}},\ \bibinfo {pages}
  {890 } (\bibinfo {year} {2010})}\BibitemShut {NoStop}%
\bibitem [{\citenamefont {Barberis}\ and\ \citenamefont
  {Peruani}(2016)}]{Barberis.2016}%
  \BibitemOpen
  \bibfield  {author} {\bibinfo {author} {\bibfnamefont {L.}~\bibnamefont
  {Barberis}}\ and\ \bibinfo {author} {\bibfnamefont {F.}~\bibnamefont
  {Peruani}},\ }\bibfield  {title} {\bibinfo {title} {{Large-Scale Patterns in
  a Minimal Cognitive Flocking Model: Incidental Leaders, Nematic Patterns, and
  Aggregates}},\ }\href {https://doi.org/10.1103/physrevlett.117.248001}
  {\bibfield  {journal} {\bibinfo  {journal} {Phys. Rev. Lett.}\ }\textbf
  {\bibinfo {volume} {117}},\ \bibinfo {pages} {248001} (\bibinfo {year}
  {2016})}\BibitemShut {NoStop}%
\bibitem [{\citenamefont {Chen}\ \emph {et~al.}(2017)\citenamefont {Chen},
  \citenamefont {Patelli}, \citenamefont {Chaté}, \citenamefont {Ma},\ and\
  \citenamefont {Shi}}]{Chen.2017lo}%
  \BibitemOpen
  \bibfield  {author} {\bibinfo {author} {\bibfnamefont {Q.-s.}\ \bibnamefont
  {Chen}}, \bibinfo {author} {\bibfnamefont {A.}~\bibnamefont {Patelli}},
  \bibinfo {author} {\bibfnamefont {H.}~\bibnamefont {Chaté}}, \bibinfo
  {author} {\bibfnamefont {Y.-q.}\ \bibnamefont {Ma}},\ and\ \bibinfo {author}
  {\bibfnamefont {X.-q.}\ \bibnamefont {Shi}},\ }\bibfield  {title} {\bibinfo
  {title} {{Fore-aft asymmetric flocking}},\ }\href
  {https://doi.org/10.1103/physreve.96.020601} {\bibfield  {journal} {\bibinfo
  {journal} {Phys. Rev. E}\ }\textbf {\bibinfo {volume} {96}},\ \bibinfo
  {pages} {020601} (\bibinfo {year} {2017})}\BibitemShut {NoStop}%
\bibitem [{\citenamefont {Bastien}\ and\ \citenamefont
  {Romanczuk}(2020)}]{Bastien.2020}%
  \BibitemOpen
  \bibfield  {author} {\bibinfo {author} {\bibfnamefont {R.}~\bibnamefont
  {Bastien}}\ and\ \bibinfo {author} {\bibfnamefont {P.}~\bibnamefont
  {Romanczuk}},\ }\bibfield  {title} {\bibinfo {title} {{A model of collective
  behavior based purely on vision}},\ }\href
  {https://doi.org/10.1126/sciadv.aay0792} {\bibfield  {journal} {\bibinfo
  {journal} {Sci. Adv.}\ }\textbf {\bibinfo {volume} {6}},\ \bibinfo {pages}
  {eaay0792} (\bibinfo {year} {2020})}\BibitemShut {NoStop}%
\bibitem [{\citenamefont {Negi}\ \emph {et~al.}(2024)\citenamefont {Negi},
  \citenamefont {Winkler},\ and\ \citenamefont {Gompper}}]{Negi.2024}%
  \BibitemOpen
  \bibfield  {author} {\bibinfo {author} {\bibfnamefont {R.~S.}\ \bibnamefont
  {Negi}}, \bibinfo {author} {\bibfnamefont {R.~G.}\ \bibnamefont {Winkler}},\
  and\ \bibinfo {author} {\bibfnamefont {G.}~\bibnamefont {Gompper}},\
  }\bibfield  {title} {\bibinfo {title} {{Collective behavior of self-steering
  active particles with velocity alignment and visual perception}},\ }\href
  {https://doi.org/10.1103/physrevresearch.6.013118} {\bibfield  {journal}
  {\bibinfo  {journal} {Phys. Rev. Res.}\ }\textbf {\bibinfo {volume} {6}},\
  \bibinfo {pages} {013118} (\bibinfo {year} {2024})}\BibitemShut {NoStop}%
\bibitem [{\citenamefont {Tsyganov}\ \emph {et~al.}(2003)\citenamefont
  {Tsyganov}, \citenamefont {Brindley}, \citenamefont {Holden},\ and\
  \citenamefont {Biktashev}}]{Tsyganov.2003}%
  \BibitemOpen
  \bibfield  {author} {\bibinfo {author} {\bibfnamefont {M.~A.}\ \bibnamefont
  {Tsyganov}}, \bibinfo {author} {\bibfnamefont {J.}~\bibnamefont {Brindley}},
  \bibinfo {author} {\bibfnamefont {A.~V.}\ \bibnamefont {Holden}},\ and\
  \bibinfo {author} {\bibfnamefont {V.~N.}\ \bibnamefont {Biktashev}},\
  }\bibfield  {title} {\bibinfo {title} {{Quasisoliton Interaction of
  Pursuit-Evasion Waves in a Predator-Prey System}},\ }\href
  {https://doi.org/10.1103/physrevlett.91.218102} {\bibfield  {journal}
  {\bibinfo  {journal} {Phys. Rev. Lett.}\ }\textbf {\bibinfo {volume} {91}},\
  \bibinfo {pages} {218102} (\bibinfo {year} {2003})}\BibitemShut {NoStop}%
\bibitem [{\citenamefont {Fruchart}\ \emph {et~al.}(2021)\citenamefont
  {Fruchart}, \citenamefont {Hanai}, \citenamefont {Littlewood},\ and\
  \citenamefont {Vitelli}}]{fruchart2021non}%
  \BibitemOpen
  \bibfield  {author} {\bibinfo {author} {\bibfnamefont {M.}~\bibnamefont
  {Fruchart}}, \bibinfo {author} {\bibfnamefont {R.}~\bibnamefont {Hanai}},
  \bibinfo {author} {\bibfnamefont {P.~B.}\ \bibnamefont {Littlewood}},\ and\
  \bibinfo {author} {\bibfnamefont {V.}~\bibnamefont {Vitelli}},\ }\bibfield
  {title} {\bibinfo {title} {{Non-reciprocal phase transitions}},\ }\href
  {https://doi.org/10.1038/s41586-021-03375-9} {\bibfield  {journal} {\bibinfo
  {journal} {Nature}\ }\textbf {\bibinfo {volume} {592}},\ \bibinfo {pages}
  {363} (\bibinfo {year} {2021})}\BibitemShut {NoStop}%
\bibitem [{\citenamefont {Woo}\ \emph {et~al.}(2026)\citenamefont {Woo},
  \citenamefont {Noh},\ and\ \citenamefont {Rieger}}]{Woo.2026}%
  \BibitemOpen
  \bibfield  {author} {\bibinfo {author} {\bibfnamefont {C.-U.}\ \bibnamefont
  {Woo}}, \bibinfo {author} {\bibfnamefont {J.~D.}\ \bibnamefont {Noh}},\ and\
  \bibinfo {author} {\bibfnamefont {H.}~\bibnamefont {Rieger}},\ }\bibfield
  {title} {\bibinfo {title} {{Extensive Spatio-Temporal Chaos in Non-reciprocal
  Flocking}},\ }\bibfield  {journal} {\bibinfo  {journal} {arXiv}\ }\href
  {https://doi.org/10.48550/arxiv.2604.06462} {10.48550/arxiv.2604.06462}
  (\bibinfo {year} {2026}),\ \bibinfo {note} {arXiv:2604.06462},\ \Eprint
  {https://arxiv.org/abs/2604.06462} {arXiv:2604.06462} \BibitemShut {NoStop}%
\bibitem [{\citenamefont {Knežević}\ \emph {et~al.}(2022)\citenamefont
  {Knežević}, \citenamefont {Welker},\ and\ \citenamefont
  {Stark}}]{Knezevic.2022}%
  \BibitemOpen
  \bibfield  {author} {\bibinfo {author} {\bibfnamefont {M.}~\bibnamefont
  {Knežević}}, \bibinfo {author} {\bibfnamefont {T.}~\bibnamefont {Welker}},\
  and\ \bibinfo {author} {\bibfnamefont {H.}~\bibnamefont {Stark}},\ }\bibfield
   {title} {\bibinfo {title} {{Collective motion of active particles exhibiting
  non-reciprocal orientational interactions}},\ }\href
  {https://doi.org/10.1038/s41598-022-23597-9} {\bibfield  {journal} {\bibinfo
  {journal} {Sci. Rep.}\ }\textbf {\bibinfo {volume} {12}},\ \bibinfo {pages}
  {19437} (\bibinfo {year} {2022})}\BibitemShut {NoStop}%
\bibitem [{\citenamefont {Maity}\ and\ \citenamefont
  {Morin}(2023)}]{Maity.2023}%
  \BibitemOpen
  \bibfield  {author} {\bibinfo {author} {\bibfnamefont {S.}~\bibnamefont
  {Maity}}\ and\ \bibinfo {author} {\bibfnamefont {A.}~\bibnamefont {Morin}},\
  }\bibfield  {title} {\bibinfo {title} {{Spontaneous Demixing of Binary
  Colloidal Flocks}},\ }\href {https://doi.org/10.1103/physrevlett.131.178304}
  {\bibfield  {journal} {\bibinfo  {journal} {Phys. Rev. Lett.}\ }\textbf
  {\bibinfo {volume} {131}},\ \bibinfo {pages} {178304} (\bibinfo {year}
  {2023})}\BibitemShut {NoStop}%
\bibitem [{\citenamefont {Kreienkamp}\ and\ \citenamefont
  {Klapp}(2024)}]{Kreienkamp.2024}%
  \BibitemOpen
  \bibfield  {author} {\bibinfo {author} {\bibfnamefont {K.~L.}\ \bibnamefont
  {Kreienkamp}}\ and\ \bibinfo {author} {\bibfnamefont {S.~H.~L.}\ \bibnamefont
  {Klapp}},\ }\bibfield  {title} {\bibinfo {title} {{Nonreciprocal Alignment
  Induces Asymmetric Clustering in Active Mixtures}},\ }\href
  {https://doi.org/10.1103/physrevlett.133.258303} {\bibfield  {journal}
  {\bibinfo  {journal} {Phys. Rev. Lett.}\ }\textbf {\bibinfo {volume} {133}},\
  \bibinfo {pages} {258303} (\bibinfo {year} {2024})}\BibitemShut {NoStop}%
\bibitem [{\citenamefont {Kreienkamp}\ and\ \citenamefont
  {Klapp}(2025)}]{Kreienkamp.2025}%
  \BibitemOpen
  \bibfield  {author} {\bibinfo {author} {\bibfnamefont {K.~L.}\ \bibnamefont
  {Kreienkamp}}\ and\ \bibinfo {author} {\bibfnamefont {S.~H.~L.}\ \bibnamefont
  {Klapp}},\ }\bibfield  {title} {\bibinfo {title} {{Synchronization and
  exceptional points in nonreciprocal active polar mixtures}},\ }\href
  {https://doi.org/10.1038/s42005-025-02220-z} {\bibfield  {journal} {\bibinfo
  {journal} {Commun. Phys.}\ }\textbf {\bibinfo {volume} {8}},\ \bibinfo
  {pages} {307} (\bibinfo {year} {2025})}\BibitemShut {NoStop}%
\bibitem [{\citenamefont {Duan}\ \emph {et~al.}(2025)\citenamefont {Duan},
  \citenamefont {Agudo-Canalejo}, \citenamefont {Golestanian},\ and\
  \citenamefont {Mahault}}]{Duan.2025}%
  \BibitemOpen
  \bibfield  {author} {\bibinfo {author} {\bibfnamefont {Y.}~\bibnamefont
  {Duan}}, \bibinfo {author} {\bibfnamefont {J.}~\bibnamefont
  {Agudo-Canalejo}}, \bibinfo {author} {\bibfnamefont {R.}~\bibnamefont
  {Golestanian}},\ and\ \bibinfo {author} {\bibfnamefont {B.}~\bibnamefont
  {Mahault}},\ }\bibfield  {title} {\bibinfo {title} {{Phase coexistence in
  nonreciprocal quorum-sensing active matter}},\ }\href
  {https://doi.org/10.1103/physrevresearch.7.013234} {\bibfield  {journal}
  {\bibinfo  {journal} {Phys. Rev. Res.}\ }\textbf {\bibinfo {volume} {7}},\
  \bibinfo {pages} {013234} (\bibinfo {year} {2025})}\BibitemShut {NoStop}%
\bibitem [{\citenamefont {Mangeat}\ \emph {et~al.}(2025)\citenamefont
  {Mangeat}, \citenamefont {Chatterjee}, \citenamefont {Noh},\ and\
  \citenamefont {Rieger}}]{Mangeat.2025}%
  \BibitemOpen
  \bibfield  {author} {\bibinfo {author} {\bibfnamefont {M.}~\bibnamefont
  {Mangeat}}, \bibinfo {author} {\bibfnamefont {S.}~\bibnamefont {Chatterjee}},
  \bibinfo {author} {\bibfnamefont {J.~D.}\ \bibnamefont {Noh}},\ and\ \bibinfo
  {author} {\bibfnamefont {H.}~\bibnamefont {Rieger}},\ }\bibfield  {title}
  {\bibinfo {title} {{Emergent complex phases in a discrete flocking model with
  reciprocal and non-reciprocal interactions}},\ }\href
  {https://doi.org/10.1038/s42005-025-02098-x} {\bibfield  {journal} {\bibinfo
  {journal} {Commun. Phys.}\ }\textbf {\bibinfo {volume} {8}},\ \bibinfo
  {pages} {186} (\bibinfo {year} {2025})}\BibitemShut {NoStop}%
\bibitem [{\citenamefont {Lardet}\ \emph {et~al.}(2024)\citenamefont {Lardet},
  \citenamefont {Voituriez}, \citenamefont {Grigolon},\ and\ \citenamefont
  {Bertrand}}]{Lardet.2024}%
  \BibitemOpen
  \bibfield  {author} {\bibinfo {author} {\bibfnamefont {E.}~\bibnamefont
  {Lardet}}, \bibinfo {author} {\bibfnamefont {R.}~\bibnamefont {Voituriez}},
  \bibinfo {author} {\bibfnamefont {S.}~\bibnamefont {Grigolon}},\ and\
  \bibinfo {author} {\bibfnamefont {T.}~\bibnamefont {Bertrand}},\ }\bibfield
  {title} {\bibinfo {title} {{Disordered Yet Directed: The Emergence of Polar
  Flocks with Disordered Interactions}},\ }\href@noop {} {\bibfield  {journal}
  {\bibinfo  {journal} {arXiv}\ } (\bibinfo {year} {2024})},\ \Eprint
  {https://arxiv.org/abs/2409.10768} {2409.10768} \BibitemShut {NoStop}%
\bibitem [{\citenamefont {Choi}\ \emph {et~al.}(2025)\citenamefont {Choi},
  \citenamefont {Noh},\ and\ \citenamefont {Rieger}}]{Choi.2025}%
  \BibitemOpen
  \bibfield  {author} {\bibinfo {author} {\bibfnamefont {J.}~\bibnamefont
  {Choi}}, \bibinfo {author} {\bibfnamefont {J.~D.}\ \bibnamefont {Noh}},\ and\
  \bibinfo {author} {\bibfnamefont {H.}~\bibnamefont {Rieger}},\ }\bibfield
  {title} {\bibinfo {title} {{Flocking with random nonreciprocal
  interactions}},\ }\href {https://doi.org/10.1103/y26b-qfym} {\bibfield
  {journal} {\bibinfo  {journal} {Phys. Rev. Res.}\ }\textbf {\bibinfo {volume}
  {7}},\ \bibinfo {pages} {L042059} (\bibinfo {year} {2025})}\BibitemShut
  {NoStop}%
\bibitem [{\citenamefont {Saha}\ \emph {et~al.}(2020)\citenamefont {Saha},
  \citenamefont {Agudo-Canalejo},\ and\ \citenamefont
  {Golestanian}}]{Saha.2020}%
  \BibitemOpen
  \bibfield  {author} {\bibinfo {author} {\bibfnamefont {S.}~\bibnamefont
  {Saha}}, \bibinfo {author} {\bibfnamefont {J.}~\bibnamefont
  {Agudo-Canalejo}},\ and\ \bibinfo {author} {\bibfnamefont {R.}~\bibnamefont
  {Golestanian}},\ }\bibfield  {title} {\bibinfo {title} {Scalar active
  mixtures: The nonreciprocal cahn-hilliard model},\ }\href
  {https://doi.org/10.1103/physrevx.10.041009} {\bibfield  {journal} {\bibinfo
  {journal} {Phys. Rev. X}\ }\textbf {\bibinfo {volume} {10}},\ \bibinfo
  {pages} {041009} (\bibinfo {year} {2020})}\BibitemShut {NoStop}%
\bibitem [{\citenamefont {Dinelli}\ \emph {et~al.}(2023)\citenamefont
  {Dinelli}, \citenamefont {O’Byrne}, \citenamefont {Curatolo}, \citenamefont
  {Zhao}, \citenamefont {Sollich},\ and\ \citenamefont
  {Tailleur}}]{Dinelli.2023}%
  \BibitemOpen
  \bibfield  {author} {\bibinfo {author} {\bibfnamefont {A.}~\bibnamefont
  {Dinelli}}, \bibinfo {author} {\bibfnamefont {J.}~\bibnamefont {O’Byrne}},
  \bibinfo {author} {\bibfnamefont {A.}~\bibnamefont {Curatolo}}, \bibinfo
  {author} {\bibfnamefont {Y.}~\bibnamefont {Zhao}}, \bibinfo {author}
  {\bibfnamefont {P.}~\bibnamefont {Sollich}},\ and\ \bibinfo {author}
  {\bibfnamefont {J.}~\bibnamefont {Tailleur}},\ }\bibfield  {title} {\bibinfo
  {title} {{Non-reciprocity across scales in active mixtures}},\ }\href
  {https://doi.org/10.1038/s41467-023-42713-5} {\bibfield  {journal} {\bibinfo
  {journal} {Nat. Commun.}\ }\textbf {\bibinfo {volume} {14}},\ \bibinfo
  {pages} {7035} (\bibinfo {year} {2023})}\BibitemShut {NoStop}%
\bibitem [{\citenamefont {Brauns}\ and\ \citenamefont
  {Marchetti}(2024)}]{Brauns.2024}%
  \BibitemOpen
  \bibfield  {author} {\bibinfo {author} {\bibfnamefont {F.}~\bibnamefont
  {Brauns}}\ and\ \bibinfo {author} {\bibfnamefont {M.~C.}\ \bibnamefont
  {Marchetti}},\ }\bibfield  {title} {\bibinfo {title} {{Nonreciprocal Pattern
  Formation of Conserved Fields}},\ }\href
  {https://doi.org/10.1103/physrevx.14.021014} {\bibfield  {journal} {\bibinfo
  {journal} {Phys. Rev. X}\ }\textbf {\bibinfo {volume} {14}},\ \bibinfo
  {pages} {021014} (\bibinfo {year} {2024})}\BibitemShut {NoStop}%
\bibitem [{\citenamefont {Parkavousi}\ \emph {et~al.}(2025)\citenamefont
  {Parkavousi}, \citenamefont {Rana}, \citenamefont {Golestanian},\ and\
  \citenamefont {Saha}}]{Parkavousi.2025}%
  \BibitemOpen
  \bibfield  {author} {\bibinfo {author} {\bibfnamefont {L.}~\bibnamefont
  {Parkavousi}}, \bibinfo {author} {\bibfnamefont {N.}~\bibnamefont {Rana}},
  \bibinfo {author} {\bibfnamefont {R.}~\bibnamefont {Golestanian}},\ and\
  \bibinfo {author} {\bibfnamefont {S.}~\bibnamefont {Saha}},\ }\bibfield
  {title} {\bibinfo {title} {{Enhanced Stability and Chaotic Condensates in
  Multispecies Nonreciprocal Mixtures}},\ }\href
  {https://doi.org/10.1103/physrevlett.134.148301} {\bibfield  {journal}
  {\bibinfo  {journal} {Phys. Rev. Lett.}\ }\textbf {\bibinfo {volume} {134}},\
  \bibinfo {pages} {148301} (\bibinfo {year} {2025})}\BibitemShut {NoStop}%
\bibitem [{\citenamefont {Saha}\ and\ \citenamefont
  {Golestanian}(2025)}]{Saha.2025}%
  \BibitemOpen
  \bibfield  {author} {\bibinfo {author} {\bibfnamefont {S.}~\bibnamefont
  {Saha}}\ and\ \bibinfo {author} {\bibfnamefont {R.}~\bibnamefont
  {Golestanian}},\ }\bibfield  {title} {\bibinfo {title} {{Effervescence in a
  binary mixture with nonlinear non-reciprocal interactions}},\ }\href
  {https://doi.org/10.1038/s41467-025-61728-8} {\bibfield  {journal} {\bibinfo
  {journal} {Nat. Commun.}\ }\textbf {\bibinfo {volume} {16}},\ \bibinfo
  {pages} {7310} (\bibinfo {year} {2025})}\BibitemShut {NoStop}%
\bibitem [{\citenamefont {Meredith}\ \emph {et~al.}(2020)\citenamefont
  {Meredith}, \citenamefont {Moerman}, \citenamefont {Groenewold},
  \citenamefont {Chiu}, \citenamefont {Kegel}, \citenamefont {Blaaderen},\ and\
  \citenamefont {Zarzar}}]{Meredith.2020}%
  \BibitemOpen
  \bibfield  {author} {\bibinfo {author} {\bibfnamefont {C.~H.}\ \bibnamefont
  {Meredith}}, \bibinfo {author} {\bibfnamefont {P.~G.}\ \bibnamefont
  {Moerman}}, \bibinfo {author} {\bibfnamefont {J.}~\bibnamefont {Groenewold}},
  \bibinfo {author} {\bibfnamefont {Y.-J.}\ \bibnamefont {Chiu}}, \bibinfo
  {author} {\bibfnamefont {W.~K.}\ \bibnamefont {Kegel}}, \bibinfo {author}
  {\bibfnamefont {A.~v.}\ \bibnamefont {Blaaderen}},\ and\ \bibinfo {author}
  {\bibfnamefont {L.~D.}\ \bibnamefont {Zarzar}},\ }\bibfield  {title}
  {\bibinfo {title} {{Predator–prey interactions between droplets driven by
  non-reciprocal oil exchange}},\ }\href
  {https://doi.org/10.1038/s41557-020-00575-0} {\bibfield  {journal} {\bibinfo
  {journal} {Nature Chemistry}\ }\textbf {\bibinfo {volume} {12}},\ \bibinfo
  {pages} {1136} (\bibinfo {year} {2020})}\BibitemShut {NoStop}%
\bibitem [{\citenamefont {Chen}\ \emph {et~al.}(2024)\citenamefont {Chen},
  \citenamefont {Lei}, \citenamefont {Xiang}, \citenamefont {Duan},
  \citenamefont {Peng},\ and\ \citenamefont {Zhang}}]{Chen.2024}%
  \BibitemOpen
  \bibfield  {author} {\bibinfo {author} {\bibfnamefont {J.}~\bibnamefont
  {Chen}}, \bibinfo {author} {\bibfnamefont {X.}~\bibnamefont {Lei}}, \bibinfo
  {author} {\bibfnamefont {Y.}~\bibnamefont {Xiang}}, \bibinfo {author}
  {\bibfnamefont {M.}~\bibnamefont {Duan}}, \bibinfo {author} {\bibfnamefont
  {X.}~\bibnamefont {Peng}},\ and\ \bibinfo {author} {\bibfnamefont {H.~P.}\
  \bibnamefont {Zhang}},\ }\bibfield  {title} {\bibinfo {title} {{Emergent
  Chirality and Hyperuniformity in an Active Mixture with Nonreciprocal
  Interactions}},\ }\href {https://doi.org/10.1103/physrevlett.132.118301}
  {\bibfield  {journal} {\bibinfo  {journal} {Phys. Rev. Lett.}\ }\textbf
  {\bibinfo {volume} {132}},\ \bibinfo {pages} {118301} (\bibinfo {year}
  {2024})}\BibitemShut {NoStop}%
\bibitem [{\citenamefont {Woo}\ \emph {et~al.}(2025)\citenamefont {Woo},
  \citenamefont {Rieger},\ and\ \citenamefont {Noh}}]{Woo.20258m}%
  \BibitemOpen
  \bibfield  {author} {\bibinfo {author} {\bibfnamefont {C.-U.}\ \bibnamefont
  {Woo}}, \bibinfo {author} {\bibfnamefont {H.}~\bibnamefont {Rieger}},\ and\
  \bibinfo {author} {\bibfnamefont {J.~D.}\ \bibnamefont {Noh}},\ }\bibfield
  {title} {\bibinfo {title} {{Nonreciprocal yet Symmetric Multi-Species Active
  Matter: Emergence of Chirality and Species Separation}},\ }\href@noop {}
  {\bibfield  {journal} {\bibinfo  {journal} {arXiv}\ } (\bibinfo {year}
  {2025})},\ \Eprint {https://arxiv.org/abs/2512.18749} {2512.18749}
  \BibitemShut {NoStop}%
\bibitem [{\citenamefont {Wu}(1982)}]{Wu.1982}%
  \BibitemOpen
  \bibfield  {author} {\bibinfo {author} {\bibfnamefont {F.~Y.}\ \bibnamefont
  {Wu}},\ }\bibfield  {title} {\bibinfo {title} {{The Potts model}},\ }\href
  {https://doi.org/10.1103/revmodphys.54.235} {\bibfield  {journal} {\bibinfo
  {journal} {Rev. Mod. Phys.}\ }\textbf {\bibinfo {volume} {54}},\ \bibinfo
  {pages} {235} (\bibinfo {year} {1982})}\BibitemShut {NoStop}%
\bibitem [{\citenamefont {Tanaka}(2006)}]{Tanaka.2007}%
  \BibitemOpen
  \bibfield  {author} {\bibinfo {author} {\bibfnamefont {D.}~\bibnamefont
  {Tanaka}},\ }\bibfield  {title} {\bibinfo {title} {{General Chemotactic Model
  of Oscillators}},\ }\href {https://doi.org/10.1103/physrevlett.99.134103}
  {\bibfield  {journal} {\bibinfo  {journal} {Phys. Rev. Lett.}\ }\textbf
  {\bibinfo {volume} {99}},\ \bibinfo {pages} {134103} (\bibinfo {year}
  {2006})}\BibitemShut {NoStop}%
\bibitem [{\citenamefont {O’Keeffe}\ \emph {et~al.}(2017)\citenamefont
  {O’Keeffe}, \citenamefont {Hong},\ and\ \citenamefont
  {Strogatz}}]{OKeeffe.2017}%
  \BibitemOpen
  \bibfield  {author} {\bibinfo {author} {\bibfnamefont {K.~P.}\ \bibnamefont
  {O’Keeffe}}, \bibinfo {author} {\bibfnamefont {H.}~\bibnamefont {Hong}},\
  and\ \bibinfo {author} {\bibfnamefont {S.~H.}\ \bibnamefont {Strogatz}},\
  }\bibfield  {title} {\bibinfo {title} {{Oscillators that sync and swarm}},\
  }\href {https://doi.org/10.1038/s41467-017-01190-3} {\bibfield  {journal}
  {\bibinfo  {journal} {Nat. Commun.}\ }\textbf {\bibinfo {volume} {8}},\
  \bibinfo {pages} {1504} (\bibinfo {year} {2017})}\BibitemShut {NoStop}%
\bibitem [{\citenamefont {Hong}\ \emph {et~al.}(2023)\citenamefont {Hong},
  \citenamefont {O'Keeffe}, \citenamefont {Lee},\ and\ \citenamefont
  {Park}}]{Hong.2023}%
  \BibitemOpen
  \bibfield  {author} {\bibinfo {author} {\bibfnamefont {H.}~\bibnamefont
  {Hong}}, \bibinfo {author} {\bibfnamefont {K.~P.}\ \bibnamefont {O'Keeffe}},
  \bibinfo {author} {\bibfnamefont {J.~S.}\ \bibnamefont {Lee}},\ and\ \bibinfo
  {author} {\bibfnamefont {H.}~\bibnamefont {Park}},\ }\bibfield  {title}
  {\bibinfo {title} {{Swarmalators with thermal noise}},\ }\href
  {https://doi.org/10.1103/physrevresearch.5.023105} {\bibfield  {journal}
  {\bibinfo  {journal} {Phys. Rev. Res.}\ }\textbf {\bibinfo {volume} {5}},\
  \bibinfo {pages} {023105} (\bibinfo {year} {2023})}\BibitemShut {NoStop}%
\bibitem [{\citenamefont {Romanczuk}\ \emph {et~al.}(2012)\citenamefont
  {Romanczuk}, \citenamefont {B{\"a}r}, \citenamefont {Ebeling}, \citenamefont
  {Lindner},\ and\ \citenamefont {Schimansky-Geier}}]{Romanczuk.2012}%
  \BibitemOpen
  \bibfield  {author} {\bibinfo {author} {\bibfnamefont {P.}~\bibnamefont
  {Romanczuk}}, \bibinfo {author} {\bibfnamefont {M.}~\bibnamefont {B{\"a}r}},
  \bibinfo {author} {\bibfnamefont {W.}~\bibnamefont {Ebeling}}, \bibinfo
  {author} {\bibfnamefont {B.}~\bibnamefont {Lindner}},\ and\ \bibinfo {author}
  {\bibfnamefont {L.}~\bibnamefont {Schimansky-Geier}},\ }\bibfield  {title}
  {\bibinfo {title} {Active brownian particles},\ }\href
  {https://doi.org/10.1140/epjst/e2012-01529-y} {\bibfield  {journal} {\bibinfo
   {journal} {Eur. Phys. J. B}\ }\textbf {\bibinfo {volume} {202}},\ \bibinfo
  {pages} {1} (\bibinfo {year} {2012})}\BibitemShut {NoStop}%
\bibitem [{\citenamefont {Martín-Gómez}\ \emph {et~al.}(2018)\citenamefont
  {Martín-Gómez}, \citenamefont {Levis}, \citenamefont {Díaz-Guilera},\ and\
  \citenamefont {Pagonabarraga}}]{Martin-Gomez.2018}%
  \BibitemOpen
  \bibfield  {author} {\bibinfo {author} {\bibfnamefont {A.}~\bibnamefont
  {Martín-Gómez}}, \bibinfo {author} {\bibfnamefont {D.}~\bibnamefont
  {Levis}}, \bibinfo {author} {\bibfnamefont {A.}~\bibnamefont
  {Díaz-Guilera}},\ and\ \bibinfo {author} {\bibfnamefont {I.}~\bibnamefont
  {Pagonabarraga}},\ }\bibfield  {title} {\bibinfo {title} {{Collective motion
  of active Brownian particles with polar alignment}},\ }\href
  {https://doi.org/10.1039/c8sm00020d} {\bibfield  {journal} {\bibinfo
  {journal} {Soft Matter}\ }\textbf {\bibinfo {volume} {14}},\ \bibinfo {pages}
  {2610} (\bibinfo {year} {2018})}\BibitemShut {NoStop}%
\bibitem [{\citenamefont {Uchida}\ and\ \citenamefont
  {Golestanian}(2009)}]{Uchida.2009}%
  \BibitemOpen
  \bibfield  {author} {\bibinfo {author} {\bibfnamefont {N.}~\bibnamefont
  {Uchida}}\ and\ \bibinfo {author} {\bibfnamefont {R.}~\bibnamefont
  {Golestanian}},\ }\bibfield  {title} {\bibinfo {title} {{Synchronization and
  Collective Dynamics in a Carpet of Microfluidic Rotors}},\ }\href
  {https://doi.org/10.1103/physrevlett.104.178103} {\bibfield  {journal}
  {\bibinfo  {journal} {Phys. Rev. Lett.}\ }\textbf {\bibinfo {volume} {104}},\
  \bibinfo {pages} {178103} (\bibinfo {year} {2009})}\BibitemShut {NoStop}%
\bibitem [{Note1()}]{Note1}%
  \BibitemOpen
  \bibinfo {note} {The discrete-time dynamics of Eq.~\protect \textup {\hbox
  {\mathsurround \z@ \protect \normalfont (\ignorespaces \ref
  {eq:model_rule}\unskip \@@italiccorr )}} can be derived from a discrete-time
  dynamics for the internal phase variable.}\BibitemShut {Stop}%
\bibitem [{\citenamefont {Chatterjee}\ \emph {et~al.}(2023)\citenamefont
  {Chatterjee}, \citenamefont {Mangeat}, \citenamefont {Woo}, \citenamefont
  {Rieger},\ and\ \citenamefont {Noh}}]{Chatterjee.2023}%
  \BibitemOpen
  \bibfield  {author} {\bibinfo {author} {\bibfnamefont {S.}~\bibnamefont
  {Chatterjee}}, \bibinfo {author} {\bibfnamefont {M.}~\bibnamefont {Mangeat}},
  \bibinfo {author} {\bibfnamefont {C.-U.}\ \bibnamefont {Woo}}, \bibinfo
  {author} {\bibfnamefont {H.}~\bibnamefont {Rieger}},\ and\ \bibinfo {author}
  {\bibfnamefont {J.~D.}\ \bibnamefont {Noh}},\ }\bibfield  {title} {\bibinfo
  {title} {{Flocking of two unfriendly species: The two-species Vicsek
  model}},\ }\href {https://doi.org/10.1103/physreve.107.024607} {\bibfield
  {journal} {\bibinfo  {journal} {Phys. Rev. E}\ }\textbf {\bibinfo {volume}
  {107}},\ \bibinfo {pages} {024607} (\bibinfo {year} {2023})}\BibitemShut
  {NoStop}%
\bibitem [{\citenamefont {Bandini}\ \emph {et~al.}(2024)\citenamefont
  {Bandini}, \citenamefont {Venturelli}, \citenamefont {Loos}, \citenamefont
  {Jelic},\ and\ \citenamefont {Gambassi}}]{Bandini.2024}%
  \BibitemOpen
  \bibfield  {author} {\bibinfo {author} {\bibfnamefont {G.}~\bibnamefont
  {Bandini}}, \bibinfo {author} {\bibfnamefont {D.}~\bibnamefont {Venturelli}},
  \bibinfo {author} {\bibfnamefont {S.~A.~M.}\ \bibnamefont {Loos}}, \bibinfo
  {author} {\bibfnamefont {A.}~\bibnamefont {Jelic}},\ and\ \bibinfo {author}
  {\bibfnamefont {A.}~\bibnamefont {Gambassi}},\ }\bibfield  {title} {\bibinfo
  {title} {{The XY model with vision cone: non-reciprocal vs. reciprocal
  interactions}},\ }\href@noop {} {\bibfield  {journal} {\bibinfo  {journal}
  {arXiv}\ } (\bibinfo {year} {2024})},\ \Eprint
  {https://arxiv.org/abs/2412.19297} {2412.19297} \BibitemShut {NoStop}%
\bibitem [{\citenamefont {Dopierala}\ \emph {et~al.}(2025)\citenamefont
  {Dopierala}, \citenamefont {Chaté}, \citenamefont {Shi},\ and\ \citenamefont
  {Solon}}]{Dopierala.2025}%
  \BibitemOpen
  \bibfield  {author} {\bibinfo {author} {\bibfnamefont {D.}~\bibnamefont
  {Dopierala}}, \bibinfo {author} {\bibfnamefont {H.}~\bibnamefont {Chaté}},
  \bibinfo {author} {\bibfnamefont {X.-q.}\ \bibnamefont {Shi}},\ and\ \bibinfo
  {author} {\bibfnamefont {A.}~\bibnamefont {Solon}},\ }\bibfield  {title}
  {\bibinfo {title} {{Inescapable anisotropy of non-reciprocal XY models}},\
  }\href@noop {} {\bibfield  {journal} {\bibinfo  {journal} {arXiv}\ }
  (\bibinfo {year} {2025})},\ \Eprint {https://arxiv.org/abs/2503.14466}
  {2503.14466} \BibitemShut {NoStop}%
\bibitem [{\citenamefont {Popli}\ \emph {et~al.}(2025)\citenamefont {Popli},
  \citenamefont {Maitra},\ and\ \citenamefont {Ramaswamy}}]{Popli.2025}%
  \BibitemOpen
  \bibfield  {author} {\bibinfo {author} {\bibfnamefont {P.}~\bibnamefont
  {Popli}}, \bibinfo {author} {\bibfnamefont {A.}~\bibnamefont {Maitra}},\ and\
  \bibinfo {author} {\bibfnamefont {S.}~\bibnamefont {Ramaswamy}},\ }\bibfield
  {title} {\bibinfo {title} {{Don't look back: Ordering and defect cloaking in
  non-reciprocal lattice XY models}},\ }\href@noop {} {\bibfield  {journal}
  {\bibinfo  {journal} {arXiv}\ } (\bibinfo {year} {2025})},\ \Eprint
  {https://arxiv.org/abs/2503.06480} {2503.06480} \BibitemShut {NoStop}%
\bibitem [{\citenamefont {Ma}\ \emph {et~al.}(2025)\citenamefont {Ma},
  \citenamefont {Meng}, \citenamefont {Cheng},\ and\ \citenamefont
  {Wang}}]{Ma.2025}%
  \BibitemOpen
  \bibfield  {author} {\bibinfo {author} {\bibfnamefont {W.}~\bibnamefont
  {Ma}}, \bibinfo {author} {\bibfnamefont {F.}~\bibnamefont {Meng}}, \bibinfo
  {author} {\bibfnamefont {R.}~\bibnamefont {Cheng}},\ and\ \bibinfo {author}
  {\bibfnamefont {J.}~\bibnamefont {Wang}},\ }\bibfield  {title} {\bibinfo
  {title} {{Rich collective behaviors in nonreciprocal multispecies systems:
  The interplay between nonreciprocity and permutation symmetry among
  species}},\ }\href {https://doi.org/10.1103/s6df-sng9} {\bibfield  {journal}
  {\bibinfo  {journal} {Phys. Rev. E}\ }\textbf {\bibinfo {volume} {112}},\
  \bibinfo {pages} {024210} (\bibinfo {year} {2025})}\BibitemShut {NoStop}%
\bibitem [{\citenamefont {Kuramoto}(1975)}]{kuramoto1975lecture}%
  \BibitemOpen
  \bibfield  {author} {\bibinfo {author} {\bibfnamefont {Y.}~\bibnamefont
  {Kuramoto}},\ }\href@noop {} {\bibinfo {title} {Lecture notes in physics vol.
  39}} (\bibinfo {year} {1975})\BibitemShut {NoStop}%
\bibitem [{\citenamefont {Sakaguchi}\ and\ \citenamefont
  {Kuramoto}(1986)}]{Sakaguchi.1986}%
  \BibitemOpen
  \bibfield  {author} {\bibinfo {author} {\bibfnamefont {H.}~\bibnamefont
  {Sakaguchi}}\ and\ \bibinfo {author} {\bibfnamefont {Y.}~\bibnamefont
  {Kuramoto}},\ }\bibfield  {title} {\bibinfo {title} {{A Soluble Active
  Rotater Model Showing Phase Transitions via Mutual Entertainment}},\ }\href
  {https://doi.org/10.1143/ptp.76.576} {\bibfield  {journal} {\bibinfo
  {journal} {Prog. Theor. Phys.}\ }\textbf {\bibinfo {volume} {76}},\ \bibinfo
  {pages} {576} (\bibinfo {year} {1986})}\BibitemShut {NoStop}%
\bibitem [{\citenamefont {Abrams}\ \emph {et~al.}(2008)\citenamefont {Abrams},
  \citenamefont {Mirollo}, \citenamefont {Strogatz},\ and\ \citenamefont
  {Wiley}}]{Abrams.2008}%
  \BibitemOpen
  \bibfield  {author} {\bibinfo {author} {\bibfnamefont {D.~M.}\ \bibnamefont
  {Abrams}}, \bibinfo {author} {\bibfnamefont {R.}~\bibnamefont {Mirollo}},
  \bibinfo {author} {\bibfnamefont {S.~H.}\ \bibnamefont {Strogatz}},\ and\
  \bibinfo {author} {\bibfnamefont {D.~A.}\ \bibnamefont {Wiley}},\ }\bibfield
  {title} {\bibinfo {title} {{Solvable Model for Chimera States of Coupled
  Oscillators}},\ }\href {https://doi.org/10.1103/physrevlett.101.084103}
  {\bibfield  {journal} {\bibinfo  {journal} {Phys. Rev. Lett.}\ }\textbf
  {\bibinfo {volume} {101}},\ \bibinfo {pages} {084103} (\bibinfo {year}
  {2008})}\BibitemShut {NoStop}%
\bibitem [{\citenamefont {Pikovsky}\ and\ \citenamefont
  {Rosenblum}(2008)}]{Pikovsky.2008}%
  \BibitemOpen
  \bibfield  {author} {\bibinfo {author} {\bibfnamefont {A.}~\bibnamefont
  {Pikovsky}}\ and\ \bibinfo {author} {\bibfnamefont {M.}~\bibnamefont
  {Rosenblum}},\ }\bibfield  {title} {\bibinfo {title} {{Partially Integrable
  Dynamics of Hierarchical Populations of Coupled Oscillators}},\ }\href
  {https://doi.org/10.1103/physrevlett.101.264103} {\bibfield  {journal}
  {\bibinfo  {journal} {Phys. Rev. Lett.}\ }\textbf {\bibinfo {volume} {101}},\
  \bibinfo {pages} {264103} (\bibinfo {year} {2008})}\BibitemShut {NoStop}%
\bibitem [{\citenamefont {Kuramoto}\ and\ \citenamefont
  {Battogtokh}(2002)}]{Kuramoto.2002}%
  \BibitemOpen
  \bibfield  {author} {\bibinfo {author} {\bibfnamefont {Y.}~\bibnamefont
  {Kuramoto}}\ and\ \bibinfo {author} {\bibfnamefont {D.}~\bibnamefont
  {Battogtokh}},\ }\bibfield  {title} {\bibinfo {title} {{Coexistence of
  Coherence and Incoherence in Nonlocally Coupled Phase Oscillators}},\
  }\href@noop {} {\bibfield  {journal} {\bibinfo  {journal} {Nonlinear Phenom.
  Complex Syst.}\ }\textbf {\bibinfo {volume} {5}},\ \bibinfo {pages} {380}
  (\bibinfo {year} {2002})}\BibitemShut {NoStop}%
\bibitem [{\citenamefont {Bertin}\ \emph {et~al.}(2009)\citenamefont {Bertin},
  \citenamefont {Droz},\ and\ \citenamefont
  {Gr{\'e}goire}}]{bertinHydrodynamic2009}%
  \BibitemOpen
  \bibfield  {author} {\bibinfo {author} {\bibfnamefont {E.}~\bibnamefont
  {Bertin}}, \bibinfo {author} {\bibfnamefont {M.}~\bibnamefont {Droz}},\ and\
  \bibinfo {author} {\bibfnamefont {G.}~\bibnamefont {Gr{\'e}goire}},\
  }\bibfield  {title} {\bibinfo {title} {Hydrodynamic equations for
  self-propelled particles: Microscopic derivation and stability analysis},\
  }\href {https://doi.org/10.1088/1751-8113/42/44/445001} {\bibfield  {journal}
  {\bibinfo  {journal} {J. Phys. A}\ }\textbf {\bibinfo {volume} {42}},\
  \bibinfo {pages} {445001} (\bibinfo {year} {2009})}\BibitemShut {NoStop}%
\bibitem [{\citenamefont {Peshkov}\ \emph {et~al.}(2014)\citenamefont
  {Peshkov}, \citenamefont {Bertin}, \citenamefont {Ginelli},\ and\
  \citenamefont {Chat\'e}}]{Peshkov.2014}%
  \BibitemOpen
  \bibfield  {author} {\bibinfo {author} {\bibfnamefont {A.}~\bibnamefont
  {Peshkov}}, \bibinfo {author} {\bibfnamefont {E.}~\bibnamefont {Bertin}},
  \bibinfo {author} {\bibfnamefont {F.}~\bibnamefont {Ginelli}},\ and\ \bibinfo
  {author} {\bibfnamefont {H.}~\bibnamefont {Chat\'e}},\ }\bibfield  {title}
  {\bibinfo {title} {{Boltzmann-Ginzburg-Landau approach for continuous
  descriptions of generic Vicsek-like models}},\ }\href
  {https://doi.org/10.1140/epjst/e2014-02193-y} {\bibfield  {journal} {\bibinfo
   {journal} {Eur. Phys. J. Spec. Top.}\ }\textbf {\bibinfo {volume} {223}},\
  \bibinfo {pages} {1315 } (\bibinfo {year} {2014})}\BibitemShut {NoStop}%
\bibitem [{\citenamefont {Mahault}(2018)}]{mahault2018outstanding}%
  \BibitemOpen
  \bibfield  {author} {\bibinfo {author} {\bibfnamefont {B.}~\bibnamefont
  {Mahault}},\ }\emph {\bibinfo {title} {Outstanding problems in the
  statistical physics of active matter}},\ \href@noop {} {Ph.D. thesis},\
  \bibinfo  {school} {Universit{\'e} Paris Saclay (COmUE)} (\bibinfo {year}
  {2018})\BibitemShut {NoStop}%
\bibitem [{Note2()}]{Note2}%
  \BibitemOpen
  \bibinfo {note} {The Hadamard product $\protect \bm {a}\circ \protect \bm
  {b}$ can be written as $\protect \bm {a} \circ \protect \bm {b} = \protect
  \bm {D}(\protect \bm {a}) \protect \bm {b}$}\BibitemShut {NoStop}%
\bibitem [{Note3()}]{Note3}%
  \BibitemOpen
  \bibinfo {note} {{The chiral frequency may deviate from the formulae in
  Eqs.~\protect \textup {\hbox {\mathsurround \z@ \protect \normalfont
  (\ignorespaces \ref {eq:Win_MF}\unskip \@@italiccorr )}} and \protect \textup
  {\hbox {\mathsurround \z@ \protect \normalfont (\ignorespaces \ref
  {eq:Wout_MF}\unskip \@@italiccorr )}} far from the stability boundaries due
  to a nonlinear correction.}}\BibitemShut {Stop}%
\bibitem [{Note4()}]{Note4}%
  \BibitemOpen
  \bibinfo {note} {We remark that Eq.~\protect \textup {\hbox {\mathsurround
  \z@ \protect \normalfont (\ignorespaces \ref {eq:HydroSym_app}\unskip
  \@@italiccorr )}} reduces to a complex Ginzburg-Landau equation~\cite
  {Aranson.2002} when $\eta =\zeta =0$ and $\rho (\protect \bm {r},t) =
  1$.}\BibitemShut {Stop}%
\bibitem [{\citenamefont {García-Morales}\ and\ \citenamefont
  {Krischer}(2012)}]{Garcia-Morales.2012}%
  \BibitemOpen
  \bibfield  {author} {\bibinfo {author} {\bibfnamefont {V.}~\bibnamefont
  {García-Morales}}\ and\ \bibinfo {author} {\bibfnamefont {K.}~\bibnamefont
  {Krischer}},\ }\bibfield  {title} {\bibinfo {title} {{The complex
  Ginzburg–Landau equation: an introduction}},\ }\href
  {https://doi.org/10.1080/00107514.2011.642554} {\bibfield  {journal}
  {\bibinfo  {journal} {Contemp. Phys.}\ }\textbf {\bibinfo {volume} {53}},\
  \bibinfo {pages} {79} (\bibinfo {year} {2012})}\BibitemShut {NoStop}%
\bibitem [{\citenamefont {Strogatz}(2024)}]{Strogatz.2024}%
  \BibitemOpen
  \bibfield  {author} {\bibinfo {author} {\bibfnamefont {S.~H.}\ \bibnamefont
  {Strogatz}},\ }\href {https://doi.org/10.1201/9780429398490} {\emph {\bibinfo
  {title} {{Nonlinear Dynamics and Chaos, With Applications to Physics,
  Biology, Chemistry, and Engineering}}}},\ \bibinfo {edition} {3rd}\ ed.\
  (\bibinfo  {publisher} {Chapman and Hall},\ \bibinfo {address} {New York},\
  \bibinfo {year} {2024})\BibitemShut {NoStop}%
\bibitem [{Note5()}]{Note5}%
  \BibitemOpen
  \bibinfo {note} {{We discretized the Boltzmann equation using a uniform grid
  spacing $\Delta x = \Delta y = 1/16$ and a time step $\Delta t = 10^{-3}$ for
  a numerical study. Time integration was performed using a fourth-order
  Runge-Kutta scheme. The nonlinear terms were evaluated using a
  pseudo-spectral method, together with the standard $2/3$ dealiasing rule to
  reduce aliasing errors~\cite {fornberg1998practical, boyd2001chebyshev}. To
  avoid numerical instability, we included a diffusion term, $4 D_t \partial
  _z\partial _{z^*} \protect \bm {f}_k$ with $D_t=0.1$~\cite
  {mahault2018outstanding}.}}\BibitemShut {Stop}%
\bibitem [{SM()}]{SM}%
  \BibitemOpen
  \href@noop {} {}\bibinfo {note} {See Supplemental Material for
  Movies.}\BibitemShut {Stop}%
\bibitem [{\citenamefont {Fornberg}(1998)}]{fornberg1998practical}%
  \BibitemOpen
  \bibfield  {author} {\bibinfo {author} {\bibfnamefont {B.}~\bibnamefont
  {Fornberg}},\ }\href@noop {} {\emph {\bibinfo {title} {A practical guide to
  pseudospectral methods}}},\ \bibinfo {number} {1}\ (\bibinfo  {publisher}
  {Cambridge university press},\ \bibinfo {year} {1998})\BibitemShut {NoStop}%
\bibitem [{\citenamefont {Boyd}(2001)}]{boyd2001chebyshev}%
  \BibitemOpen
  \bibfield  {author} {\bibinfo {author} {\bibfnamefont {J.~P.}\ \bibnamefont
  {Boyd}},\ }\href@noop {} {\emph {\bibinfo {title} {Chebyshev and Fourier
  spectral methods}}}\ (\bibinfo  {publisher} {Courier Corporation},\ \bibinfo
  {year} {2001})\BibitemShut {NoStop}%
\bibitem [{\citenamefont {Berezinskii}(1971)}]{Berezinskii.1971}%
  \BibitemOpen
  \bibfield  {author} {\bibinfo {author} {\bibfnamefont {V.}~\bibnamefont
  {Berezinskii}},\ }\bibfield  {title} {\bibinfo {title} {Destruction of
  long-range order in one-dimensional and two-dimensional systems having a
  continuous symmetry group i. classical systems},\ }\href@noop {} {\bibfield
  {journal} {\bibinfo  {journal} {Sov. Phys. JETP}\ }\textbf {\bibinfo {volume}
  {32}},\ \bibinfo {pages} {493} (\bibinfo {year} {1971})}\BibitemShut
  {NoStop}%
\bibitem [{\citenamefont {Jos{\'e}}\ \emph {et~al.}(1977)\citenamefont
  {Jos{\'e}}, \citenamefont {Kadanoff}, \citenamefont {Kirkpatrick},\ and\
  \citenamefont {Nelson}}]{Jose.1977}%
  \BibitemOpen
  \bibfield  {author} {\bibinfo {author} {\bibfnamefont {J.~V.}\ \bibnamefont
  {Jos{\'e}}}, \bibinfo {author} {\bibfnamefont {L.~P.}\ \bibnamefont
  {Kadanoff}}, \bibinfo {author} {\bibfnamefont {S.}~\bibnamefont
  {Kirkpatrick}},\ and\ \bibinfo {author} {\bibfnamefont {D.~R.}\ \bibnamefont
  {Nelson}},\ }\bibfield  {title} {\bibinfo {title} {Renormalization, vortices,
  and symmetry-breaking perturbations in the two-dimensional planar model},\
  }\href@noop {} {\bibfield  {journal} {\bibinfo  {journal} {Phys. Rev. B}\
  }\textbf {\bibinfo {volume} {16}},\ \bibinfo {pages} {1217} (\bibinfo {year}
  {1977})}\BibitemShut {NoStop}%
\bibitem [{\citenamefont {Izyumov}\ and\ \citenamefont
  {Skryabin}(1988)}]{Izyumov.1988}%
  \BibitemOpen
  \bibfield  {author} {\bibinfo {author} {\bibfnamefont {Y.~A.}\ \bibnamefont
  {Izyumov}}\ and\ \bibinfo {author} {\bibfnamefont {Y.~N.}\ \bibnamefont
  {Skryabin}},\ }\href@noop {} {\emph {\bibinfo {title} {{Statistical Mechanics
  of Magnetically Ordered Systems}}}}\ (\bibinfo  {publisher}
  {Springer-Verlag},\ \bibinfo {year} {1988})\BibitemShut {NoStop}%
\bibitem [{\citenamefont {Toner}\ and\ \citenamefont {Tu}(1995)}]{Toner.1995}%
  \BibitemOpen
  \bibfield  {author} {\bibinfo {author} {\bibfnamefont {J.}~\bibnamefont
  {Toner}}\ and\ \bibinfo {author} {\bibfnamefont {Y.}~\bibnamefont {Tu}},\
  }\bibfield  {title} {\bibinfo {title} {{Long-Range Order in a Two-Dimensional
  Dynamical XY Model: How Birds Fly Together}},\ }\href
  {https://doi.org/10.1103/physrevlett.75.4326} {\bibfield  {journal} {\bibinfo
   {journal} {Phys. Rev. Lett.}\ }\textbf {\bibinfo {volume} {75}},\ \bibinfo
  {pages} {4326} (\bibinfo {year} {1995})}\BibitemShut {NoStop}%
\bibitem [{\citenamefont {Woo}\ and\ \citenamefont {Noh}(2024)}]{Woo.2024}%
  \BibitemOpen
  \bibfield  {author} {\bibinfo {author} {\bibfnamefont {C.-U.}\ \bibnamefont
  {Woo}}\ and\ \bibinfo {author} {\bibfnamefont {J.~D.}\ \bibnamefont {Noh}},\
  }\bibfield  {title} {\bibinfo {title} {{Nonequilibrium phase transitions in a
  Brownian p-state clock model}},\ }\href
  {https://doi.org/10.1103/physreve.109.014105} {\bibfield  {journal} {\bibinfo
   {journal} {Phys. Rev. E}\ }\textbf {\bibinfo {volume} {109}},\ \bibinfo
  {pages} {014105} (\bibinfo {year} {2024})}\BibitemShut {NoStop}%
\bibitem [{\citenamefont {Rouzaire}\ \emph {et~al.}(2025)\citenamefont
  {Rouzaire}, \citenamefont {Rahmani}, \citenamefont {Pagonabarraga},
  \citenamefont {Peruani},\ and\ \citenamefont {Levis}}]{Rouzaire.2025}%
  \BibitemOpen
  \bibfield  {author} {\bibinfo {author} {\bibfnamefont {Y.}~\bibnamefont
  {Rouzaire}}, \bibinfo {author} {\bibfnamefont {P.}~\bibnamefont {Rahmani}},
  \bibinfo {author} {\bibfnamefont {I.}~\bibnamefont {Pagonabarraga}}, \bibinfo
  {author} {\bibfnamefont {F.}~\bibnamefont {Peruani}},\ and\ \bibinfo {author}
  {\bibfnamefont {D.}~\bibnamefont {Levis}},\ }\bibfield  {title} {\bibinfo
  {title} {{Activity Leads to Topological Phase Transition in 2D Populations of
  Heterogeneous Oscillators}},\ }\href
  {https://doi.org/10.1103/physrevlett.134.188301} {\bibfield  {journal}
  {\bibinfo  {journal} {Phys. Rev. Lett.}\ }\textbf {\bibinfo {volume} {134}},\
  \bibinfo {pages} {188301} (\bibinfo {year} {2025})}\BibitemShut {NoStop}%
\bibitem [{\citenamefont {Mahault}\ \emph {et~al.}(2018)\citenamefont
  {Mahault}, \citenamefont {Jiang}, \citenamefont {Bertin}, \citenamefont {Ma},
  \citenamefont {Patelli}, \citenamefont {Shi},\ and\ \citenamefont
  {Chaté}}]{Mahault.2018z9b}%
  \BibitemOpen
  \bibfield  {author} {\bibinfo {author} {\bibfnamefont {B.}~\bibnamefont
  {Mahault}}, \bibinfo {author} {\bibfnamefont {X.-c.}\ \bibnamefont {Jiang}},
  \bibinfo {author} {\bibfnamefont {E.}~\bibnamefont {Bertin}}, \bibinfo
  {author} {\bibfnamefont {Y.-q.}\ \bibnamefont {Ma}}, \bibinfo {author}
  {\bibfnamefont {A.}~\bibnamefont {Patelli}}, \bibinfo {author} {\bibfnamefont
  {X.-q.}\ \bibnamefont {Shi}},\ and\ \bibinfo {author} {\bibfnamefont
  {H.}~\bibnamefont {Chaté}},\ }\bibfield  {title} {\bibinfo {title}
  {{Self-Propelled Particles with Velocity Reversals and Ferromagnetic
  Alignment: Active Matter Class with Second-Order Transition to
  Quasi-Long-Range Polar Order}},\ }\href
  {https://doi.org/10.1103/physrevlett.120.258002} {\bibfield  {journal}
  {\bibinfo  {journal} {Phys. Rev. Lett.}\ }\textbf {\bibinfo {volume} {120}},\
  \bibinfo {pages} {258002} (\bibinfo {year} {2018})}\BibitemShut {NoStop}%
\bibitem [{\citenamefont {Banerjee}\ \emph {et~al.}(2017)\citenamefont
  {Banerjee}, \citenamefont {Souslov}, \citenamefont {Abanov},\ and\
  \citenamefont {Vitelli}}]{Banerjee.2017}%
  \BibitemOpen
  \bibfield  {author} {\bibinfo {author} {\bibfnamefont {D.}~\bibnamefont
  {Banerjee}}, \bibinfo {author} {\bibfnamefont {A.}~\bibnamefont {Souslov}},
  \bibinfo {author} {\bibfnamefont {A.~G.}\ \bibnamefont {Abanov}},\ and\
  \bibinfo {author} {\bibfnamefont {V.}~\bibnamefont {Vitelli}},\ }\bibfield
  {title} {\bibinfo {title} {{Odd viscosity in chiral active fluids}},\ }\href
  {https://doi.org/10.1038/s41467-017-01378-7} {\bibfield  {journal} {\bibinfo
  {journal} {Nat. Commun.}\ }\textbf {\bibinfo {volume} {8}},\ \bibinfo {pages}
  {1573} (\bibinfo {year} {2017})}\BibitemShut {NoStop}%
\bibitem [{\citenamefont {Rozman}\ and\ \citenamefont
  {Yeomans}(2024)}]{Rozman.2024}%
  \BibitemOpen
  \bibfield  {author} {\bibinfo {author} {\bibfnamefont {J.}~\bibnamefont
  {Rozman}}\ and\ \bibinfo {author} {\bibfnamefont {J.~M.}\ \bibnamefont
  {Yeomans}},\ }\bibfield  {title} {\bibinfo {title} {{Cell Sorting in an
  Active Nematic Vertex Model}},\ }\href
  {https://doi.org/10.1103/physrevlett.133.248401} {\bibfield  {journal}
  {\bibinfo  {journal} {Phys. Rev. Lett.}\ }\textbf {\bibinfo {volume} {133}},\
  \bibinfo {pages} {248401} (\bibinfo {year} {2024})}\BibitemShut {NoStop}%
\bibitem [{\citenamefont {Graham}\ \emph {et~al.}(2024)\citenamefont {Graham},
  \citenamefont {Zhang},\ and\ \citenamefont {Yeomans}}]{Graham.2024}%
  \BibitemOpen
  \bibfield  {author} {\bibinfo {author} {\bibfnamefont {J.~N.}\ \bibnamefont
  {Graham}}, \bibinfo {author} {\bibfnamefont {G.}~\bibnamefont {Zhang}},\ and\
  \bibinfo {author} {\bibfnamefont {J.~M.}\ \bibnamefont {Yeomans}},\
  }\bibfield  {title} {\bibinfo {title} {{Cell sorting by active forces in a
  phase-field model of cell monolayers}},\ }\href
  {https://doi.org/10.1039/d3sm01033c} {\bibfield  {journal} {\bibinfo
  {journal} {Soft Matter}\ }\textbf {\bibinfo {volume} {20}},\ \bibinfo {pages}
  {2955} (\bibinfo {year} {2024})}\BibitemShut {NoStop}%
\bibitem [{\citenamefont {Riedel}\ \emph {et~al.}(2005)\citenamefont {Riedel},
  \citenamefont {Kruse},\ and\ \citenamefont {Howard}}]{Riedel.2005}%
  \BibitemOpen
  \bibfield  {author} {\bibinfo {author} {\bibfnamefont {I.~H.}\ \bibnamefont
  {Riedel}}, \bibinfo {author} {\bibfnamefont {K.}~\bibnamefont {Kruse}},\ and\
  \bibinfo {author} {\bibfnamefont {J.}~\bibnamefont {Howard}},\ }\bibfield
  {title} {\bibinfo {title} {{A Self-Organized Vortex Array of Hydrodynamically
  Entrained Sperm Cells}},\ }\href {https://doi.org/10.1126/science.1110329}
  {\bibfield  {journal} {\bibinfo  {journal} {Science}\ }\textbf {\bibinfo
  {volume} {309}},\ \bibinfo {pages} {300} (\bibinfo {year}
  {2005})}\BibitemShut {NoStop}%
\bibitem [{\citenamefont {Sumino}\ \emph {et~al.}(2012)\citenamefont {Sumino},
  \citenamefont {Nagai}, \citenamefont {Shitaka}, \citenamefont {Tanaka},
  \citenamefont {Yoshikawa}, \citenamefont {Chaté},\ and\ \citenamefont
  {Oiwa}}]{Sumino.2012}%
  \BibitemOpen
  \bibfield  {author} {\bibinfo {author} {\bibfnamefont {Y.}~\bibnamefont
  {Sumino}}, \bibinfo {author} {\bibfnamefont {K.~H.}\ \bibnamefont {Nagai}},
  \bibinfo {author} {\bibfnamefont {Y.}~\bibnamefont {Shitaka}}, \bibinfo
  {author} {\bibfnamefont {D.}~\bibnamefont {Tanaka}}, \bibinfo {author}
  {\bibfnamefont {K.}~\bibnamefont {Yoshikawa}}, \bibinfo {author}
  {\bibfnamefont {H.}~\bibnamefont {Chaté}},\ and\ \bibinfo {author}
  {\bibfnamefont {K.}~\bibnamefont {Oiwa}},\ }\bibfield  {title} {\bibinfo
  {title} {{Large-scale vortex lattice emerging from collectively moving
  microtubules}},\ }\href {https://doi.org/10.1038/nature10874} {\bibfield
  {journal} {\bibinfo  {journal} {Nature}\ }\textbf {\bibinfo {volume} {483}},\
  \bibinfo {pages} {448} (\bibinfo {year} {2012})}\BibitemShut {NoStop}%
\bibitem [{\citenamefont {Han}\ \emph {et~al.}(2020)\citenamefont {Han},
  \citenamefont {Kokot}, \citenamefont {Tovkach}, \citenamefont {Glatz},
  \citenamefont {Aranson},\ and\ \citenamefont {Snezhko}}]{Han.2020}%
  \BibitemOpen
  \bibfield  {author} {\bibinfo {author} {\bibfnamefont {K.}~\bibnamefont
  {Han}}, \bibinfo {author} {\bibfnamefont {G.}~\bibnamefont {Kokot}}, \bibinfo
  {author} {\bibfnamefont {O.}~\bibnamefont {Tovkach}}, \bibinfo {author}
  {\bibfnamefont {A.}~\bibnamefont {Glatz}}, \bibinfo {author} {\bibfnamefont
  {I.~S.}\ \bibnamefont {Aranson}},\ and\ \bibinfo {author} {\bibfnamefont
  {A.}~\bibnamefont {Snezhko}},\ }\bibfield  {title} {\bibinfo {title}
  {{Emergence of self-organized multivortex states in flocks of active
  rollers}},\ }\href {https://doi.org/10.1073/pnas.2000061117} {\bibfield
  {journal} {\bibinfo  {journal} {Proc. Natl. Acad. Sci.}\ }\textbf {\bibinfo
  {volume} {117}},\ \bibinfo {pages} {9706} (\bibinfo {year}
  {2020})}\BibitemShut {NoStop}%
\bibitem [{\citenamefont {Wioland}\ \emph {et~al.}(2013)\citenamefont
  {Wioland}, \citenamefont {Woodhouse}, \citenamefont {Dunkel}, \citenamefont
  {Kessler},\ and\ \citenamefont {Goldstein}}]{Wioland.2013}%
  \BibitemOpen
  \bibfield  {author} {\bibinfo {author} {\bibfnamefont {H.}~\bibnamefont
  {Wioland}}, \bibinfo {author} {\bibfnamefont {F.~G.}\ \bibnamefont
  {Woodhouse}}, \bibinfo {author} {\bibfnamefont {J.}~\bibnamefont {Dunkel}},
  \bibinfo {author} {\bibfnamefont {J.~O.}\ \bibnamefont {Kessler}},\ and\
  \bibinfo {author} {\bibfnamefont {R.~E.}\ \bibnamefont {Goldstein}},\
  }\bibfield  {title} {\bibinfo {title} {{Confinement Stabilizes a Bacterial
  Suspension into a Spiral Vortex}},\ }\href
  {https://doi.org/10.1103/physrevlett.110.268102} {\bibfield  {journal}
  {\bibinfo  {journal} {Phys. Rev. Lett.}\ }\textbf {\bibinfo {volume} {110}},\
  \bibinfo {pages} {268102} (\bibinfo {year} {2013})}\BibitemShut {NoStop}%
\bibitem [{\citenamefont {Opathalage}\ \emph {et~al.}(2019)\citenamefont
  {Opathalage}, \citenamefont {Norton}, \citenamefont {Juniper}, \citenamefont
  {Langeslay}, \citenamefont {Aghvami}, \citenamefont {Fraden},\ and\
  \citenamefont {Dogic}}]{Opathalage.2019}%
  \BibitemOpen
  \bibfield  {author} {\bibinfo {author} {\bibfnamefont {A.}~\bibnamefont
  {Opathalage}}, \bibinfo {author} {\bibfnamefont {M.~M.}\ \bibnamefont
  {Norton}}, \bibinfo {author} {\bibfnamefont {M.~P.~N.}\ \bibnamefont
  {Juniper}}, \bibinfo {author} {\bibfnamefont {B.}~\bibnamefont {Langeslay}},
  \bibinfo {author} {\bibfnamefont {S.~A.}\ \bibnamefont {Aghvami}}, \bibinfo
  {author} {\bibfnamefont {S.}~\bibnamefont {Fraden}},\ and\ \bibinfo {author}
  {\bibfnamefont {Z.}~\bibnamefont {Dogic}},\ }\bibfield  {title} {\bibinfo
  {title} {{Self-organized dynamics and the transition to turbulence of
  confined active nematics}},\ }\href {https://doi.org/10.1073/pnas.1816733116}
  {\bibfield  {journal} {\bibinfo  {journal} {Proc. Natl. Acad. Sci.}\ }\textbf
  {\bibinfo {volume} {116}},\ \bibinfo {pages} {4788} (\bibinfo {year}
  {2019})}\BibitemShut {NoStop}%
\bibitem [{\citenamefont {Nishiguchi}\ \emph {et~al.}(2025)\citenamefont
  {Nishiguchi}, \citenamefont {Shiratani}, \citenamefont {Takeuchi},\ and\
  \citenamefont {Aranson}}]{Nishiguchi.2025}%
  \BibitemOpen
  \bibfield  {author} {\bibinfo {author} {\bibfnamefont {D.}~\bibnamefont
  {Nishiguchi}}, \bibinfo {author} {\bibfnamefont {S.}~\bibnamefont
  {Shiratani}}, \bibinfo {author} {\bibfnamefont {K.~A.}\ \bibnamefont
  {Takeuchi}},\ and\ \bibinfo {author} {\bibfnamefont {I.~S.}\ \bibnamefont
  {Aranson}},\ }\bibfield  {title} {\bibinfo {title} {{Vortex reversal is a
  precursor of confined bacterial turbulence}},\ }\href
  {https://doi.org/10.1073/pnas.2414446122} {\bibfield  {journal} {\bibinfo
  {journal} {Proc. Natl. Acad. Sci.}\ }\textbf {\bibinfo {volume} {122}},\
  \bibinfo {pages} {e2414446122} (\bibinfo {year} {2025})}\BibitemShut
  {NoStop}%
\bibitem [{\citenamefont {Denk}\ \emph {et~al.}(2015)\citenamefont {Denk},
  \citenamefont {Huber}, \citenamefont {Reithmann},\ and\ \citenamefont
  {Frey}}]{Denk.2015}%
  \BibitemOpen
  \bibfield  {author} {\bibinfo {author} {\bibfnamefont {J.}~\bibnamefont
  {Denk}}, \bibinfo {author} {\bibfnamefont {L.}~\bibnamefont {Huber}},
  \bibinfo {author} {\bibfnamefont {E.}~\bibnamefont {Reithmann}},\ and\
  \bibinfo {author} {\bibfnamefont {E.}~\bibnamefont {Frey}},\ }\bibfield
  {title} {\bibinfo {title} {{Active Curved Polymers Form Vortex Patterns on
  Membranes}},\ }\href {https://doi.org/10.1103/physrevlett.116.178301}
  {\bibfield  {journal} {\bibinfo  {journal} {Phys. Rev. Lett.}\ }\textbf
  {\bibinfo {volume} {116}},\ \bibinfo {pages} {178301} (\bibinfo {year}
  {2015})}\BibitemShut {NoStop}%
\bibitem [{\citenamefont {Faluweki}\ \emph {et~al.}(2023)\citenamefont
  {Faluweki}, \citenamefont {Cammann}, \citenamefont {Mazza},\ and\
  \citenamefont {Goehring}}]{Faluweki.2023}%
  \BibitemOpen
  \bibfield  {author} {\bibinfo {author} {\bibfnamefont {M.~K.}\ \bibnamefont
  {Faluweki}}, \bibinfo {author} {\bibfnamefont {J.}~\bibnamefont {Cammann}},
  \bibinfo {author} {\bibfnamefont {M.~G.}\ \bibnamefont {Mazza}},\ and\
  \bibinfo {author} {\bibfnamefont {L.}~\bibnamefont {Goehring}},\ }\bibfield
  {title} {\bibinfo {title} {{Active Spaghetti: Collective Organization in
  Cyanobacteria}},\ }\href {https://doi.org/10.1103/physrevlett.131.158303}
  {\bibfield  {journal} {\bibinfo  {journal} {Phys. Rev. Lett.}\ }\textbf
  {\bibinfo {volume} {131}},\ \bibinfo {pages} {158303} (\bibinfo {year}
  {2023})}\BibitemShut {NoStop}%
\bibitem [{\citenamefont {Cammann}\ \emph {et~al.}(2024)\citenamefont
  {Cammann}, \citenamefont {Faluweki}, \citenamefont {Dambacher}, \citenamefont
  {Goehring},\ and\ \citenamefont {Mazza}}]{Cammann.2024}%
  \BibitemOpen
  \bibfield  {author} {\bibinfo {author} {\bibfnamefont {J.}~\bibnamefont
  {Cammann}}, \bibinfo {author} {\bibfnamefont {M.~K.}\ \bibnamefont
  {Faluweki}}, \bibinfo {author} {\bibfnamefont {N.}~\bibnamefont {Dambacher}},
  \bibinfo {author} {\bibfnamefont {L.}~\bibnamefont {Goehring}},\ and\
  \bibinfo {author} {\bibfnamefont {M.~G.}\ \bibnamefont {Mazza}},\ }\bibfield
  {title} {\bibinfo {title} {{Topological transition in filamentous
  cyanobacteria: from motion to structure}},\ }\href
  {https://doi.org/10.1038/s42005-024-01866-5} {\bibfield  {journal} {\bibinfo
  {journal} {Commun. Phys.}\ }\textbf {\bibinfo {volume} {7}},\ \bibinfo
  {pages} {376} (\bibinfo {year} {2024})}\BibitemShut {NoStop}%
\bibitem [{Note6()}]{Note6}%
  \BibitemOpen
  \bibinfo {note} {The bimodality coefficient $\beta _x$ of a random variable
  $x$ is defined as $(1+m_3^2(x))/m_4(x)$ where $m_3$ and $m_4$ are the
  skewness and the kurtosis~\cite {Knapp.2007}.}\BibitemShut {Stop}%
\bibitem [{\citenamefont {Knapp}(2007)}]{Knapp.2007}%
  \BibitemOpen
  \bibfield  {author} {\bibinfo {author} {\bibfnamefont {T.~R.}\ \bibnamefont
  {Knapp}},\ }\bibfield  {title} {\bibinfo {title} {{Bimodality Revisited}},\
  }\href {https://doi.org/10.22237/jmasm/1177992120} {\bibfield  {journal}
  {\bibinfo  {journal} {J. Mod. Appl. Stat. Methods}\ }\textbf {\bibinfo
  {volume} {6}},\ \bibinfo {pages} {8} (\bibinfo {year} {2007})}\BibitemShut
  {NoStop}%
\bibitem [{\citenamefont {Speck}\ \emph {et~al.}(2014)\citenamefont {Speck},
  \citenamefont {Bialké}, \citenamefont {Menzel},\ and\ \citenamefont
  {Löwen}}]{Speck.2014}%
  \BibitemOpen
  \bibfield  {author} {\bibinfo {author} {\bibfnamefont {T.}~\bibnamefont
  {Speck}}, \bibinfo {author} {\bibfnamefont {J.}~\bibnamefont {Bialké}},
  \bibinfo {author} {\bibfnamefont {A.~M.}\ \bibnamefont {Menzel}},\ and\
  \bibinfo {author} {\bibfnamefont {H.}~\bibnamefont {Löwen}},\ }\bibfield
  {title} {\bibinfo {title} {Effective cahn-hilliard equation for the phase
  separation of active brownian particles},\ }\href
  {https://doi.org/10.1103/physrevlett.112.218304} {\bibfield  {journal}
  {\bibinfo  {journal} {Phys. Rev. Lett.}\ }\textbf {\bibinfo {volume} {112}},\
  \bibinfo {pages} {218304} (\bibinfo {year} {2014})}\BibitemShut {NoStop}%
\bibitem [{\citenamefont {Wittkowski}\ \emph {et~al.}(2014)\citenamefont
  {Wittkowski}, \citenamefont {Tiribocchi}, \citenamefont {Stenhammar},
  \citenamefont {Allen}, \citenamefont {Marenduzzo},\ and\ \citenamefont
  {Cates}}]{Wittkowski.2014}%
  \BibitemOpen
  \bibfield  {author} {\bibinfo {author} {\bibfnamefont {R.}~\bibnamefont
  {Wittkowski}}, \bibinfo {author} {\bibfnamefont {A.}~\bibnamefont
  {Tiribocchi}}, \bibinfo {author} {\bibfnamefont {J.}~\bibnamefont
  {Stenhammar}}, \bibinfo {author} {\bibfnamefont {R.~J.}\ \bibnamefont
  {Allen}}, \bibinfo {author} {\bibfnamefont {D.}~\bibnamefont {Marenduzzo}},\
  and\ \bibinfo {author} {\bibfnamefont {M.~E.}\ \bibnamefont {Cates}},\
  }\bibfield  {title} {\bibinfo {title} {Scalar $\phi^4$ field theory for
  active-particle phase separation},\ }\href
  {https://doi.org/10.1038/ncomms5351} {\bibfield  {journal} {\bibinfo
  {journal} {Nat. Comm.}\ }\textbf {\bibinfo {volume} {5}},\ \bibinfo {pages}
  {4351} (\bibinfo {year} {2014})}\BibitemShut {NoStop}%
\bibitem [{\citenamefont {Tjhung}\ \emph {et~al.}(2018)\citenamefont {Tjhung},
  \citenamefont {Nardini},\ and\ \citenamefont {Cates}}]{Tjhung.2018}%
  \BibitemOpen
  \bibfield  {author} {\bibinfo {author} {\bibfnamefont {E.}~\bibnamefont
  {Tjhung}}, \bibinfo {author} {\bibfnamefont {C.}~\bibnamefont {Nardini}},\
  and\ \bibinfo {author} {\bibfnamefont {M.~E.}\ \bibnamefont {Cates}},\
  }\bibfield  {title} {\bibinfo {title} {Cluster phases and bubbly phase
  separation in active fluids: Reversal of the ostwald process},\ }\href
  {https://doi.org/10.1103/physrevx.8.031080} {\bibfield  {journal} {\bibinfo
  {journal} {Phys. Rev. X}\ }\textbf {\bibinfo {volume} {8}},\ \bibinfo {pages}
  {031080} (\bibinfo {year} {2018})}\BibitemShut {NoStop}%
\bibitem [{\citenamefont {Chen}\ \emph {et~al.}(2026)\citenamefont {Chen},
  \citenamefont {Lee},\ and\ \citenamefont {Toner}}]{Chen.2026}%
  \BibitemOpen
  \bibfield  {author} {\bibinfo {author} {\bibfnamefont {L.}~\bibnamefont
  {Chen}}, \bibinfo {author} {\bibfnamefont {C.~F.}\ \bibnamefont {Lee}},\ and\
  \bibinfo {author} {\bibfnamefont {J.}~\bibnamefont {Toner}},\ }\bibfield
  {title} {\bibinfo {title} {{Only the ambidextrous can flock: Two-dimensional
  chiral Malthusian flocks, time cholesterics, and the Kardar-Parisi-Zhang
  equation}},\ }\href {https://doi.org/10.1103/sj5y-8dsd} {\bibfield  {journal}
  {\bibinfo  {journal} {Phys. Rev. E}\ }\textbf {\bibinfo {volume} {113}},\
  \bibinfo {pages} {045419} (\bibinfo {year} {2026})}\BibitemShut {NoStop}%
\bibitem [{\citenamefont {Lauter}\ \emph {et~al.}(2017)\citenamefont {Lauter},
  \citenamefont {Mitra},\ and\ \citenamefont {Marquardt}}]{Lauter.2017}%
  \BibitemOpen
  \bibfield  {author} {\bibinfo {author} {\bibfnamefont {R.}~\bibnamefont
  {Lauter}}, \bibinfo {author} {\bibfnamefont {A.}~\bibnamefont {Mitra}},\ and\
  \bibinfo {author} {\bibfnamefont {F.}~\bibnamefont {Marquardt}},\ }\bibfield
  {title} {\bibinfo {title} {{From Kardar-Parisi-Zhang scaling to explosive
  desynchronization in arrays of limit-cycle oscillators}},\ }\href
  {https://doi.org/10.1103/physreve.96.012220} {\bibfield  {journal} {\bibinfo
  {journal} {Phys. Rev. E}\ }\textbf {\bibinfo {volume} {96}},\ \bibinfo
  {pages} {012220} (\bibinfo {year} {2017})}\BibitemShut {NoStop}%
\bibitem [{\citenamefont {Daviet}\ \emph {et~al.}(2025)\citenamefont {Daviet},
  \citenamefont {Zelle}, \citenamefont {Asadollahi},\ and\ \citenamefont
  {Diehl}}]{Daviet.2025}%
  \BibitemOpen
  \bibfield  {author} {\bibinfo {author} {\bibfnamefont {R.}~\bibnamefont
  {Daviet}}, \bibinfo {author} {\bibfnamefont {C.~P.}\ \bibnamefont {Zelle}},
  \bibinfo {author} {\bibfnamefont {A.}~\bibnamefont {Asadollahi}},\ and\
  \bibinfo {author} {\bibfnamefont {S.}~\bibnamefont {Diehl}},\ }\bibfield
  {title} {\bibinfo {title} {{Kardar-Parisi-Zhang Scaling in Time-Crystalline
  Matter}},\ }\href {https://doi.org/10.1103/pbtn-wsgv} {\bibfield  {journal}
  {\bibinfo  {journal} {Phys. Rev. Lett.}\ }\textbf {\bibinfo {volume} {135}},\
  \bibinfo {pages} {047101} (\bibinfo {year} {2025})}\BibitemShut {NoStop}%
\bibitem [{git()}]{github}%
  \BibitemOpen
  \href@noop {} {}\bibinfo {note}
  {\href{http://github.com/jdnoh/Q-NRVM}{http://github.com/jdnoh/Q-NRVM}}\BibitemShut
  {NoStop}%
\bibitem [{\citenamefont {Aranson}\ and\ \citenamefont
  {Kramer}(2002)}]{Aranson.2002}%
  \BibitemOpen
  \bibfield  {author} {\bibinfo {author} {\bibfnamefont {I.~S.}\ \bibnamefont
  {Aranson}}\ and\ \bibinfo {author} {\bibfnamefont {L.}~\bibnamefont
  {Kramer}},\ }\bibfield  {title} {\bibinfo {title} {{The world of the complex
  Ginzburg-Landau equation}},\ }\href
  {https://doi.org/10.1103/revmodphys.74.99} {\bibfield  {journal} {\bibinfo
  {journal} {Rev. Mod. Phys.}\ }\textbf {\bibinfo {volume} {74}},\ \bibinfo
  {pages} {99} (\bibinfo {year} {2002})}\BibitemShut {NoStop}%
\end{thebibliography}%

\end{document}